\pdfoutput=1
\pdfoutput=1
\documentclass[structabstract]{aa}  
\pdfoutput=1                                   
\usepackage{graphicx}
\usepackage{graphicx,color,layout}
\usepackage{epstopdf}
\usepackage{txfonts}
\usepackage{natbib}
\bibpunct{(}{)}{;}{a}{}{,} % to follow the A&A style
\usepackage{lscape}

\begin{document}
   \title{The Gaia-ESO Survey: the chemical structure of the Galactic discs from the first internal data release \thanks{Based on observations collected at ESO telescopes under 
Gaia-ESO survey programme.
}
}

   \author{\v{S}. Mikolaitis
          \inst{1,2}
          \and
	  V.~Hill
	  \inst{1}
          \and
	  A.~Recio--Blanco
          \inst{1}
          \and
          P.~de~Laverny
	  \inst{1}
	  C. Allende Prieto
	  \inst{9}
          \and	
      G. Kordopatis
      \inst{3}
          \and	
      G. Tautvai\v{s}iene
      \inst{2}        
D.~Romano	 \inst{10}  \and
G.~Gilmore       \inst{3}  \and
S.~Randich       \inst{4}  \and
S.~Feltzing      \inst{5}  \and
G.~Micela        \inst{6}  \and
A.~Vallenari     \inst{7}  \and
E.~J.~Alfaro     \inst{8}  \and
% C.~Allende~Prieto\inst{8}  \and
T.~Bensby        \inst{5}  \and
A.~Bragaglia     \inst{10}  \and
E.~Flaccomio     \inst{6}  \and
A.~C.~Lanzafame  \inst{11}  \and
E.~Pancino       \inst{10,12}  \and
R.~Smiljanic     \inst{13,14}  \and
M.~Bergemann     \inst{3}  \and
G.~Carraro       \inst{15}  \and
M.~T. Costado    \inst{8}  \and
F.~Damiani       \inst{6}  \and
A.~Hourihane     \inst{3}  \and
P.~Jofr\'e       \inst{3}  \and
C.~Lardo         \inst{10}  \and
L.~Magrini       \inst{4}  \and
E.~Maiorca       \inst{4}  \and
L.~Morbidelli    \inst{4}  \and
L.~Sbordone      \inst{16}  \and
S.~G.~Sousa      \inst{17,18}  \and
C.~C.~Worley     \inst{3}  \and
S.~Zaggia        \inst{7} 
      %\and
	  %ESO Members
	  %\inst{2}
	  %\fnmsep\thanks{Thanks to ESO,OCA,CNES}
          }
\institute{Laboratoire Lagrange (UMR7293), Universit\'e de Nice Sophia Antipolis, CNRS, Observatoire de la C\^ote d'Azur, BP 4229, F-06304 Nice Cedex 04, France  
              \email{Sarunas.Mikolaitis@oca.eu}   
\and        Institute of Theoretical Physics and Astronomy, Vilnius University, A. Go\v{s}tauto 12, LT-01108 Vilnius, Lithuania                        
\and        Institute of Astronomy, University of Cambridge, Madingley Road, Cambridge CB3 0HA, United Kingdom 
\and        INAF - Osservatorio Astrofisico di Arcetri, Largo E. Fermi 5, 50125, Florence, Italy 
\and        Lund Observatory, Department of Astronomy and Theoretical Physics, Box 43, SE-221 00 Lund, Sweden 
\and        INAF - Osservatorio Astronomico di Palermo, Piazza del Parlamento 1, 90134, Palermo, Italy 
\and        INAF - Padova Observatory, Vicolo dell'Osservatorio 5, 35122 Padova, Italy 
\and        Instituto de Astrof\'{i}sica de Andaluc\'{i}a-CSIC, Apdo. 3004, 18080, Granada, Spain 
\and        Instituto de Astrof\'{\i}sica de Canarias, E-38205 La Laguna, Tenerife, Spain
\and        INAF - Osservatorio Astronomico di Bologna, via Ranzani 1, 40127, Bologna, Italy 
\and        Dipartimento di Fisica e Astronomia, Sezione Astrofisica, Universit\^{a} di Catania, via S. Sofia 78, 95123, Catania, Italy 
\and        ASI Science Data Center, Via del Politecnico SNC, 00133 Roma, Italy 
\and        Department for Astrophysics, Nicolaus Copernicus Astronomical Center, ul. Rabia\'{n}ska 8, 87-100 Toru\'{n}, Poland 
\and        European Southern Observatory, Karl-Schwarzschild-Str. 2, 85748 Garching bei M\"unchen, Germany 
\and        European Southern Observatory, Alonso de Cordova 3107 Vitacura, Santiago de Chile, Chile 
\and        ZAH - Landessternwarte Heidelberg, K\~{a}nigstuhl 12, D-69117, Heidelberg, Germany 
\and        Centro de Astrof\'isica, Universidade do Porto, Rua das Estrelas, 4150-762 Porto, Portugal 
\and        Departamento de F\'isica e Astronomia, Faculdade de Ci\^encias, Universidade do Porto, Rua do Campo Alegre, 4169-007 Porto, Portugal 
             }

   \date{Accepted to A\&A}

\authorrunning{Name Surname}

% \abstract{}{}{}{}{} 
% 5 {} token are mandatory
 
  \abstract
  {} 
   {Until recently, most high-resolution spectroscopic studies of the Galactic thin and thick discs were mostly confined to objects in the solar vicinity. Here we aim at enlarging the volume in which individual chemical abundances are used to characterise the thin and thick discs, using the first internal data release of the Gaia-ESO survey (GES iDR1). 
   }
   {We used the spectra of around 2\,000 FGK dwarfs and giants from the GES iDR1, obtained at resolutions of up to R$\sim$20\,000  with the FLAMES/GIRAFFE spectrograph. We derive and discuss the abundances of eight elements (Mg, Al, Si, Ca, Ti, Fe, Cr, Ni, and Y).
   }
     {
We show that the trends of these elemental abundances with iron are very similar to those in the solar neighbourhood. We find a natural division between $\alpha$-rich and $\alpha$-poor stars, best seen in the bimodality of the [Mg/M] distributions in bins of metallicity, which we attribute to thick- and thin-disc sequences, respectively. This separation is visible for most $\alpha$-elements and for aluminium. With the possible exception of Al, the observed dispersion around the trends is well described by the expected errors, leaving little room for astrophysical dispersion.
Using previously derived distances from Recio-Blanco et al. (2014b) for our sample, we further find that the thick-disc is more extended vertically and is more centrally concentrated towards the inner Galaxy than the thin-disc, which indicates a shorter scale-length. We derive the radial (4~to~12~kpc) and vertical (0~to~3.5~kpc) gradients in metallicity, iron, four $\alpha$-element abundances, and aluminium for the two populations, taking into account the identified correlation between $R_{\rm GC}$ and $|Z|$.
Similarly to other works, a radial metallicity gradient is found in the thin disc. The positive radial individual [$\alpha$/M] gradients found are at variance from the gradients observed in the RAVE survey. The thin disc also hosts a negative vertical metallicity gradient in the solar cylinder, accompanied by positive individual [$\alpha$/M] and [Al/M] gradients. 
The thick-disc, on the other hand, presents no radial metallicity gradient, a shallower vertical metallicity gradient than the thin-disc, an $\alpha$-elements-to-iron radial gradient in the opposite sense than that of the thin disc, and positive vertical individual [$\alpha$/M] and [Al/M] gradients.
We examine several thick-disc formation scenarii in the light of these radial and vertical trends.
   }
   {}


 \titlerunning{}
 \authorrunning{}              
   \maketitle
%________________________________________________________________

\section{Introduction}\label{sec:introduction}

The overall chemical composition of stars tracks the chemical composition of the interstellar matter from which they were born. Coupling this information with kinematics for large samples of stars in our Milky Way is a powerful tool for distinguishing the various stellar populations that compose our Galaxy, and most importantly, to understand their origin. The formation of our Galactic disc, or rather our Galactic discs, despite being the most massive visible component, is still a matter of vibrant debate, and the origin of the thick disc is one of the  most important questions that remain to be answered.

Our Galaxy has been suggested from the results of star counts to host a thick disc in addition to its thin disc (\citealt{Gilmore1983}). Thick discs seem to be ubiquitous in late-type galaxies \citep[e.g.][]{Yoachim2006}, but their dominant formation mechanism is not known, and a variety of processes have been proposed. 
The formation scenarios can be categorized into four broad categories: (i) the heating of a pre-existing thin disc by a violent merger was proposed by  \citet{Quinn1993} and further studied by many authors with variations \citep[e.g.][]{Kazantzidis2008, Villalobos2008, Qu2011}; (ii) the merger of small satellites that deposit their stars into a thick disc was proposed by \citet{Abadi2003}; (iii) the formation of a thick disc in situ following a strong accretion of gas, either from a wet merger \citep{Brook2004} or from gas filaments forming a turbulent clumpy disc 
at high redshifts \citep{Bournaud2009}; (iv) the mere radial rearrangement of the disc via radial mixing \citep[e.g.][]{Schonrich2009a, Schonrich2009b}, triggered by resonant scattering with transient spiral arms \citep[e.g.][]{Roskar2008} or by a resonance overlap of the bar and spiral structure \citep{Minchev2010}, or even triggered by a merger \citep{Minchev2013}. These various mechanisms leave behind specific structural, chemical, and kinematical signatures that can be deciphered by large enough (statistically significant) and broad enough (Galaxy-wide as opposed to located in the solar vicinity) stellar samples are accessible. 

Using detailed chemical abundances in stellar samples in the solar vicinity to probe the Galactic disc(s) evolution is not a new topic; today it has reached exquisite accuracies on statistically significant samples (see for example the very recent works by \citealt{Fuhrmann2011}, \citealt{Adibekyan2012} and \citealt{Bensby2014} who used samples of several hundred to almost a thousand stars each). From these works, a particularly striking result is the clear distinction of thick- and thin-disc stars, which either have been identified kinematically (e.g.~\citealt{Bensby2005}) and then shown to be 
chemically distinct, or identified chemically and then shown to present different kinematical properties \citep[e.g.][]{Adibekyan2012}. In the solar neighbourhood, 
the thick disc is thus demonstrated to be a kinematically hotter population \citep[e.g.][]{Casetti-Dinescu2011}, lagging the thin-disc rotation by about 50 km/s, with a lower mean 
metallicity \citep[$\rm{[Fe/H]}\simeq -0.58$ dex to be compared with $\simeq -0.03$~dex for the thin disc,][]{Fuhrmann2011} and enhanced in $\alpha$-elements with respect to the thin disc 
\citep[e.g.][]{Fuhrmann2011, Adibekyan2012, Bensby2014}. Furthermore, this population is older than that of the thin disc and shows a relatively tight age-metallicity relation 
\citep{Fuhrmann1998,Haywood2013,Bensby2014, Bergemann2014}. Constraints from the solar neighbourhood already sketch a detailed picture of the nature of the thin and thick discs, but the large-scale constraints are also of importance to understand the origin of populations. 

Large spectroscopic surveys of stars thus play an increasingly important role in our understanding of Galactic populations, and in particular the thin and thick discs. 
The first very large spectro-photometric survey was the Geneva Copenhagen Survey (\citealt{Nordstrom2004}), where radial velocities of stars were analysed jointly with photometric metallicities for about 16\,000 FGK stars observed by the Hipparcos satellite data. This survey is mostly local, but with very good statistics. 
The Sloan Extension for Galactic Understanding and Exploration (SEGUE) was the first very large scale spectroscopic survey reaching far from the solar neighbourhood (g$\sim$14-19~mag), where 
stellar parameters and metallicities are derived from R=2\,000 spectra in a wide wavelength range for about 240\,000 stars \citep{Yanny2009}. Moreover, \citet{Lee2011} provided [$\alpha$/Fe] and kinematics of  part of the  SEGUE sample (17\,277 G-type dwarfs). At the moment, SEGUE:DR10 includes 735,484 stellar spectra (see www.sdss3.org/dr10).
The Radial Velocity Experiment survey (RAVE) concentrated on a narrow region around the \ion{Ca}{II} triplet at slightly higher resolution (R$\simeq$7500) 
and obtained spectra of half a million relatively bright stars \citep[I$\sim$9-13,][]{Steinmetz2012, Kordopatis2013}, deriving stellar parameters and metallicities. Moreover, RAVE \citep[see][]{Boeche2013} derived individual chemical elemental abundances for such fraction of their sample where signal-to-noise ratio (S/N) is higher than~40.
The Apache Point Observatory Galactic Evolution Experiment (APOGEE) is a very large high-resolution (R$\simeq$22\,500) survey, concentrated on atmospheric parameters and abundances of giant stars \citep{Anders2013, Schlesinger2012, Hayden2013}.
From these surveys, a large-scale characteristics of the thin and thick discs started to emerge: the thick disc occupies (in agreement with its hotter kinematics) a larger vertical volume around the midplane \citep{Juric2008, Ivezic2008}, perhaps at a shorter scale-length than the thin disc \citep[][]{Bensby2011, Cheng2012b}, and shows no radial metallicity gradient \citep[][]{Cheng2012a}. The scale heights and scale lengths are suggested to be independent of metallicity \citep{Kordopatis2011}. 

The Gaia-ESO survey (GES) is a further step in collecting large spectroscopic data, obtaining high-resolution spectra of field and open cluster stars across the Galaxy \citep[see][]{Gilmore2012, Randich2013}. GES takes advantage of the VLT~FLAMES multi-fiber facility, which simultaneously feeds the GIRAFFE spectrograph (R$\sim$20\,000) and the UVES spectrograph (R$\sim$43\,000). Because of the multiplexing capabilities of each instrument (130 vs 8), the larger part of the sample is obtained with GIRAFFE, aiming at a total sample of $\sim$100\,000 stars across the main stellar populations of our Galaxy. From these high-resolution spectra radial velocities, stellar parameters, and detailed abundances are derived with GIRAFFE \citep{Blanco2014b} and UVES \citep{Smiljanic2014}. 

A first paper using the UVES part of the GES first internal data relase (iDR1) examined the age-metallicity relations beyond the solar neighbourhood \citep{Bergemann2014}. The GIRAFFE field stars of iDR1 have been used in \citet{Blanco2014a} (the first paper from this series) to show that the sample naturally separates chemically into an $\alpha$-rich and an $\alpha$-poor population that display different kinematical behaviours. In the present paper we investigate this route in more detail using elemental abundances, and also examine Galactic disc(s) gradients.

The iDR1 of the Gaia-ESO survey \citep{Blanco2014b} contains $\sim$10\,000 GIRAFFE spectra (R$\sim$20\,000 and R$\sim$16\,000). 
We here we present detailed abundances based on part of the iDR1, selecting only stars 
observed in HR10 and HR21 setups with a S/N higher than 15.
Using a carefully selected line list, 
we determined the abundances of nine elements (Mg, Al, Si, Ca, Ti, Fe, Cr, Ni, and Y) for spectra of field F, G, K stars. We studied $\sim$2\,000 individual stars located at a wide range of galactocentric distances and heights above the mid-plane. 
The paper is organised as follows:
in Section~\ref{sec:sample_method}, we provide the main characteristics of the observational data and discuss the method of chemical abundance determination and possible errors of analysis. 
In Section~\ref{sec:properties_of_disc} we discuss the results of the analysis and compare them with other studies.
In Section~\ref{sec:metalpoor} we concentrate on metal-poor stars in the sample.
Section~\ref{sec:gradients} presents radial and vertical gradients in the discs and discusses the results in the context of Galactic evolution and disc-formation scenarii. 
Section~\ref{sec:conclusions} concludes this paper.
Together with \citet{Blanco2014a}, this is the first effort to describe the detailed chemical structure of the Galactic disc based on the Gaia-ESO Survey first internal data-release of GIRAFFE spectra.

\section{Stellar sample: atmospheric parameters and chemical abundances}
\label{sec:sample_method}

\begin{figure}[!htb]
 \centering
 \includegraphics[scale=0.31]{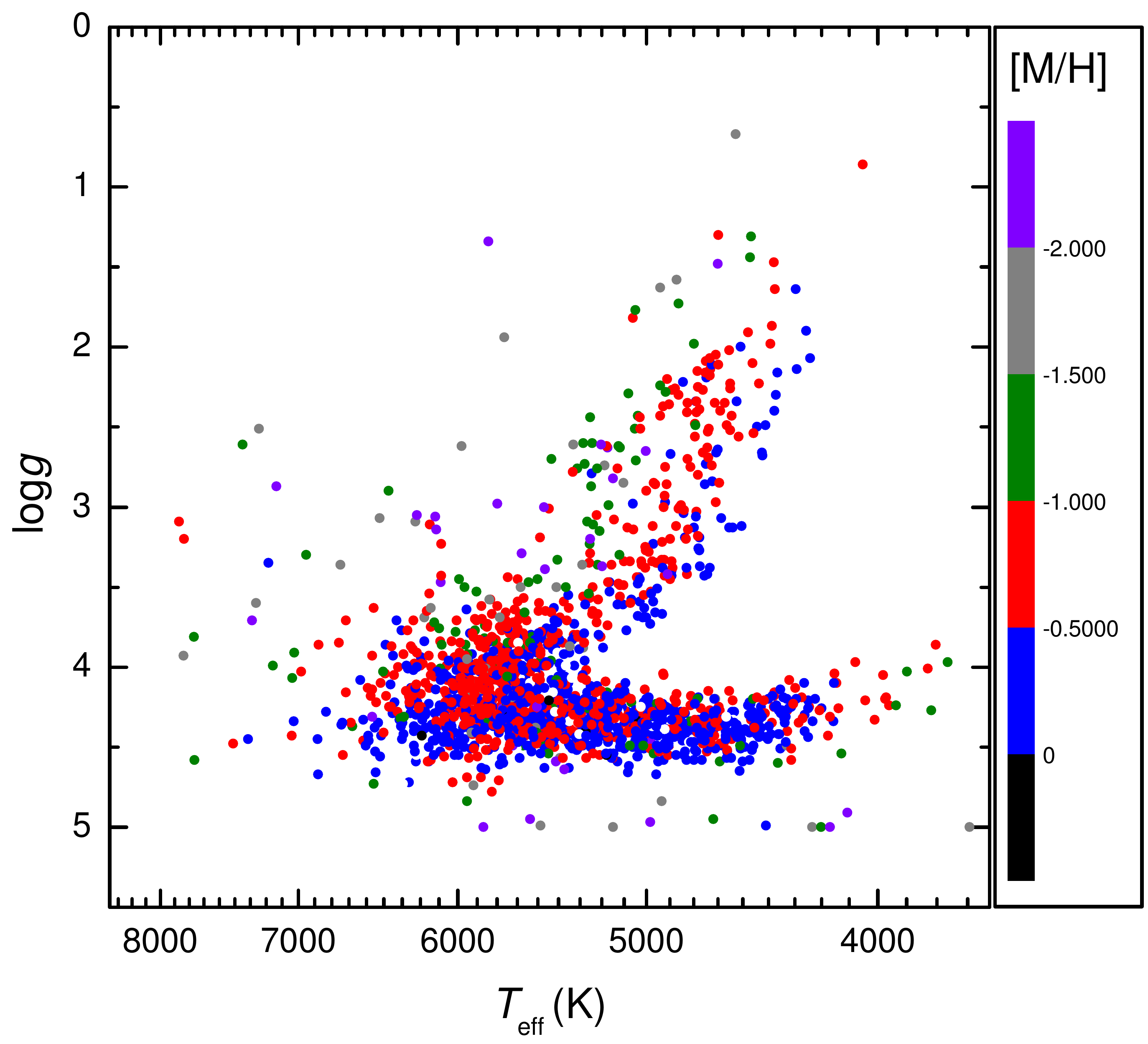}
 \caption{
  Temperature - gravity diagram of the \textit{main} sample stars. The atmospheric
parameters are taken from \cite{Blanco2014b}. The metallicity is coded in colours.
 }
 \label{fig:HR_diagramme_color}
 \end{figure}

This paper is based on a large part of iDR1 of the Gaia-ESO survey.
The full iDR1 description is provided by \citet{Blanco2014b}; we here only recall a few details for clarity. 
The iDR1 sample consists of about 10\,000 spectra of FGK-type stars
observed with the VLT/GIRAFFE instrument and mainly its HR10 (5339~--~5619~\AA), and/or HR21 (8484~--~9001~\AA) setups.
The HR03, HR05A, HR06,HR09B, HR14A, and HR15 setups were also used for other goals of iDR1.
Several Galactic populations have been targeted, such as the Galactic bulge, discs, and halo,
together with several open and globular clusters.
Two HR10 and HR21 spectra were observed in 4\,534 stars.
Among these data, we first selected the stars observed both in HR10 and HR21 that had a S/N higher than 15 per pixel in HR10 (the S/N in HR21 is about twice as high).
Throughout, we refer to these stars as the \textit{main} subsample.
These criteria were chosen to precisely describe the chemical properties of the target.
In total, this {\it main} subsample is composed of 1\,916 stars whose stellar atmospheric 
parameters (effective temperature, surface gravity, mean metallicity, global enrichment
in $\alpha$-elements versus iron, and elemental abundances) have been derived by the GES consortium \citep[see][]{Blanco2014b}.
The target location and the indication of the stars with a distance smaller than 600~pc from the Sun (Solar neighbourhood) is provided in the Fig.~14 of \cite{Blanco2014a}.
The typical errors on the stellar parameters are $140\pm107~K$ for $T_{\rm eff}$, $0.26\pm0.16$~dex 
for ${\rm log}~g$ and $0.13\pm0.12$ dex for $\rm{\rm [M/H]}$ (averaged errors and standard deviation over the whole \textit{main} sample).
The effective temperature vs. gravity diagram of this selected sample is shown in Fig.~\ref{fig:HR_diagramme_color}.
We note that about 12\% (237 stars) of this {\it main} subsample are giant stars (defined as ${\rm log}~g {\rm <} $ 3.5).
Finally, a $T_{\rm eff}$ and ${\rm log}~g$ progressively degenerate in the cool part of the main sequence for the effective temperatures lower than 5000~K. \citet{Blanco2014a} showed that the residual bias in effective temperatures caused by this degeneracy is about 100~K for the $4500>~T_{\rm eff}>4200$~K regime and is lower than the typical error (see \citealt{Blanco2014a, Blanco2014b}). The stellar photospheric elemental abundances recommended by the Gaia-ESO survey  are listed in Table~\ref{tab:TABLE1EXAMPLE} \citep[see also][]{Blanco2014b}.
%Note that our abundances were used as the Gaia-ESO survey iDR1 recommended abundances.

The stellar parameters and elemental abundances are determined within GES in two steps. (i) The final recommended effective $T_{\rm eff}$,${\rm log}~g$, [M/H] and [$\alpha$/M]
\footnote{\citet{Blanco2014a} write [Fe/H] and [$\alpha$/Fe] for the recommended metallicity and $\alpha$-to-metallicity ratios. We write [M/H], [$\alpha$/M], and [Fe/H] for the recommended metallicity,  $\alpha$-to-metallicity ratio, and iron abundances (i.e.~[M/H]=[Fe/H]$_{Recio-Blanco}$ and [$\alpha$/M]=[$\alpha$/Fe]$_{Recio-Blanco}$).} are estimated thanks to the combination of three different procedures: MATISSE (\citealt{Blanco2006}), FERRE (\citealt{APrieto2006}), and SME (\citealt{Valenti1996}). (ii) These final recommended parameters are then used to derive the abundances of individual elements. For iDR1, the final recommended elemental abundances were derived with only one method which we describe in more detail below.

Using the final adopted atmospheric parameters, we estimated the abundances of several chemical species through spectrum synthesis. 
For that purpose, we adopted the MARCS (\citealt{Gustafsson2008}) model atmospheres, which were also adopted by the GES consortium analysis,
and the recent version (v12.1.1) of the spectrum synthesis code TURBOSPECTRUM \citep{Alvarez1998}. We also used the atomic and molecular data provided by the linelist group of GES  (version~3), who
collected the most recent and complete experimental and theoretical data sources (GES linelist group, \citet{Heiter2014} in preparation).
We then developed a specific algorithm
to derive the individual chemical
abundances of the selected lines by searching for the best fit to the observed spectrum.
The main steps of this automated algorithm are following:
\begin{itemize}
\item Perform the radial velocity correction (provided by the consortium) on the roughly normalized GES spectra.
\item Extract the portion of the atomic and molecular linelist around the analysed line ($\pm 5~\AA$ around weak lines or $\pm 15~\AA$ around a broader line).
\item Compute the synthetic spectrum over the selected wavelength range  using the recommended physical atmospheric parameters 
and adopting a solar mixture as a starting point for the chemical abundance [El/M]~=~0.0.
\item Correct the local continuum of the selected region for the observed spectrum.
We corrected the local continuum corresponding to the possible line-free zones of the synthetic spectrum, defined as regions where the intensity of the synthetic spectrum is depressed by less than 0.02.
If the possible line-free zones are too small or do not exist for certain types of stars,
we iteratively searched for the possible less contaminated zones in the synthetic spectrum.
\item Create a small array of synthetic spectra assuming different elemental abundances and compute the $\chi^2$ of the fit ($\pm$1.2~\AA~around regular lines and $\pm$4.5~\AA~around broad calcium-triplet lines for each of them).
\item Vary the abundances around the minima of the previous quadratic function and adopt
the final abundance as the one corresponding to the lowest value of the $\chi^2$ parameter.
\end{itemize}
About 110 seconds per spectrum are required to estimate the abundance of a selected line.

With this technique, we were able to derive the individual chemical abundances
of the $\alpha$-elements Mg, Ca, Si, and Ti, and of Al, Ni, Cr, and Y, which present
several weak and strong lines in GIRAFFE HR10 and HR21 spectra of late-type stars.
We also adopted the DR1 mean metallicty [M/H] for this elemental abundance analysis.
This assumption was checked by independently measuring the
iron abundance from two iron lines around 5\,543 and 8\,514~$\AA$. They correlate well, with a small bias ($-0.03$~dex $\pm$ 0.10) with respect to metallicity recommended by the iDR1 (see \citealt{Blanco2014b} for a plot). We checked also the mean of the [Mg/M], [Ca/M], [Si/M], and [Ti/M] abundances vs. the recommended global [$\alpha$/M] and again found a very low bias ($-0.012$~dex $\pm$ 0.06).
The full line list is provided in Table~\ref{tab:TABLE_LINES}.
All atomic parameters were provided by \citet{Heiter2014} and the adopted reference solar abundances are those of \citet{Grevesse2007}.

The magnesium (\ion{Mg}{I}) abundances were determined from the two most prominent lines at 5\,528.4~$\AA$ in the HR10 and 8\,806.7~$\AA$ in the HR21 spectrum. These two lines are strong enough to be present in all the selected spectra. However, two additional weaker lines of \ion{Mg}{I} in HR21 were also analysed when possible. The cores of the \ion{Ca}{II} triplet lines cannot be well synthesized, probably because
of departures from the assumed equilibria and chromospheric effects. 
Therefore, we restricted our fitting procedure to the wings of these strong lines. 
We disregarded $\pm1.7~\AA$ around the centre of each of these lines.
Aluminum (\ion{Al}{I}) abundances were derived from the 5\,557.0~$\AA$ line and the region around the doublet at 8\,772.8-8\,773.8~$\AA$. 
However, since this doublet has a significant probability to be contaminated by 
telluric lines, we rejected the spectra for which a telluric contamination might occur because of the 
stellar radial velocity ($-191~\pm~17$~km/s and $254~\pm~17$~km/s).

\begin{table}

\caption{$\lambda(\AA)$ lines.
}
%\vspace{0.2cm}

\begin{tabular}{llccccccccccccccccccccc}
\hline\hline
\ion{Mg}{I}  	&	 \ion{Al}{I}  	&	 \ion{Si}{I}  	&	 \ion{Ca}{I}  	&	 \ion{Ca}{II}	\\
\hline
5528.4	&	5557.0	&	5488.9	&	8717.8	&	8498.0	\\
8717.8	&	8772.8	&	8556.7	&	5590.1	&	8542.0	\\
8736.0	&	8773.8	&	8648.4	&	5601.2	&	8662.1	\\
8806.7	&		&	8728.0	&		&	8912.0	\\
	&		&	8742.4	&		&		\\
	&		&	8892.7	&		&		\\
\hline\hline
\ion{Ti}{I}  	&	 \ion{Ti}{II} 	&	 \ion{Cr}{I}  	&	 \ion{Ni}{I}  	&	 \ion{Y}{II}	\\
\hline
5351.1	&	5336.7	&	5345.7	&	5587.8	&	5544.6	\\
5426.2	&	5381.0	&	5348.3	&	5593.7	&	5546.0	\\
5471.1	&	5418.7	&	5409.7	&	5462.4	&	5509.8	\\
5477.6	&		&		&	5435.8	&		\\
5490.1	&		&		&		&		\\
5514.3	&		&		&		&		\\
8518.3	&		&		&		&		\\

\hline
% & & Total Sample & 198,000 \\
 \label{tab:TABLE_LINES}

\end{tabular}

\end{table}

\begin{table*}

\caption{Chemical abundances of Gaia-ESO survey first data release. This is a small part of the table$^{***}$. The full information about Gaia-ESO survey and access to advanced databases is located online$^{****}$.
}
%\vspace{0.2cm}
\resizebox{\textwidth}{!}{%
\begin{tabular}{llccccccccccccccccccccc}
\hline\hline
Star 	 & 	 Star 	 & 	 Mg\,I$^*$ 	 & 	 $\sigma^{**}$ 	 & 	 Al\,I 	 & 	 $\sigma$ 	 & 	 SI\,I 	 & 	 $\sigma$ 	 & 	 Ca\,I 	 & 	 $\sigma$ 	 & 	 Ca\,II 	 & 	 $\sigma$ 	 & 	 Ti\,I 	 & 	 $\sigma$ 	 & 	 Ti\,II 	 & 	 $\sigma$ 	 & 	 Cr\,I 	 & 	 $\sigma$ 	 & 	 Ni\,I 	 & 	 $\sigma$ 	 & 	 Y\,II 	&	 $\sigma$ 	 \\
\hline
 00190657-4702076	&	             G\_2\_r\_70\_45 	&	7.30	&	0.17	&	6.33	&	0.00	&	7.26	&	0.10	&	6.05	&	0.09	&	6.07	&	0.05	&	4.85	&	0.11	&	4.77	&	0.11	&	5.22	&	0.08	&	5.75	&	0.16	&		&		\\
 00191267-4656479	&	            G\_2\_b\_70\_154 	&	7.00	&	0.11	&	5.97	&	0.03	&	6.75	&	0.28	&	5.61	&	0.05	&	5.65	&	0.15	&		&		&	4.47	&	0.14	&	4.83	&	0.08	&		&		&		&		\\
 00191811-4656554	&	            G\_2\_b\_70\_153 	&	6.86	&	0.12	&	5.94	&	0.03	&	6.73	&	0.28	&	5.38	&	0.07	&	5.58	&	0.22	&		&		&	4.28	&	0.06	&	4.22	&	0.07	&	5.15	&	0.05	&		&		\\
 00193307-4654268	&	            G\_2\_r\_70\_162 	&	7.18	&	0.10	&	6.12	&	0.02	&	7.15	&	0.17	&	6.01	&	0.10	&	6.03	&	0.15	&	4.70	&	0.03	&	4.54	&	0.29	&	5.24	&	0.12	&	5.71	&	0.23	&	1.83	&		\\
 00195573-4649434	&	            G\_2\_r\_70\_196 	&	7.36	&	0.02	&	6.22	&	0.08	&	7.36	&	0.04	&	6.02	&	0.02	&	6.17	&	0.11	&	4.67	&	0.02	&	4.76	&	0.04	&	5.20	&	0.06	&	5.87	&	0.04	&	1.74	&	0.08	\\
 00200267-4655523	&	            G\_2\_r\_70\_161 	&	7.27	&	0.08	&	6.20	&	0.01	&	7.15	&	0.17	&	6.16	&	0.03	&	6.01	&	0.02	&	4.69	&	0.03	&		&	0.21	&	5.27	&	0.13	&	6.19	&	0.05	&	2.26	&	0.24	\\
 00322798-4402100	&	           G\_2\_r\_108\_112 	&	7.27	&	0.11	&	6.16	&	0.04	&	7.24	&	0.10	&	5.98	&	0.03	&	6.12	&	0.19	&	4.60	&	0.01	&	4.55	&	0.09	&	5.17	&	0.12	&	5.90	&	0.19	&	1.82	&		\\
 ...			&	           ...		 	&	...	&	...	&	...	&	...	&	...	&	...	&	...	&	...	&	...	&	...	&	...	&	...	&	...	&	...	&	...	&	...	&	...	&	...	&	...	&	...	\\

\hline
% & & Total Sample & 198,000 \\
 \label{tab:TABLE1EXAMPLE}

\end{tabular}}
$^*$ The customary astronomical scale for logarithmic abundances is adopted in this table.
Hydrogen is defined to be $\log \epsilon_{\rm H} = 12.00$, i.e. $\log \epsilon_{\rm X} = \log (N_{\rm X}/N_{\rm H}) + 12$, where
$N_{\rm X}$ and $N_{\rm H}$ are the number densities of element X and hydrogen. 
\\
$^{**}$ $\sigma$ is the quadratic sum of the effects on the abundances caused by the errors of $T_{\rm eff}$, ${\rm log}~g$, ${\rm [M/H]}$, and $v_{\rm t}$.
\\
$^{***}$ The full table is available at the CDS via anonymous ftp
 to cdsarc.u-strasbg.fr (130.79.128.5) or via
http://cdsarc.u-strasbg.fr/viz-bin/qcat?J/A+A/
\\
$^{****}$ http://www.gaia-eso.eu
\end{table*}

\begin{table}
\caption{Errors due to uncertain continuum placement and $\chi^2$ fitting from Monte Carlo simulations.}
%\vspace{0.2cm}
\begin{tabular}{lcccccc}
\hline\hline
Line & S/N=15* & S/N=25* & S/N=50* & S/N=80* \\
\hline
\multicolumn{5}{c}{HD~107328 ($T_{\rm eff}=4590 K$, ${\rm log}~g=2.20$, ${\rm [M/H]}=-0.34$)}\\
${\rm	[\ion{Mg}{I}/H]	}$&	0.05	&	0.03	&	0.02	&	0.01	\\
${\rm	[\ion{Al}{I}/H]	}$&	0.07	&	0.04	&	0.03	&	0.02	\\
${\rm	[\ion{Si}{I}/H]	}$&	0.08	&	0.05	&	0.03	&	0.02	\\
${\rm	[\ion{Ca}{I}/H]	}$&	0.11	&	0.07	&	0.04	&	0.02	\\
${\rm	[\ion{Ca}{II}/H]	}$&	0.03	&	0.02	&	0.01	&	0.01	\\
${\rm	[\ion{Ti}{I}/H]	}$&	0.04	&	0.02	&	0.02	&	0.01	\\
${\rm	[\ion{Ti}{II}/H]	}$&	0.15	&	0.10	&	0.05	&	0.03	\\
${\rm	[\ion{Cr}{I}/H]	}$&	0.09	&	0.06	&	0.03	&	0.03	\\
${\rm	[\ion{Fe}{I}/H]	}$&	0.10	&	0.06	&	0.02	&	0.02	\\
${\rm	[\ion{Ni}{I}/H]	}$&	0.13	&	0.09	&	0.05	&	0.03	\\
${\rm	[\ion{Y}{II}/H]	}$&	0.15	&	0.10	&	0.06	&	0.04	\\

\hline
\multicolumn{5}{c}{G\_1\_r\_25\_269 ($T_{\rm eff}=4547 K$, ${\rm log}~g=4.50$, ${\rm [M/H]}=-0.44$)}\\
${\rm	[\ion{Mg}{I}/H]	}$&	0.03	&	0.02	&	0.01	&	0.01	\\
${\rm	[\ion{Al}{I}/H]	}$&	0.05	&	0.03	&	0.02	&	0.01	\\
${\rm	[\ion{Si}{I}/H]	}$&	0.11	&	0.09	&	0.06	&	0.05	\\
${\rm	[\ion{Ca}{I}/H]	}$&	0.05	&	0.03	&	0.02	&	0.01	\\
${\rm	[\ion{Ca}{II}/H]	}$&	0.03	&	0.02	&	0.02	&	0.01	\\
${\rm	[\ion{Ti}{I}/H]	}$&	0.04	&	0.02	&	0.01	&	0.01	\\
${\rm	[\ion{Ti}{II}/H]	}$&	0.14	&	0.08	&	0.04	&	0.03	\\
${\rm	[\ion{Cr}{I}/H]	}$&	0.05	&	0.03	&	0.02	&	0.01	\\
${\rm	[\ion{Fe}{I}/H]	}$&	0.09	&	0.06	&	0.03	&	0.03	\\
${\rm	[\ion{Ni}{I}/H]	}$&	0.10	&	0.06	&	0.04	&	0.03	\\
${\rm	[\ion{Y}{II}/H]	}$&	0.16	&	0.14	&	0.09	&	0.06	\\

\hline
\multicolumn{5}{c}{18 Sco ($T_{\rm eff}=5747 K$, ${\rm log}~g=4.43$, ${\rm [M/H]}=0.02$)}\\
${\rm	[\ion{Mg}{I}/H]	}$&	0.04	&	0.02	&	0.01	&	0.01	\\
${\rm	[\ion{Al}{I}/H]	}$&	0.07	&	0.05	&	0.03	&	0.02	\\
${\rm	[\ion{Si}{I}/H]	}$&	0.06	&	0.04	&	0.02	&	0.01	\\
${\rm	[\ion{Ca}{I}/H]	}$&	0.07	&	0.04	&	0.03	&	0.02	\\
${\rm	[\ion{Ca}{II}/H]}$&	0.02	&	0.01	&	0.01	&	0.01	\\
${\rm	[\ion{Ti}{I}/H]	}$&	0.06	&	0.04	&	0.02	&	0.02	\\
${\rm	[\ion{Ti}{II}/H]}$&	0.14	&	0.09	&	0.05	&	0.03	\\
${\rm	[\ion{Cr}{I}/H]	}$&	0.10	&	0.06	&	0.04	&	0.02	\\
${\rm	[\ion{Fe}{I}/H]	}$&	0.08	&	0.06	&	0.03	&	0.02	\\
${\rm	[\ion{Ni}{I}/H]	}$&	0.09	&	0.07	&	0.04	&	0.03	\\
${\rm	[\ion{Y}{II}/H]	}$&	0.16	&	0.14	&	0.09	&	0.07	\\

\hline
% & & Total Sample & 198,000 \\
 \label{tab:Montecarlo}

\end{tabular}

$^*$ The S/N in the HR10 spectrum. For the HR21 spectrum the S/N is twice as high.
\end{table}

\begin{table}
\caption{Mean line-to-line scatter for the \textit{main} and \textit{clean}$^*$ subsamples, where two or more lines are available.
$N_{max}$ is the maximum number of available lines.
}
%\vspace{0.2cm}
\begin{tabular}{lcccccc}
\hline\hline
Line & $\langle \sigma \rangle_{main}$ & $\pm$ & $\langle \sigma \rangle_{clean}$ & $\pm$ & $N_{max}$ \\
\hline
${\rm	[\ion{Mg}{I}/H]	}$&	0.11	&	0.06	&	0.09	&	0.05	&	4\\
${\rm	[\ion{Al}{I}/H]	}$&	0.05	&	0.05	&	0.04	&	0.04	&	3\\
${\rm	[\ion{Si}{I}/H]	}$&	0.14	&	0.06	&	0.10	&	0.05	&	6\\
${\rm	[\ion{Ca}{I}/H]	}$&	0.13	&	0.08	&	0.10	&	0.06	&	3\\
${\rm	[\ion{Ca}{II}/H]}$&	0.11	&	0.06	&	0.08	&	0.05	&	4\\
${\rm	[\ion{Ti}{I}/H]	}$&	0.10	&	0.08	&	0.08	&	0.07	&	7\\
${\rm	[\ion{Ti}{II}/H]}$&	0.17	&	0.10	&	0.15	&	0.09	&	3\\
${\rm	[\ion{Cr}{I}/H]	}$&	0.11	&	0.07	&	0.09	&	0.06	&	3\\
${\rm	[\ion{Cr}{I}/H]	}$&	0.11	&	0.07	&	0.09	&	0.06	&	3\\
${\rm	[\ion{Fe}{I}/H]	}$&	0.11	&	0.08	&	0.09	&	0.05	&	2\\
${\rm	[\ion{Ni}{I}/H]	}$&	0.15	&	0.08	&	0.13	&	0.08	&	4\\
${\rm	[\ion{Y}{II}/H]	}$&	0.15	&	0.10	&	0.14	&	0.10	&	3\\
\hline
% & & Total Sample & 198,000 \\
 \label{tab:Scatter}

\end{tabular}

$^*$ The \textit{clean} subsample will be introduced in Sect.~\ref{sec:properties_of_disc} with strict selection criteria on atmospheric parameters errors
($T_{\rm eff} < 200$~K, the error in ${\rm log}~g < 0.15$ and the error in {\rm [M/H]}~$<$~0.15).
\end{table}  

   \begin{table}
\begin{center}
      \caption{Effects on the derived abundances, resulting from the atmospheric parameters uncertainties
for the four groups of stars, selected by $T_{\rm eff}$, and ${\rm log}~g$.
The table shows the median of the propagated errors due to $\Delta T_{\rm eff}$, $\Delta {\rm log}~g$, $\Delta {\rm [M/H]}$, $\Delta v_{\rm t}$. 
The $ \sigma_{\rm total([X/H])} $ stands for the median of the quadratic sum of all four effects on [X/H] ratios. 
The $ \sigma_{\rm total([X/M])} $ stands for the median of the quadratic sum  of all four effects on [X/M] ratios. 
The $ \sigma_{\rm all([X/M])} $ is the combined effect of $ \sigma_{\rm total([X/M])} $ and the line-to-line scatter (see \ref{tab:Scatter}). 
}
        \label{tab:Sensitivity}
      \[
      \resizebox{\columnwidth}{!}{%
         \begin{tabular}{lrrccccc}
            \hline
%	    \hline
            \noalign{\smallskip}
	    El & 
	    ${ \Delta T_{\rm eff} }$ & 
            ${ \Delta \log g }$ & 
            $\Delta {\rm [M/H]}$ & 
            ${ \Delta v_{\rm t} }$ 
	     & $ \sigma_{\rm total\left[\frac{X}{H}\right]} $  
	     & $ \sigma_{\rm total\left[\frac{X}{M}\right]} $  
	     & $ \sigma_{\rm all\left[\frac{X}{M}\right]} $  
	     \\ 
            \noalign{\smallskip}
            \hline
            \noalign{\smallskip}
\multicolumn{5}{c}{$T_{\rm eff}<5000    K$, ${\rm log}~g<3.5   $}\\
Mg\,{\sc i} 	&	0.112	&	0.028	&	0.020	&	0.020	&	0.119	&   0.037 &	0.048	\\
Al\,{\sc i} 	&	0.074	&	0.007	&	0.033	&	0.017	&	0.084	&   0.048 &	0.061	\\
Si\,{\sc i} 	&	0.082	&	0.090	&	0.020	&	0.070	&	0.151	&   0.119 &	0.155	\\
Ca\,{\sc i} 	&	0.062	&	0.012	&	0.011	&	0.020	&	0.067	&   0.051 &	0.112	\\
Ca\,{\sc ii}	&	0.013	&	0.076	&	0.060	&	0.017	&	0.099	&   0.087 &	0.065	\\
Ti\,{\sc i} 	&	0.204	&	0.035	&	0.010	&	0.017	&	0.208	&   0.044 &	0.091	\\
Ti\,{\sc ii}	&	0.124	&	0.067	&	0.060	&	0.020	&	0.158	&   0.048 &	0.063	\\
Cr\,{\sc i} 	&	0.202	&	0.024	&	0.020	&	0.020	&	0.205	&   0.026 &	0.094	\\
Fe\,{\sc i} 	&	0.196	&	0.020	&	0.020	&	0.020	&	0.200	&   0.033 &	0.085	\\
Ni\,{\sc i} 	&	0.285	&	0.026	&	0.020	&	0.020	&	0.290	&   0.038 &	0.097	\\
Y\,{\sc ii} 	&	0.118	&	0.171	&	0.133	&	0.017	&	0.257	&   0.084 &	0.158	\\

                 \noalign{\smallskip}
\hline
                 \noalign{\smallskip}
\multicolumn{5}{c}{$T_{\rm eff}>5000    K$, ${\rm log}~g<3.5   $}\\
Mg\,{\sc i} 	&	0.093	&	0.093	&	0.017	&	0.020	&	0.128	&  0.031  &	0.050	\\
Al\,{\sc i} 	&	0.087	&	0.046	&	0.020	&	0.065	&	0.112	&  0.039  &	0.054	\\
Si\,{\sc i} 	&	0.044	&	0.061	&	0.020	&	0.030	&	0.080	&  0.062  &	0.069	\\
Ca\,{\sc i} 	&	0.212	&	0.077	&	0.160	&	0.017	&	0.272	&  0.047  &	0.080	\\
Ca\,{\sc ii}	&	0.049	&	0.007	&	0.030	&	0.017	&	0.058	&  0.053  &	0.061	\\
Ti\,{\sc i} 	&	0.205	&	0.007	&	0.050	&	0.017	&	0.182	&  0.024  &	0.070	\\
Ti\,{\sc ii}	&	0.143	&	0.082	&	0.050	&	0.020	&	0.117	&  0.035  &	0.044	\\
Cr\,{\sc i} 	&	0.246	&	0.033	&	0.050	&	0.017	&	0.247	&  0.019  &	0.070	\\
Fe\,{\sc i} 	&	0.232	&	0.020	&	0.041	&	0.020	&	0.237	&  0.039 &	0.087	\\
Ni\,{\sc i} 	&	0.338	&	0.027	&	0.059	&	0.020	&	0.298	&  0.046  &	0.076	\\
Y\,{\sc ii} 	&	0.191	&	0.141	&	0.113	&	0.017	&	0.221	&  0.046  &	0.120	\\

                 \noalign{\smallskip}
\hline
                 \noalign{\smallskip}\multicolumn{5}{c}{$T_{\rm eff}<5000    K$, ${\rm log}~g>3.5   $}\\
Mg\,{\sc i} 	&	0.116	&	0.031	&	0.020	&	0.020	&	0.123	&  0.029  &	0.049	\\
Al\,{\sc i} 	&	0.078	&	0.009	&	0.033	&	0.017	&	0.086	&  0.039  &	0.065	\\
Si\,{\sc i} 	&	0.090	&	0.064	&	0.020	&	0.070	&	0.132	&  0.095  &	0.152	\\
Ca\,{\sc i} 	&	0.063	&	0.009	&	0.011	&	0.020	&	0.068	&  0.036  &	0.111	\\
Ca\,{\sc ii}	&	0.013	&	0.054	&	0.060	&	0.017	&	0.084	&  0.073  &	0.070	\\
Ti\,{\sc i} 	&	0.217	&	0.025	&	0.010	&	0.017	&	0.212	&  0.043  &	0.079	\\
Ti\,{\sc ii}	&	0.133	&	0.048	&	0.060	&	0.020	&	0.154	&  0.088  &	0.054	\\
Cr\,{\sc i} 	&	0.214	&	0.013	&	0.020	&	0.020	&	0.214	&  0.006  &	0.087	\\
Fe\,{\sc i} 	&	0.230	&	0.021	&	0.020	&	0.020	&	0.232	&  0.039 &	0.087	\\
Ni\,{\sc i} 	&	0.303	&	0.014	&	0.020	&	0.020	&	0.295	&  0.031  &	0.088	\\
Y\,{\sc ii} 	&	0.129	&	0.120	&	0.133	&	0.017	&	0.228	&  0.084  &	0.128	\\

                 \noalign{\smallskip}
\hline
                 \noalign{\smallskip}\multicolumn{5}{c}{$T_{\rm eff}>5000    K$, ${\rm log}~g>3.5   $}\\
Mg\,{\sc i} 	&	0.065	&	0.041	&	0.017	&	0.020	&	0.087	&  0.024  &   0.052	\\
Al\,{\sc i} 	&	0.052	&	0.014	&	0.020	&	0.065	&	0.086	&  0.035  &   0.060	\\
Si\,{\sc i} 	&	0.032	&	0.030	&	0.020	&	0.030	&	0.061	&  0.031  &   0.064	\\
Ca\,{\sc i} 	&	0.135	&	0.037	&	0.160	&	0.017	&	0.207	&  0.057  &   0.075	\\
Ca\,{\sc ii}	&	0.039	&	0.009	&	0.030	&	0.017	&	0.053	&  0.046  &   0.057	\\
Ti\,{\sc i} 	&	0.131	&	0.008	&	0.050	&	0.017	&	0.132	&  0.042  &   0.075	\\
Ti\,{\sc ii}	&	0.091	&	0.046	&	0.050	&	0.020	&	0.092	&  0.050  &   0.057	\\
Cr\,{\sc i} 	&	0.149	&	0.017	&	0.050	&	0.017	&	0.155	&  0.014  &   0.096	\\
Fe\,{\sc i} 	&	0.120	&	0.014	&	0.041	&	0.020	&	0.129	&  0.022  &   0.080	\\
Ni\,{\sc i} 	&	0.196	&	0.016	&	0.059	&	0.020	&	0.184	&  0.031  &   0.087	\\
Y\,{\sc ii} 	&	0.116	&	0.079	&	0.113	&	0.017	&	0.159	&  0.044  &   0.125	\\

            \hline
         \end{tabular}}
      \]
%\begin{list}{}{}
  \end{center}
   \end{table}

\subsection{Error estimation for the chemical abundances}
\label{sec:ERROR}

Using these techniques, we also estimated the different sources of possible uncertainties when deriving the chemical abundances. 
Typically, the  sources of uncertainty can be divided into two categories.  
The first category includes the errors that affect a single line 
(e.g. random errors of the line fitting or continuum placement).
The second category includes the errors that affect all the lines, which are
mainly the errors caused by the atmospheric parameter uncertainties (such as errors in the effective temperature, surface gravity, and microturbulent velocity).
First, we studied the errors that might be 
caused by a possible incorrect continuum placement and $\chi^2$ fitting.
There are several sources of information about these errors.
One way (item 1 below) is to perform Monte Carlo simulations, selecting a statistically significant 
set of spectra and adding noise artificially. This simulation can show the sensitivity of the method  to noise ($\chi^2$ fitting and continuum placement).
Another way is to follow the line-to-line scatter (item 2 below). If there is a statistically significant number of the lines of a given element,
the scatter informs about the combined effect of the erroneous $\chi^2$ fitting, continuum placement, and uncertain atomic parameters for different lines.
Finally (item 3), we studied the propagation of errors from the model parameters ($T_{\rm eff}, {\rm log}~g$, ${\rm [M/H]}$, and $v_{\rm t}$)
to abundances. The analysis of these errors is presented below.

\begin{enumerate}
\item
For the Monte Carlo simulations, we took the spectra of three stars (cool and warm dwarfs and a cool giant) that are representative of the main target types. These stars were:
18~Sco ($T_{\rm eff}=5747$, ${\rm log}~g=4.43$, ${\rm [M/H]}=0.02$), 
G\_1\_r\_25\_269 ($T_{\rm eff}=4547 K$, ${\rm log}~g=4.50$, ${\rm [M/H]}=-0.44$),
 and HD~107328 ($T_{\rm eff}=4590 K$, ${\rm log}~g=2.20$, ${\rm [M/H]}=-0.34$).
The spectra of these bright stars have a S/N of 196 per pixel in HR10 and 280 in HR21 for 18 Sco, 84 and 211 for the G\_1\_r\_25\_269 and 170 and 230 for HD~107328. 
We point out that HD~107328 and 18~Sco are benchmark\footnote{The Gaia benchmark stars are defined as 
well-known bright stars for which well-determined $T_{\rm eff}$ and ${\rm log}~g$ values are available from direct methods, independently of spectroscopy (see Heiter et al. 2014a, in prep.). 
Their metallicities are well constrained from a careful spectroscopic study conducted by \citet{Jofre2014}, and they were chosen for this test because of the high S/N of their spectra.
On the other hand, there were no cool-dwarf benchmark stars observed by the survey at the time of the analysis described here, therefore we chose one star of the field for this test. The results of the test might be a slightly more pessimistic for this star since its spectra have a lower initial S/N.
The median of the S/N of the sample is $\sim$25 in HR10 and $\sim$51 in HR21, i.e. there is a significant difference in S/N between the spctra of HR10 and HR21 (a factor of $\sim 2$). This factor was adopted in our test.} 
stars.
To investigate the noise effect, we degraded these
spectra with a white Gaussian noise to a S/N= 15/30, 25/50, 50/100, and 80/160 for HR10 and HR21. 
The S/N=15 (HR10) is the lower limit of our sample.
We then generated 100 spectra around each line and for each SNR value and estimated the corresponding 100 abundances for each line at each SNR.
In that way, we determined sensitivity of our abundance estimates to the S/N.
These uncertainties are provided in the form of the standard deviation in Table~\ref{tab:Montecarlo}.
\item
The other way to estimate the random errors is to evaluate the line-to-line scatter.
It is customary to provide this as the typical error estimation because of
the continuum placement, the fitting, and uncertain atomic parameters. However, this error estimate is only robust when there are enough lines. We were able to use only 1-7 lines to derive the abundances of
a specific species. Because of a limited spectral interval and S/N in many cases.
We provide these error estimates for all elements in Table~\ref{tab:Scatter}, 
where $\langle\sigma\rangle$ is the mean over the \textit{main} subsample of the standard deviation for a given element.
\item
The third source of error is the propagation of the uncertainties on the main atmospheric parameters.
The errors on the atmospheric parameters are provided for each star by the Gaia-ESO survey consortium
\citep[see][]{Blanco2014b} and have been propagated into the chemical analysis.
The median errors over the sample due to errors on $T_{\rm eff}$, ${\rm log}~g$, ${\rm [M/H]}$, and $v_{\rm t}$ separately, and the combination of all four summed in quadrature, are presented in Table~\ref{tab:Sensitivity}. 
This is the only source of error presented in iDR1 database.
\end{enumerate}

The microturbulent velocity was derived using a fixed function of $T_{\rm eff}$, ${\rm log}~g$, and ${\rm [M/H]}$ for most of the stars in GES \citep[][in preparation]{Bergemann2014}, 
and therefore does not carry an error estimate on an individual star basis. We therefore adopted $\pm0.3$~km/s as the error of the microturbulent velocity.

As we mentioned before, the final estimate of the abundance uncertainty for a given star and a given 
element should include the contributions of the random errors and of the propagation of the errors on the stellar parameters.
In the sixth column of Table~\ref{tab:Sensitivity}, we list the total ($ \sigma_{\rm total([X/H])} $)  contribution
of the atmospheric parameter uncertainties 
to the abundances ([X/H]). However, this effect differs for the [X/M] ratios 
since in many cases the effect of changing stellar parameters is similar for lines of 
different elements and the [X/M] ratio is thus less sensitive to stellar parameter uncertainties.
We indeed found that $ \sigma_{\rm total([X/H])} $ is higher than $ \sigma_{\rm total([X/M])} $ for many elements.
We adopted the mean metallicity recommended by GES, but to determine the effect 
of uncertain atmospheric parameters on the [X/M] ratio, we also directly followed [\ion{Fe}{I}/H] based on two iron lines in HR10 and HR21.
The total error budget takes into account errors due to stellar parameters as well as random errors.
We thus calculated $\sigma_{\rm all}$ for every given element of every given star as\\ \\
$\sigma_{\rm all} = \sqrt{\sigma^2_{\rm total([X/M])}+\left(\frac{\sigma_{\rm N}}{\sqrt{N}}\right)^2   } $,\\ \\
where $\sigma_{\rm N}$ is the line-to-line scatter and $N$ is the number of analysed atomic lines. 
If the number of lines is $N=1$, we adopted $\langle \sigma \rangle_{main} $ from Table~\ref{tab:Scatter} 
for a given element as a $\sigma_{\rm N}$.
The total $\sigma_{\rm all}$ uncertainties for a given abundance are reported in the last column of Table~\ref{tab:Sensitivity}.

\begin{figure*}[!htb]
 \centering
 \includegraphics[scale=0.55]{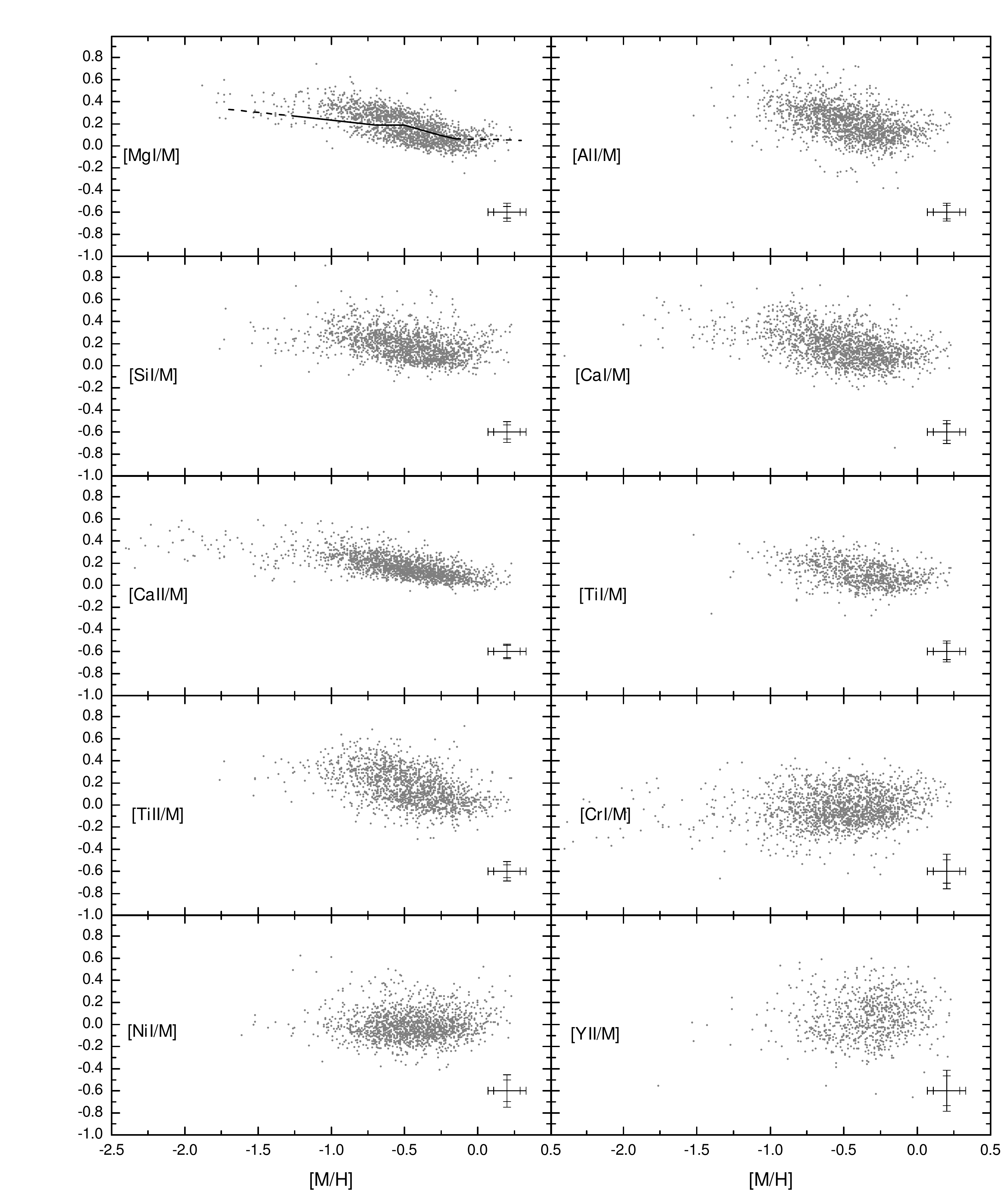}
 \caption{
  Observed elemental abundances as a function of metallicity for the \textit{main} subsample of stars.
  The names of the chemical elements are written in each panel. The error bar in the right lower corner of each panel indicate the typical error over the sample: the smaller error bar is the median error over the whole \textit{main} subsample, while the larger errorbar is the median error over abundances derived from spectra with S/N$<$25 in HR10. }
 \label{fig:XFE}
 \end{figure*}

\section{Chemical properties of the Galactic disc as revealed by GES}
\label{sec:properties_of_disc}

In this section, we present and discuss the elemental abundances derived for the Galactic disc stars.
First, we describe the chemical properties of the \textit{main} sample of stars (1\,916 objects)
as defined in the previous section.
Then, to clarify the picture, we define a \textit{clean} subsample of dwarfs with a stricter selection on the 
derived stellar parameter errors. 
For that purpose, we rejected all giant stars ($\rm{log}~g < 3.5$) and stars with 
high errors in effective temperature, surface gravity, and metallicity,
and only kept stars with
$\sigma (T_{\rm eff}) < 200$~K, $\sigma({\rm log}~g) < 0.15$~dex, and $\sigma {\rm ([M/H])} < 0.15$~dex.
Chromium, nickel, and especially yttrium suffer from large measurement
errors because of the weakness of the available lines (see Sec.\ref{sec:sample_method}).
Therefore we excluded in this \textit{clean} subsample the abundances of these species derived from S/N$<$40 spectra.
This \textit{clean} subsample contains 
about 700~stars (i.e. about 35\% of the \textit{main} sample).

We then compared the chemical characterization of the stellar populations
with results obtained by previous high-resolution spectroscopic studies devoted
to the chemical properties of the Galactic disc 
(\citealt{Chen2000, APrieto2004, Bensby2003, Bensby2005, Neves2009, Adibekyan2012}). 
Most of these studies are based on wider wavelength ranges,
higher S/N and resolution spectra, 
so that keeping track of our uncertainties is important when performing this comparison.
The propagated errors described in Sec.~\ref{sec:ERROR} are reported in Figures~\ref{fig:XFE} and \ref{fig:ALL_CLEANED}
alongside the abundances.
In addition, we compare in Fig.~\ref{fig:SCATTER} the propagated measurement errors to the 
observed scatter of each abundance ratio in bins of metallicities and Galactic populations
to evaluate the true astrophysical scatter.

For the \textit{main} sample of $\sim$2\,000 stars, 
the abundance ratios of magnesium, aluminium, silicon, calcium, titanium, chromium, nickel, and yttrium with respect to iron are shown in Fig.~\ref{fig:XFE}.
Typical error bars are shown in the figure.
Each of these abundance ratios has a very large scatter of results, as expected from the estimated error bars.
Despite the scatter, Fig.~\ref{fig:XFE} shows the typical 
trends of the $\alpha$-elements ([\ion{Mg}{I}/M], [\ion{Si}{I}/M], [\ion{Ca}{I}/M], [\ion{Ca}{II}/M], [\ion{Ti}{I}/M], and [\ion{Ti}{II}/M]) versus metallicity and even the subtle thin-to-thick disc separation is detectable in the [\ion{Mg}{I}/M] plot.
Moreover, the [\ion{Al}{I}/M] trend looks similar to those of the $\alpha$-elements.
This similarity supports the evidence shown by \citet{McWilliam1997}, and \citet{Bensby2003}: from a phenomenological point of view: aluminium could be classified as a mild $\alpha$-element, even though its nuclei have an odd number of protons. On the other hand, the iron group elements ([\ion{Cr}{I}/M], [\ion{Ni}{I}/M]) and the only $s$-process element (yttrium, [\ion{Y}{II}/M]) exhibit a scatter that blurs a detailed understanding of their trends. For this reason we made a strict selection based on S/N for the \textit{clean} subsample for these three elements.

 \begin{figure}[!htb]
 \centering
 \includegraphics[scale=0.25]{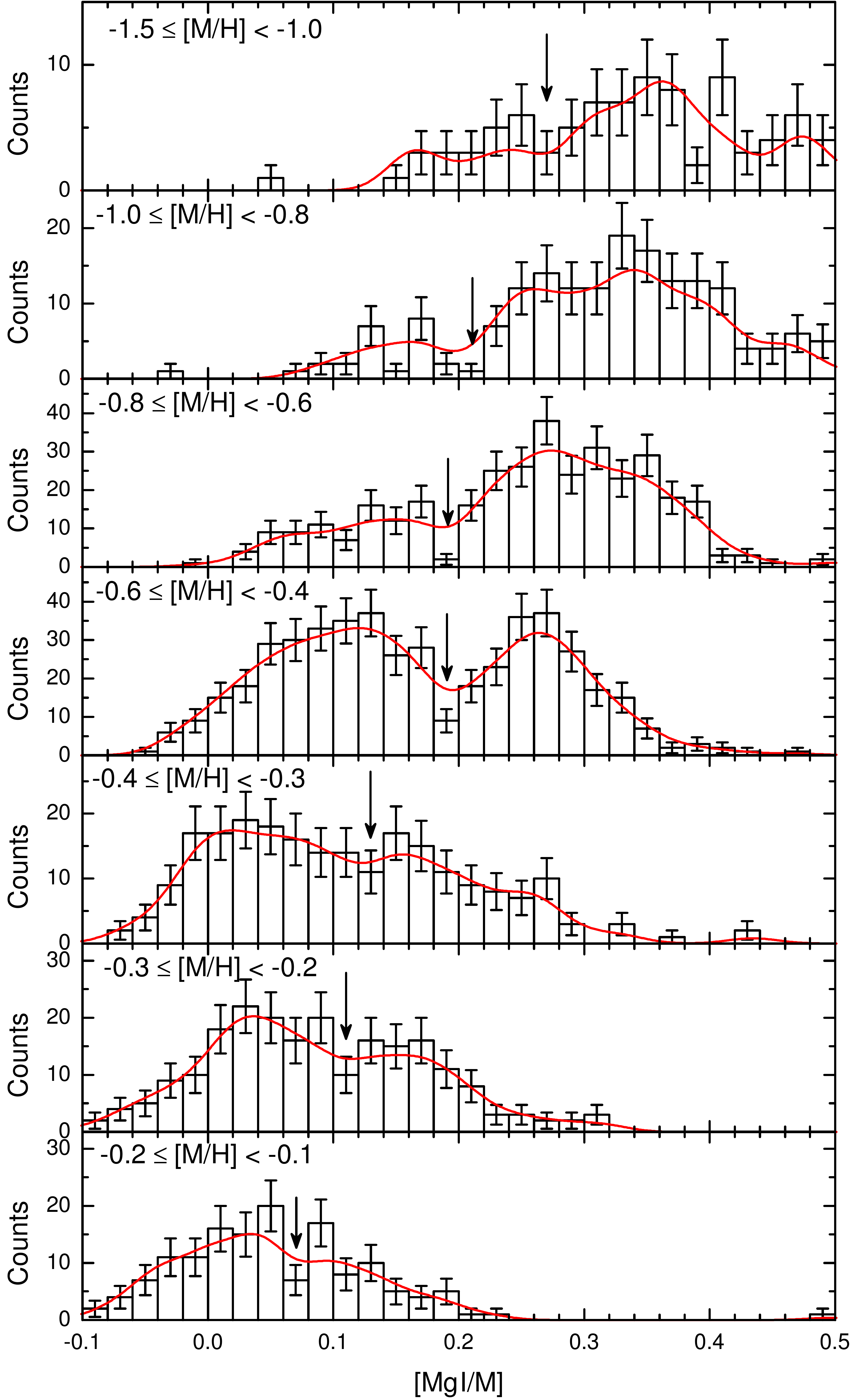}
 \caption{
  Histograms of [\ion{Mg}{I}/M] abundance values of the \textit{main} subsample with the error bars that indicate the Poisson noise. The red lines represent generalised histograms.
  The arrow marks the lower density regions that possibly indicate the separation between thin and thick discs.
%    The dashed lines shows possible chemical division between thin and thick disc stars. 
 }
 \label{fig:SEPARATION_HISTOGRAMS}
 \end{figure}

 \subsection{Chemically defining the thin and thick disc}
 \label{sec:Chemical_definition}

For clarity, we identify 
the $\alpha$-rich and $\alpha$-poor populations with the thick and thin discs, respectively.
The [\ion{Mg}{I}/M] abundance is the best probe of the chemical difference 
between thin and thick discs because of our lower observational uncertainties for this element and because it has the highest sensitivity to the population history (see \citealt{Fuhrmann1998, Gratton2000} who where the first to discuss the usefulness of Mg for studies aimed at separating different Galactic populations).

\begin{figure*}[!htb]
 \centering
 \includegraphics[scale=0.26]{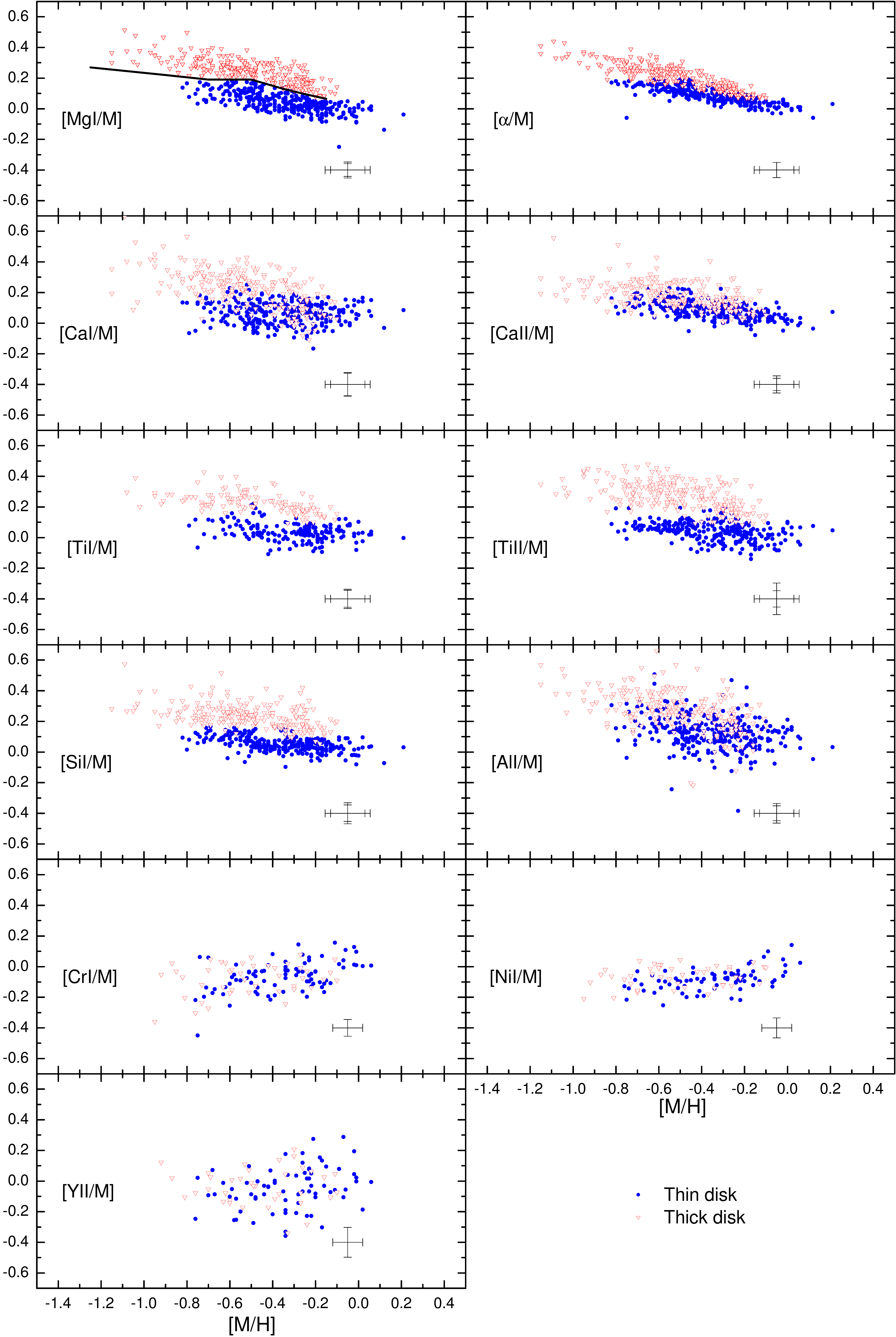}
 \caption{
 Abundance trends for our \textit{clean} subsample of stars for all elements of our analysis.
 The data are tagged into two groups called thin (blue dots) and thick disc (red triangles),
 according to the dividing line in [\ion{Mg}{I}/M] of Fig.~\ref{fig:XFE}. The error bars in the lower right corner of each panel indicate the typical error over the sample: 
 the smaller error bar is the median error over the whole \textit{clean} subsample, the larger error bar  the median error over abundances derived from spectra with S/N$<$25 in HR10.
  The \textit{clean} subsample of \ion{Cr}{I}, \ion{Ni}{I}, and \ion{Y}{II} is restricted to S/N~$>$~40. 
  }
 \label{fig:ALL_CLEANED}
 \end{figure*}

\begin{figure*}[!htb]
 \centering
 \includegraphics[scale=0.35]{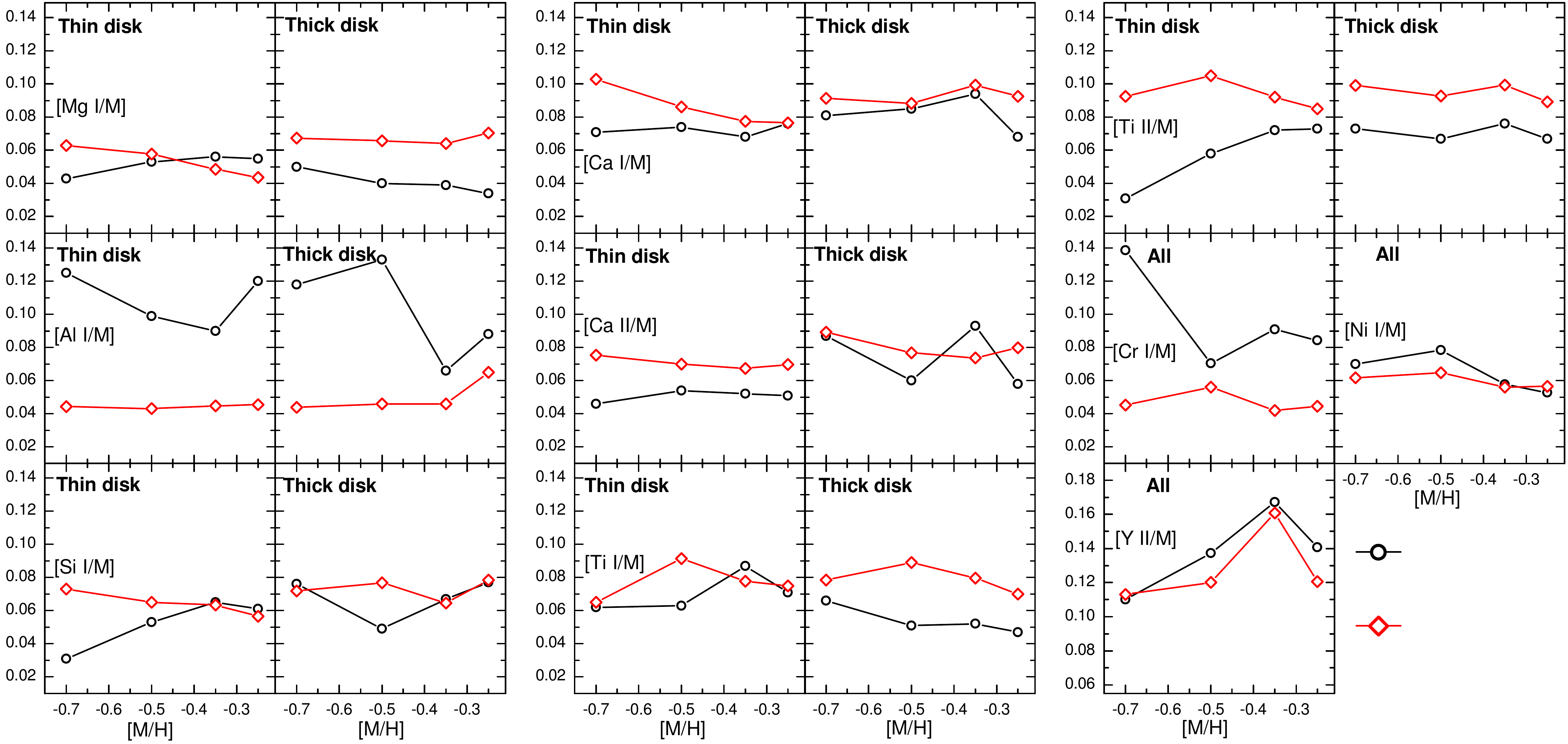}
%  \put (-305, 57){\makebox[0.7\textwidth][r]{$\sigma_{\rm rez}$}}%I want the text to go in the top right corner
%  \put (-285, 30){\makebox[0.7\textwidth][r]{$\sqrt{{\frac{\sum{\sigma^2_{\rm all\left[\frac{X}{Fe}\right]}}}{N_{\rm rez}}}}$}}%I want the text to go in the top right corner
 \put (-400, 57){\makebox[0.7\textwidth][r]{$\sigma_{\rm rez}$}}%I want the text to go in the top right corner
 \put (-380, 30){\makebox[0.7\textwidth][r]{$\sqrt{{\frac{\sum{\sigma^2_{\rm all\left[\frac{X}{Fe}\right]}}}{N_{\rm rez}}}}$}}%I want the text to go in the top right corner
 \caption{
 Observed \textit{rms} scatter around the mean abundance (black open circles) and expected errors (red open diamonds) in four metallicity bins. 
 For $\alpha$-elements the thin (left columns) and thick (right columns) discs are treated separately, whereas for the other elements (Cr, Ni, Y) the sample is treated as a whole, because of lower statistics.}
 \label{fig:SCATTER}
 \end{figure*}

Our main sample clearly separates into two distinct [\ion{Mg}{I}/M] populations.
We defined the separation between the thin and thick disc as the minimum in the [\ion{Mg}{I}/M] distributions in several intervals of metallicity in the \textit{main} subsample (see Fig.~\ref{fig:SEPARATION_HISTOGRAMS}).
We chose a bin of [\ion{Mg}{I}/M]$=$0.02~dex. We confirm that the separation is robust to changes in bin size and bin position.
The position of the separation is marked by an arrow in Fig.~\ref{fig:SEPARATION_HISTOGRAMS}.
We also plot generalized histograms in each panel of the Fig.~\ref{fig:SEPARATION_HISTOGRAMS} and show that the separation position corresponds to local minima of every generalized histogram.
This separation is most clearly detected in the metallicity range $-1.0$ to $-0.4$ or $-0.3$, which corresponds to the range where the kinematically separated thin and thick discs overlap in metallicity (e.g. \citealt{Fuhrmann2011}).
We then used this information to produce a line of chemical division
between thin and thick discs in Fig.~\ref{fig:XFE} ([\ion{Mg}{I}/M] panel: \textit{main} subsample).
Fig.~\ref{fig:ALL_CLEANED} ([\ion{Mg}{I}/M] panel) also shows the validity of the selection 
in the \textit{clean} subsample.
We also found some low-$\alpha$ metal-poor stars, (see upper panel of Fig.~\ref{fig:SEPARATION_HISTOGRAMS}), which are too metal-poor to be a part of the thin disc (see section~\ref{sec:metalpoor}).
The black solid part of the line (Fig.~\ref{fig:XFE}) 
is based on the results from the histograms of Fig.~\ref{fig:SEPARATION_HISTOGRAMS}
and the red dashed part of the line is extrapolated 
for the metal-poor end, and extrapolated assuming a negligible slope (as in \citealt{Adibekyan2011}) for the metal-rich end. 
The separation agrees with that reposted by others (e.g. \citealt{Fuhrmann2004, Fuhrmann2011, Adibekyan2012}).

According to this [\ion{Mg}{I}/M] separation, the stars of the \textit{clean} subsample were classified into two different groups (thin and thick discs) whose other elemental abundances are described in the following subsections
and illustrated in Fig.~\ref{fig:ALL_CLEANED}.
We also confirm that thin- and thick-disc separation is still visible when we bin in Galactocentric distance.

\subsection{$\alpha$-element distribution of disc stars}
        \label{secalpha}

The $\alpha$-elements Mg, Si, Ca, and Ti are mainly produced by successive
capture of $\alpha$-particles in connection with carbon, oxygen, and neon-burning in massive, short-lived stars. The are dispersed into the interstellar medium by type-II supernovae 
explosions on a time-scale of $\sim 10^7$ years. 
Iron is also produced by type-II supernovae explosions.
The type-Ia supernovae do not produce $\alpha$-elements, but they are the main producers of iron with a much longer time-scale.
Therefore, $\alpha$-to-iron abundance ratios provide information about the chemical properties of
the interstellar medium from which the observed stars were formed and the time-scale
of the chemical evolution history.

We present the abundances of the chemically tagged thin- and thick-disc stars in Fig.~\ref{fig:ALL_CLEANED}.
The thin-disc sequence approximatively extends from $-0.8$ to $0.2$~dex, the thick disc from $-1.0$ to $-0.2$~dex.
The abundance distributions of \ion{Si}{I}, \ion{Ti}{I}, \ion{Ti}{II}, \ion{Ca}{I}, and \ion{Ca}{II} 
follow the same distribution as \ion{Mg}{I}.
The thin- and thick-disc populations are clearly distinct (although with some overlaps),
with the thick-disc stars situated above the thin-disc stars in [El/M] vs. [M/H] plots.
The observed dispersions for magnesium, silicon, and calcium are compatible with the measurement errors (see Fig.~\ref{fig:SCATTER}, which compares the expected errors with the measured dispersion in bins of metallicity). Errors may be overestimated in some metallicity metallicity bins, however, most strikingly so for titanium.

There is evidence, that silicon and titanium can be significantly affected by NLTE effects (see \citealt{Bergemann2013} and \citealt{Bergemann2011}, respectively).
The silicon abundances are significantly affected in supergiant stars (see \citealt{Bergemann2013}), and it has been shown that the effect becomes weaker with increasing temperature. Thus, we expect mild NLTE effects for our sample which consists mostly of dwarf stars.
For giant stars, \citet{Bergemann2011} strongly recommended to compute [Ti/Fe] exclusively from the \ion{Ti}{II} lines.
However, keeping in mind that NLTE effects on \ion{Ti}{I} abundances mostly concerns metal-poor and giant stars, we conclude that NLTE effects for the most part of our sample stars are mild, lower than the typical observational errors on these abundances. Nevertheless, we did not use the neutral titanium in the following description of the metal-poor part of our sample, and we did not use titanium in our analysis of radial and vertical gradients.

There are two classical ways to isolate thick- disc from thin-disc stars in local samples: using their kinematics or the chemical abundances themselves (as we did here).
Following \citet{Bensby2003}, a number of works \citep[see e.g.][and references therein]{Bensby2014, Neves2009} used a kinematical criterion to separate thin- and thick-disc stars and subsequently found that thin-disc stars have lower [$\alpha$/Fe] than thick-disc stars. Gaps between thin- and thick-disc stars are observed for various $\alpha$-elemental abundance-to-iron ratios in many of these works.
\citet{Neves2009} argued for a bimodality, but not for a separation of thin- and thick-disc populations in [Mg/Fe], [Si/Fe], [Ti/Fe], and [Ca/Fe].
Their sample of stars bifurcates into two groups that contain members from the thin- and thick-disc.
On the other hand, \citet{Adibekyan2012} separated the two discs chemically and then checked the kinematics of the two populations.
Each of the chemically separated disc contained some degree of contamination from the kinematically classified other component.
However, even if the way of separating the two populations varies, the final picture in terms of
abundance ratios is similar (see discussion in \citealt{Adibekyan2012}).

According to our results (and in agreement with most other published works), the thin- and thick-disc populations can be better separated in some elements than in others.
Our two populations are better separated in silicon and titanium than was observed by \citet{Bensby2005}. \citet{Bensby2014} found that the populations are best separated in titanium (where we also found a good separation), and less well separated in magnesium, which is our best chemical separator.
According to our results, the thin- and thick-disc populations overlap more in
calcium (as traced by [\ion{Ca}{I}/M] and [\ion{Ca}{II}/M]) than in other $\alpha$-elements, in agreement with other studies, for example \citet{Bensby2005} and \citet{Neves2009}.
\citet{Neves2009} showed that the two stellar populations have chemical differences in [Ca/Fe], but these not as prominent as those of [O/Fe] or [Mg/Fe], for example. 
\citet{Ramirez2012,Ramirez2013} also showed that the separation between thin- and thick-disc stars is prominently detected by their NLTE oxygen abundances.
Furthermore, our results mostly agree with those of \citet{Adibekyan2012}, 
who chemically selected thin- and thick-disc stars and demonstrated a clear separation between the two discs in [Mg/Fe], [Si/Fe], [Ti/Fe], and [Ca/Fe]. 
However, different authors differ in their opinion as which elements are better chemical discriminants, which perhaps reflects the spectral features used to constrain the abundances of these elements and the associated errors (see Fig.~\ref{fig:SCATTER} for our own uncertainties on the different species).

\subsection{Distribution of aluminium in disc stars}

Aluminium together with sodium are mostly created in the core of massive stars
during the neon- and carbon-burning process (\citealt{Chen2000, Neves2009}).
They are spread throughout the interstellar medium by type-II supernovae explosions.
Thus, aluminium has a similar production site as $\alpha$-elements, 
which is why some authors consider aluminium as as mild $\alpha$-element 
even if it has an odd number of protons and is not the result of an $\alpha$ capture (\citealt{McWilliam1997, Neves2009}).
Aluminium was derived from three lines, one in HR10 and a strong doublet in HR21, which
is mostly detectable in the spectra of dwarf stars with metallicities down to -1.
The S/N ratio in the HR21 setup is twice as high as in HR10,
which led to a higher precision of the \ion{Al}{I} measurements. 

As expected, the trend of the aluminium-to-iron ratio (see Fig.~\ref{fig:ALL_CLEANED})  
is similar to those found for $\alpha$-elements.
The thin- and thick-disc populations overlap, but the distributions
are clearly distinct, with the thick-disc aluminium abundances generally higher than those of the thin-disc. There is a significant scatter in both thick- and thin-disc sequences.
The observed dispersions are three to four times higher than the expected measurement errors (see Fig.~\ref{fig:SCATTER}).
It is known that aluminium abundances can be sensitive to NLTE effects (\citealt{Baumueller1997, Gehren2004, Gehren2006, Andrievsky2008}).
%We made the test to check 
\citet{Baumueller1997} computed LTE and NLTE abundances from the aluminium doublet that we used for our analysis. For stars with metallicities, temperatures, and gravities similar to our sample, the NLTE abundances are 0.02~dex to 0.08~dex higher than the corresponding LTE abundances, hence a week effect.
To further investigate whether the scatter in [Al/M] is the result of NLTE affecting differently the different spectral types in the sample, we 
 checked the abundances of high S/N spectra (over 40) in three temperature regimes (5350--5750~K, 5350--6050~K, 5750--6050~K) 
 and found very similar dispersions for each temperature range, which confirms that NLTE is not the main effect driving the dispersion. 
 There is evidence that the spread in aluminium abundances might be partly caused by CN contamination. We analysed the region and found that CN contamination might be an issue for stars closer to RGB-tip (e.g. \citealt{Carretta2013}). However, our sample only has a few low ${\rm log}~g$ stars.
It is thus possible that the higher dispersion observed around the aluminium trend has an astrophysical origin, although it is still possible that our error estimates are underestimated. The scatter decreases faster than the expected error when selecting stars of higher S/N, an effect that could be attributed to optimistically low error estimates at low S/N, or to a smaller astrophysical scatter for stars close to the solar neighbourhood (and hence brighter and with higher S/N).

In any case, the trends already revealed by \citet{Bensby2005, Bensby2014, Neves2009} and \citet{Adibekyan2012} agree with our results.
Moreover, \citet{Neves2009} were able to detect 
bimodal distributions of their thin- and thick-disc sequences, although they admited that this bimodality is poorly visible since their results are strongly scattered.
\citet{Adibekyan2012} found a large scatter of [Al/Fe], 
but a quite clear separation between thin- and thick-discs. They also observed a downward trend for thick-disc stars, which our data support.

\subsection{Distribution of iron-peak elements in disc stars}

Iron-group elements behave similarly to iron: type-Ia supernovae are  
major contributors of iron-group elements, whereas the contribution of type-II supernovae
only dominates at low metallicities (\citealt{Edvardsson1993}).
Unfortunately, only few lines of these iron peak elements are included in the GIRAFFE HR10 or HR21 setups, and they are weak. 
This leads to very large uncertainties in their measurements and a relatively large scatter around the observed trends of chromium and nickel, as seen in Fig.~\ref{fig:XFE}. 
When the sample is restricted to S/N higher than~40 (Fig.~\ref{fig:ALL_CLEANED}),
the errors become small enough to reveal well-defined trends.

The trends for chromium and nickel are well defined and are similar to the solar value with
[\ion{Cr}{I}/M]=$-0.06$ (\textit{rms} 0.13~dex) and [\ion{Ni}{I}/M]=$-0.05$ (\textit{rms} 0.11~dex).
The observed dispersions are compatible with the expected measurement errors (see Fig.~\ref{fig:SCATTER}).
Chromium at low metallicities ([M/H]$\leq-0.6$) is a possible exception, 
which probably only results from an underestimate of the observational errors 
in this regime where only one line is measurable.
Furthermore, the thin- and thick-disc populations are indistinguishable in [\ion{Cr}{I}/M] and [\ion{Ni}{I}/M].

In agreement with these results, previous works have found that the [Cr/Fe] and [Ni/Fe] ratios were similar to the solar value in disc populations (for instance \citealt{Chen2000}).
In addition, the [Ni/Fe] abundances were found by several authors to increase for
metallicities higher than [Fe/H]$>$0.0 (\citealt{Chen2000, Bensby2005, APrieto2004, Neves2009, Adibekyan2012}).
Our sample supports this increase in the thin disc (see Fig.~\ref{fig:ALL_CLEANED}).

 \begin{figure*}[!htb]
 \centering
 \includegraphics[scale=0.28]{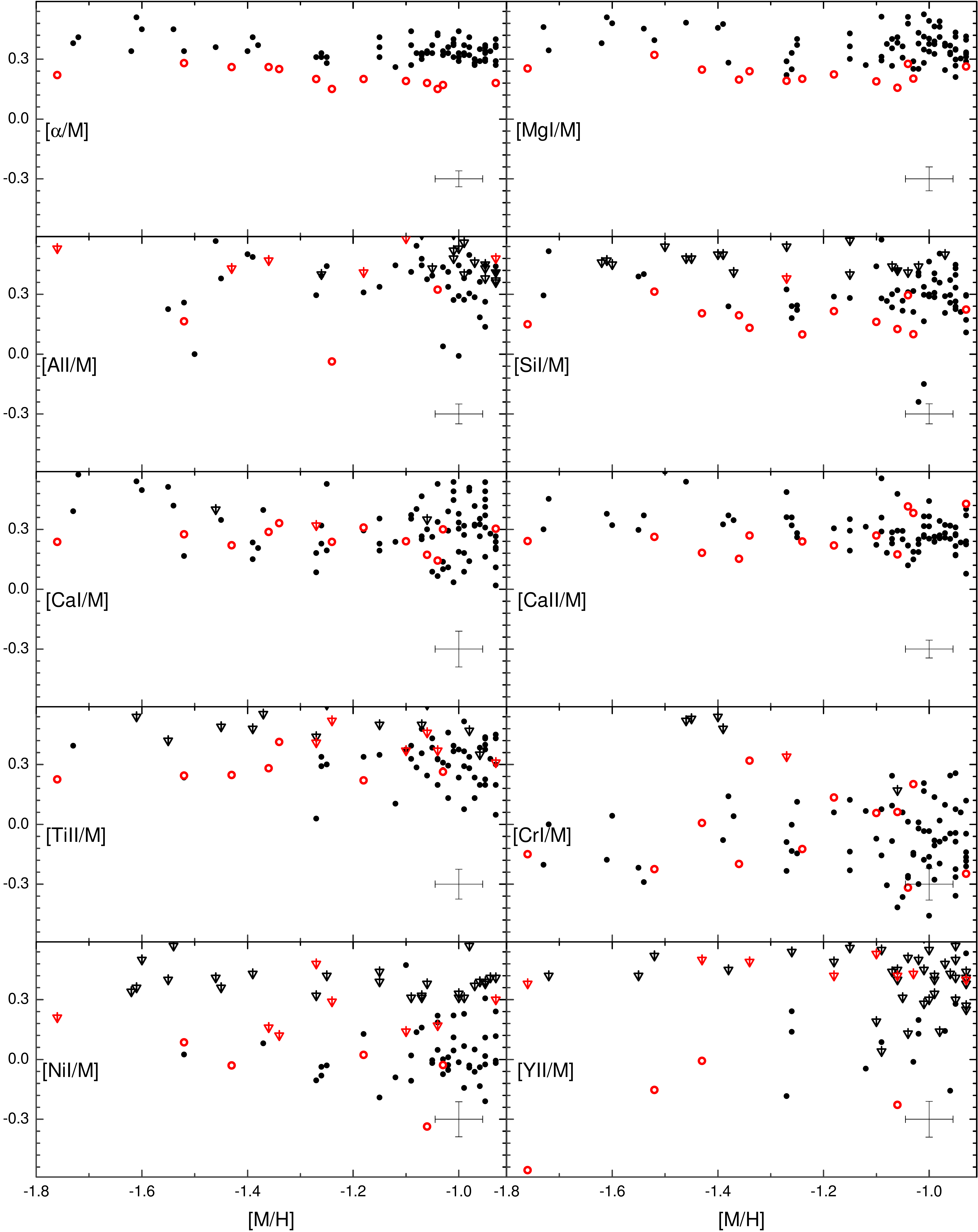}
 \caption{
 Derived chemical abundances for the \textit{metal-poor} subsample. The red circles mark low-$\alpha$ halo stars and black dots high-$\alpha$ halo stars.
 The red and black arrows mark upper limits of the abundance detection of undetected abundance values for low- and high-$\alpha$ halo stars.
  }
 \label{fig:HALO}
 \end{figure*}

\subsection{Distribution $s$-process elements in disc stars}

The $s$-process elements are mostly delivered into the interstellar medium by asymptotic giant branch (AGB) stars. 
These elements are synthesized in the H-burning shells of AGB stars.
Low-mass, longer lived (1--3 $\rm{M_{\odot}}$) stars dominate the synthesis of Y, Zr, Ba, and Nd.
These chemical species are then ejected by strong stellar winds that characterize this evolutionary phase.
Yttrium is the only  $s$-process element present in our spectra.
We were able to estimate \ion{Y}{II} abundances for only half of all targets,
because there are only a few weak lines of \ion{Y}{II} in HR10 and none in HR21.
The number of results of \ion{Y}{II} strongly depends on the quality of the spectrum, and weakly on the atmospheric parameters themselves.
We chose the results from the best quality-spectra (S/N$>$40) for the \textit{clean}
sample. As a consequence, only about 8\% of all stars are investigated in this study of Y. 

[\ion{Y}{II}/M] abundances are scattered around the solar value,
and thin- and thick-disc stars overlap completely (see Fig.~\ref{fig:ALL_CLEANED}).
As seen in Fig.~\ref{fig:SCATTER}, the observed dispersions 
are compatible with the expected measurement errors.

Our result is consistent with those reported by the few previous studies.
The star-to-star [Y/Fe] ratio scatters around zero and 
shows a weak increase towards higher [Fe/H] in the works by 
\citet{Edvardsson1993, Prochaska2000, Bensby2004, Bensby2014, Venn2004, Mashonkina2007}.

On the other hand, \citet{Mashonkina2012} stated, that the [Fe/H] is a poor evolutionary indicator for $s$-process elements. This is due to the increase of [Y/Fe], which in turn is related to the production of yttrium and other $s$-process elements in very-low mass AGB stars. Thus \citet{Mashonkina2012} found that the plot age~vs.~[Y/Fe] shows much smaller scatter. However, it is not possible to derive ages with satisfactory accuracy by the method described in \citet{Blanco2014a}.

\begin{table}
\caption{Mean abundances of high-$\alpha$ and low-$\alpha$ stars, standard deviations, and difference between two means.
}
%\vspace{0.2cm}
\begin{tabular}{lcccccc}
\hline \hline
\smallskip
Species & $ \rm \langle [\frac{X}{M}]_{l} \rangle$ & $\sigma$ & $\rm \langle [\frac{X}{M}]_{h} \rangle$ & $\sigma$ & $ \rm \langle [\frac{X}{M}]_{h} \rangle- \langle[\frac{X}{M}]_{l} \rangle$\\
\hline
${\rm 	[\alpha/M]		}$ &	0.21	&	0.04	&	0.35	&	0.06	&	0.15	\\
${\rm 	[\ion{Mg}{I}/M]	}$ &	0.23	&	0.04	&	0.37	&	0.09	&	0.15	\\
${\rm 	[\ion{Al}{I}/M]	}$ &	0.15	&	0.18	&	0.38	&	0.16	&	0.23	\\
${\rm 	[\ion{Si}{I}/M]	}$ &	0.18	&	0.07	&	0.32	&	0.12	&	0.14	\\
${\rm 	[\ion{Ca}{I}/M]	}$ &	0.27	&	0.09	&	0.31	&	0.15	&	0.04	\\
${\rm 	[\ion{Ca}{II}/M]	}$ &	0.27	&	0.09	&	0.30	&	0.09	&	0.03	\\
${\rm 	[\ion{Ti}{II}/M]	}$ &	0.27	&	0.07	&	0.31	&	0.12	&	0.04	\\
${\rm 	[\ion{Cr}{I}/M]	}$ &	-0.04	&	0.20	&	-0.09	&	0.17	&	-0.05	\\
${\rm 	[\ion{Ni}{I}/M]	}$ &	-0.06	&	0.16	&	0.02	&	0.13	&	0.08	\\
${\rm 	[\ion{Y}{II}/M]	}$ &	-0.24	&	0.23	&	0.11	&	0.20	&	0.35	\\
\hline
% & & Total Sample & 198,000 \\
 \label{tab:HALO}

\end{tabular}
\end{table}  

\section{Metal-poor stars}
\label{sec:metalpoor}

Because uncertainties on stellar parameters tend to increase with decreasing metallicity at a given SNR level, 
our \textit{clean} subsample only includes very few stars with [M/H]$\leq -1.0$, 
whereas the \textit{main} sample contained more than 100. We therefore relaxed our selection 
criterion on the stellar parameter errors for metal-poor stars in the same 
way as \citet{Blanco2014a} (sample \textit{e} in their Table~3) to investigate the elemental 
abundances in the metal-poor part of the survey (which we define as our \textit{metal-poor} subsample). This subsample contains 102 stars. 
We take the definition of the low- and high-$\alpha$ star grouping from \citet{Blanco2014a} (their Fig.~12).
However, the metal-poor high-$\alpha$ group can contain both halo and thick-disc stars because these populations are chemically indistinguishable (see e.g. \citealt{Nissen2010}). 

While the majority of stars with [M/H]$\leq -1.0$ are classified to be high-$\alpha$, about 10\% of the \textit{metal-poor} subsample are classified to be low-$\alpha$ objects.
In Fig.~\ref{fig:HALO}, we show the stars of possible high- and low-$\alpha$ sequences. First, we plot the [$\alpha$/M] from the recommended Gaia-ESO values, which were derived as one of the atmospheric parameters (see \citealt{Blanco2014a, Blanco2014b}).
Then we plot other elements (including the individual $\alpha$-elements) using the same subsample and tagging system.
The ability of measuring (weak) lines in metal-poor stars strongly depends on the S/N of the spectrum, but also directly on the metallicity or $T_{\rm eff}$ of the stars. 
As a result, several intrinsically weak lines could not be measured, 
and we were unable to provide abundances of some elements (mostly for Al, Cr, Ni, and Y) for some stars. In these cases we derived upper limits.
Some upper limits are higher than the upper boundaries  (+0.6 dex) of Fig.~\ref{fig:HALO} and therefore do not appear in the plots. We also provide the mean abundances of the high- and low-$\alpha$ stars in Table~\ref{tab:HALO}, together with standard deviations and difference between the two means.

The [\ion{Mg}{I}/M], [\ion{Si}{I}/M] and [\ion{Ca}{II}/M] abundances of the high- and low-$\alpha$ samples (see Fig.~\ref{fig:HALO}) follow the [$\alpha$/M] classification nicely.
In particular that our [$\alpha$/M] separation produces a separation in [\ion{Mg}{I}/M] that is completely compatible.
Since errors in [\ion{Ca}{I}/M] and [\ion{Ti}{II}/M] are higher than for other $\alpha$-elements, 
the distinction between populations in the distribution of these species is not as clear.
All three low-$\alpha$ population stars with aluminium detections fall below the high-$\alpha$ [\ion{Al}{I}/M] distribution. Furthermore, the [\ion{Al}{I}/M] upper limits of four low-$\alpha$ population stars fall directly on top of the high-$\alpha$ [\ion{Al}{I}/M] distribution, which shows that, if Al had been detected in these stars, they would also fall below the [\ion{Al}{I}/M] of high-$\alpha$ stars. This indicates a [\ion{Al}{I}/M] ratio that is lower in the low-$\alpha$ than in the high-$\alpha$ samples, although more firm detections of \ion{Al}{I} would help to strengthen this conclusion.

The high- and low-$\alpha$ sequences are the same in [\ion{Cr}{I}/M], while a distinction, if be present, might possibly be hidden in the distributions of [\ion{Ni}{I}/M]] abundances, which unfortunately contain many upper limits and few measurements for metallicities below $-1.2$. The situation is even worse for the [\ion{Y}{II}/M]] abundances, since the detectability of \ion{Y}{II} lines is already a challenge below a metallicity of $-0.9$, which hampers any conclusion on the $\alpha$-separated populations for this \textit{s}-process element.

High- and low-$\alpha$ sequences among halo stars were first observed by \citet{Nissen1997,Nissen2010} who showed that the halo stars fall into two populations, clearly separated in [$\alpha$/Fe]. 
Although our kinematical information \citep[see][]{Blanco2014a} is not sufficient to establish this whereas our metal-poor sample is indeed dominated by halo stars or also includes a significant portion of disc stars, the metallicity regime probed here matches these samples perfectly.
\citet{Nissen2010} observed that these populations showed clear differences in the [Mg/Fe], [Si/Fe] and, [Ti/Fe] ratios, and that they were not clearly distinct in [Ca/Fe]. Additionally, the two populations separate in [Ni/Fe] \citep{Nissen2010} and in [Y/Fe] \citep{Nissen2011}. The metal-poor low and high-$\alpha$ division can also be distinguished in the works by \citet{Gratton2003a, Gratton2003b} and \citet{Smiljanic2009}. \citet{Smiljanic2009} showed, that Be abundances vs. [$\alpha$/Fe] apparently facilitate visibility of the halo division. This was also confirmed by \citet{Tan2011}.
Our data agree with this picture, and we furthermore suggest that [\ion{Al}{I}/M] is distinct between the two populations, high-$\alpha$ stars being also [\ion{Al}{I}/M] rich, and vice versa. 
Unfortunately, it is impossible at this stage to verify whether the [\ion{Al}{I}/M] abundance correlates with [\ion{Ni}{I}/M], as [Na/Fe] seems to \citep{Nissen2010},
which could perhaps be expected if  the nucleosynthetic channels of Na and Al are indeed related for halo stars.

\section{Chemical gradients in the Galactic disc}
\label{sec:gradients}
The observed radial and vertical stellar abundance distributions are interesting tools for studying the chemical enrichment history of the Galactic disc because they are sensitive to its formation process: inside-out disc formation leaves a radial gradient 
that may be partially erased by radial mixing; vertical gradients can be created by several secular processes such as the 
heating of discs, intrinsic or violent (mergers), and radial migration turning a radial gradient into a vertical one.
Several types of Milky Way objects are used in the search for chemical gradients.
The \ion{H}{II} regions, planetary nebulae, hot stars, cepheid variables, open clusters, and field stars are used to determine various gradients in the Galaxy (see \citealt{Maciel2010} and references therein).
Cepheid variables are good distance indicators, but they are massive and thus very young objects, therefore 
they provide the present-day abundance gradients.
Open clusters are very promising tools. Their distance and age are derived precisely (although model-dependent),
and the distance and age ranges they cover are wide, which is why open clusters are used in spatial and temporal investigation of the gradients.
However, only few open clusters have been analysed so far.
There are about 80 clusters that have spectroscopically determined metallicities (see. \citet{Heiter2014} and references therein).
Thanks to the GES, the number of clusters with spectroscopically derived parameters will increase significantly.
On the other hand, errors in distances and ages of field stars are significantly higher than for cepheids or open clusters, but the high number of objects allows better statistics.
To study the radial and vertical gradients of our sample, we adopted the distances, radial distance to the Galactic centre ($R_{\rm GC}$), and height above the plane (Z) from \citet{Blanco2014a}.
Since we aim at studying the radial and vertical gradients of the thin and thick discs separately, 
we adopted a stricter criterion to minimize accidental contamination from one component to the other.
To this effect, we introduce an artificial gap that removes stars that fall 
within $\pm$0.05~dex (according to the errors in Fig.~\ref{sec:ERROR}) of 
our separation line (see~Subsection~\ref{sec:Chemical_definition} and Appendix~\ref{sec:appendix1} for the effects of introducing this strict criterion).
Furthermore, to include more distant stars in our \textit{clean} sample we supplemented it 
with giant stars that meet the same parameter quality criteria as the dwarfs in the \textit{clean} sample; 28 giants were added this way.
We checked that the abundances of these 28 giant stars do not show any observable difference from the rest of the sample and remark that the main sample contains 237 giants by construction.
Moreover, that the intrinsic scatter is low (see Fig.~\ref{fig:SCATTER}), thus it can provide significant constraints to the gradients derived from our data.

\subsection{Geometrical extent of the thin- and thick-disc samples} 
\label{sec:geometrical_extent_discs}

   \begin{figure*}[!htb]
 \centering
 \includegraphics[scale=0.28]{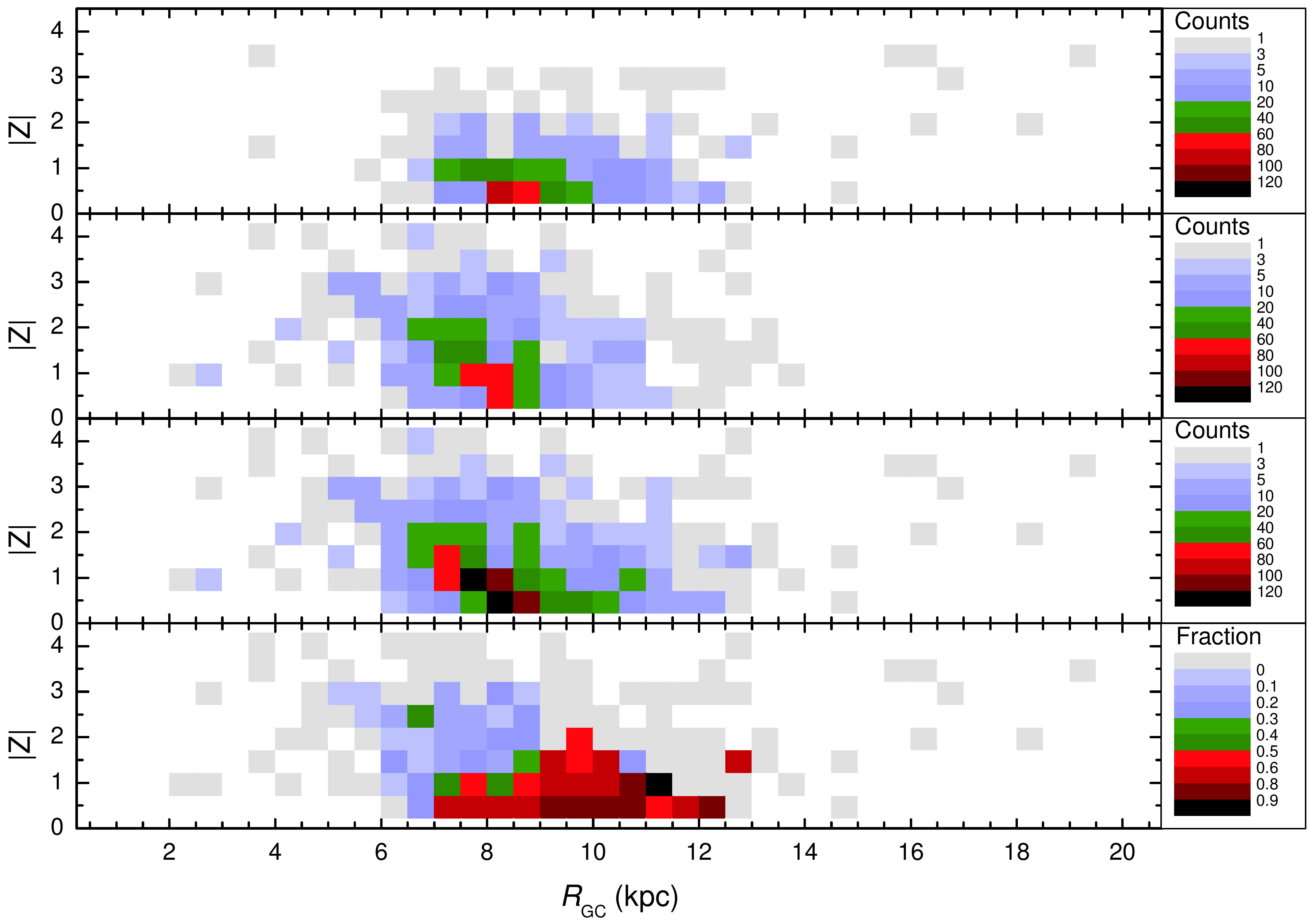}
%  \put (-330,202){\makebox[0.7\textwidth][r]{$\rm N_{thin}$}}%I want the text to go in the top right corner
%  \put (-330,145){\makebox[0.7\textwidth][r]{$\rm N_{thick}$}}%I want the text to go in the top right corner
%  \put (-330,90){\makebox[0.7\textwidth][r]{$\rm N_{thick}+\rm N_{thick}$}}%I want the text to go in the top right corner
%  \put (-330,35){\makebox[0.7\textwidth][r]{$\frac{\rm N_{thin}}{\rm N_{thin}+\rm N_{thick}}$}}%I want the text to go in the top right corner
 \put (-420,202){\makebox[0.7\textwidth][r]{$\rm N_{thin}$}}%I want the text to go in the top right corner
 \put (-420,145){\makebox[0.7\textwidth][r]{$\rm N_{thick}$}}%I want the text to go in the top right corner
 \put (-420,90){\makebox[0.7\textwidth][r]{$\rm N_{thick}+\rm N_{thick}$}}%I want the text to go in the top right corner
 \put (-420,35){\makebox[0.7\textwidth][r]{$\frac{\rm N_{thin}}{\rm N_{thin}+\rm N_{thick}}$}}%I want the text to go in the top right corner
 \caption{
 Distribution of thin-disc (\emph{top} panel), thick-disc (\emph{second} panel) and total (\emph{third} panel) samples in the in the $R_{\rm GC}$ vs. $|Z|$ plane. The \emph{bottom} panel shows the distribution of the ratio of thin-disc stars to total number of stars. 
  }
 \label{fig:R__Z_color_maps}
 \end{figure*}

    \begin{figure}[!htb]
 \centering
 \includegraphics[scale=0.24]{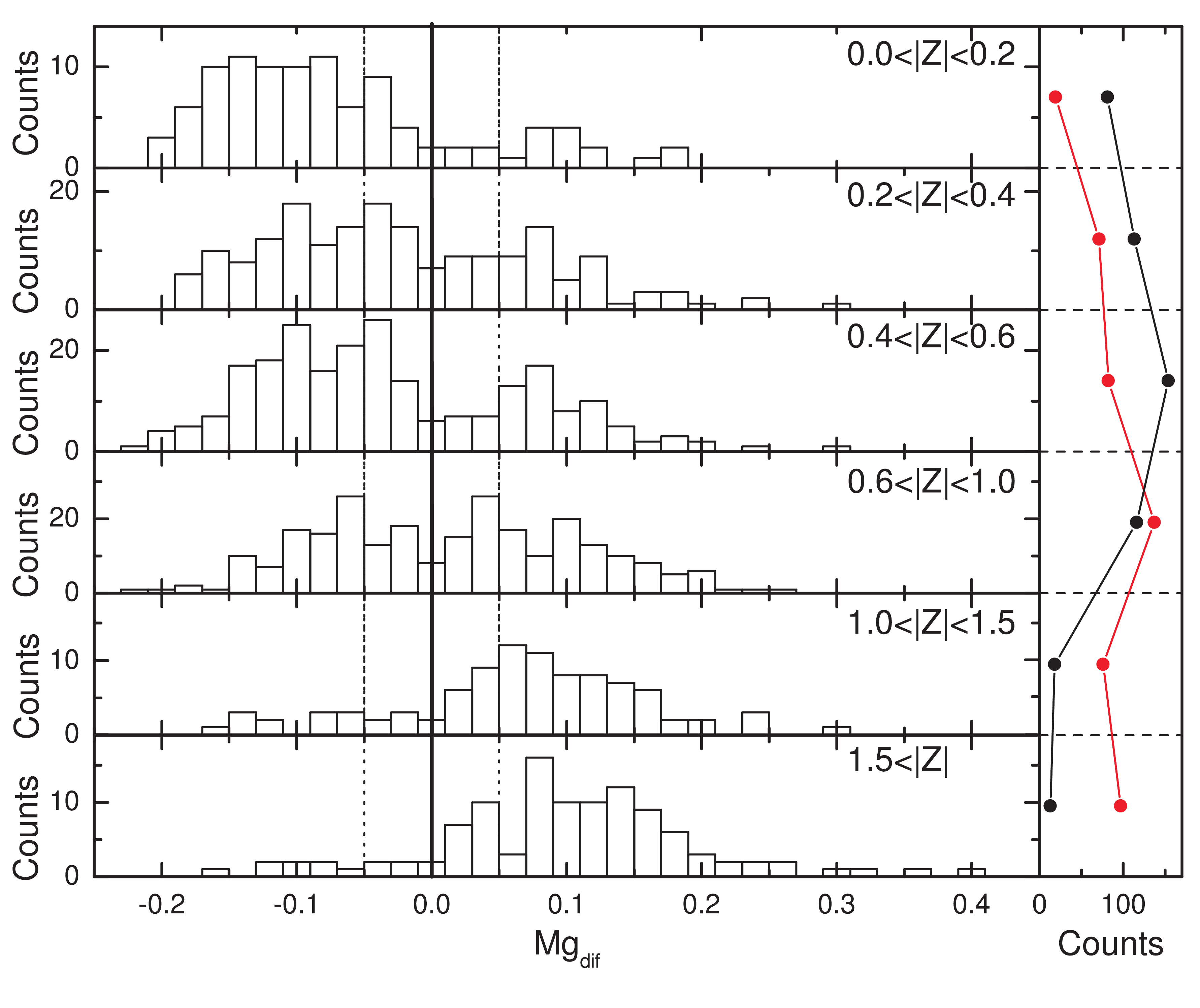}
 \caption{
  Distribution of $\rm \Delta[\ion{Mg}{I}/M] = [\ion{Mg}{I}/M] - [\ion{Mg}{I}/M]_{\rm separation}$ in $|Z|$ bins.
  $\rm \Delta[\ion{Mg}{I}/M]$ quantifies the distance in [\ion{Mg}{I}/M] of a given star to the thin- and thick-disc separating line (see Fig.~\ref{fig:XFE}).
  The black solid \textit{zero} vertical line marks the position of the separating line. The two vertical dotted lines at $\rm \Delta[\ion{Mg}{I}/M] \pm$0.05 dex mark the zone from which 
stars were excluded for the analysis of radial and vertical gradients. 
  The plot at the right side shows the total number of thin-disc (black dots) and thick-disc (red dots) stars in the same six $|Z|$ bins.
  }
 \label{fig:Z_HISTOGRAM}
 \end{figure}

To  understand the location of our sample in $R_{\rm GC}$ and $Z$, we plot in Fig.~\ref{fig:R__Z_color_maps} a density map of the targets in the {\em main} sample. The sample shows a clear extent in $R_{\rm GC}$ and $Z$ beyond the classical solar neighbourhood, up to 3\,kpc above or below the Galactic plane, and from galacto-centric radii 3 to 13\,kpc. 
This figure, also shows that our sample is dominated by stars farther away from the Galactic plane in the inner parts of the disc with respect to the typical heights probed in regions outside of the solar radius. Similarly, stars high above the plane tend to be located in regions farther away in the Galaxy than stars at low $|Z|$. Because of this effect, we restrict our thin-disc sample to $|Z|<$0.608\,kpc (see Sec.\ref{sec:radial_gradient}) in the following sections to probe the thin-disc radial gradients, and to stars within 1\,kpc of the solar radius ($R_{\rm GC} < 8\pm 1$\,kpc) to probe vertical gradients.

Furthermore, we also plot in Fig.~\ref{fig:R__Z_color_maps} the ratio of the chemically selected thin- and thick-disc samples (see Sec.~\ref{sec:properties_of_disc}) as a function of $Z$ and $R_{\rm GC}$. 
The thin-disc sample is clearly confined towards the plane of the Galaxy, while the thick-disc sample becomes dominant at large distances from the plane, 
as expected if the two populations have different scale heights. This is also shown in Fig.~\ref{fig:Z_HISTOGRAM} where we plot 
the distributions of $\rm \Delta[\ion{Mg}{I}/M] = [\ion{Mg}{I}/M] - [\ion{Mg}{I}/M]_{\rm separation}$, the distance of the star with respect to the 
separating line between the thin and thick disc in the [\ion{Mg}{I}/M] vs. [M/H] plane, as a function of $|Z|$ for all stars within 1\,kpc of 
the solar radius. The gradual appearance of the Mg-rich population with increasing distance from the Galactic plane is evident. 
At $|Z|\leq$0.608\,kpc, the Mg-poor population outnumbers the Mg-rich population.
The fact that $\alpha-poor$ stars are more confined towards the Galactic plane 
is has previously been described in the literature (see e.g. Fig.~7 by \citet{Boeche2013} and Fig.~10 by \citet{Blanco2014a}) as a thin- vs. thick-disc effect.
 
Another interesting fact becomes apparent in Fig.~\ref{fig:R__Z_color_maps}: if the chemically selected thin- and-thick disc populations have different scale heights but the same scale lengths, lines of constant $N_{\rm thin}/(N_{\rm thin}+N_{\rm thick})$ ratio would be expected to appear as horizontal lines in the representation of Fig.\ref{fig:R__Z_color_maps}. Instead, towards the inner parts of the Galaxy, the chemically selected thick-disc population remains dominant all the way to the Galactic plane, whereas at the solar radius or beyond, the thin-disc population dominates up to ~1\,kpc above the plane.
A larger the survey is desirable to better populate the upper right part of these figures (stars in the outer Galaxy with large Z) before drawing too strong a conclusion, but the distinctive excess of of Mg-rich stars close to the plane in the inner Galaxy and the lack of Mg-rich stars in the outer Galaxy suggests a shorter scale length for this population than for the thin-disc population. This agrees with the short scale length of the thick disc suggested by \citet{CLavers2005}, \citet{Juric2008}, \citet{Bensby2011},\citet{Cheng2012b}, and \citet{Bovy2012b}.

   \begin{figure*}[!htb]
 \centering
 \includegraphics[scale=0.32]{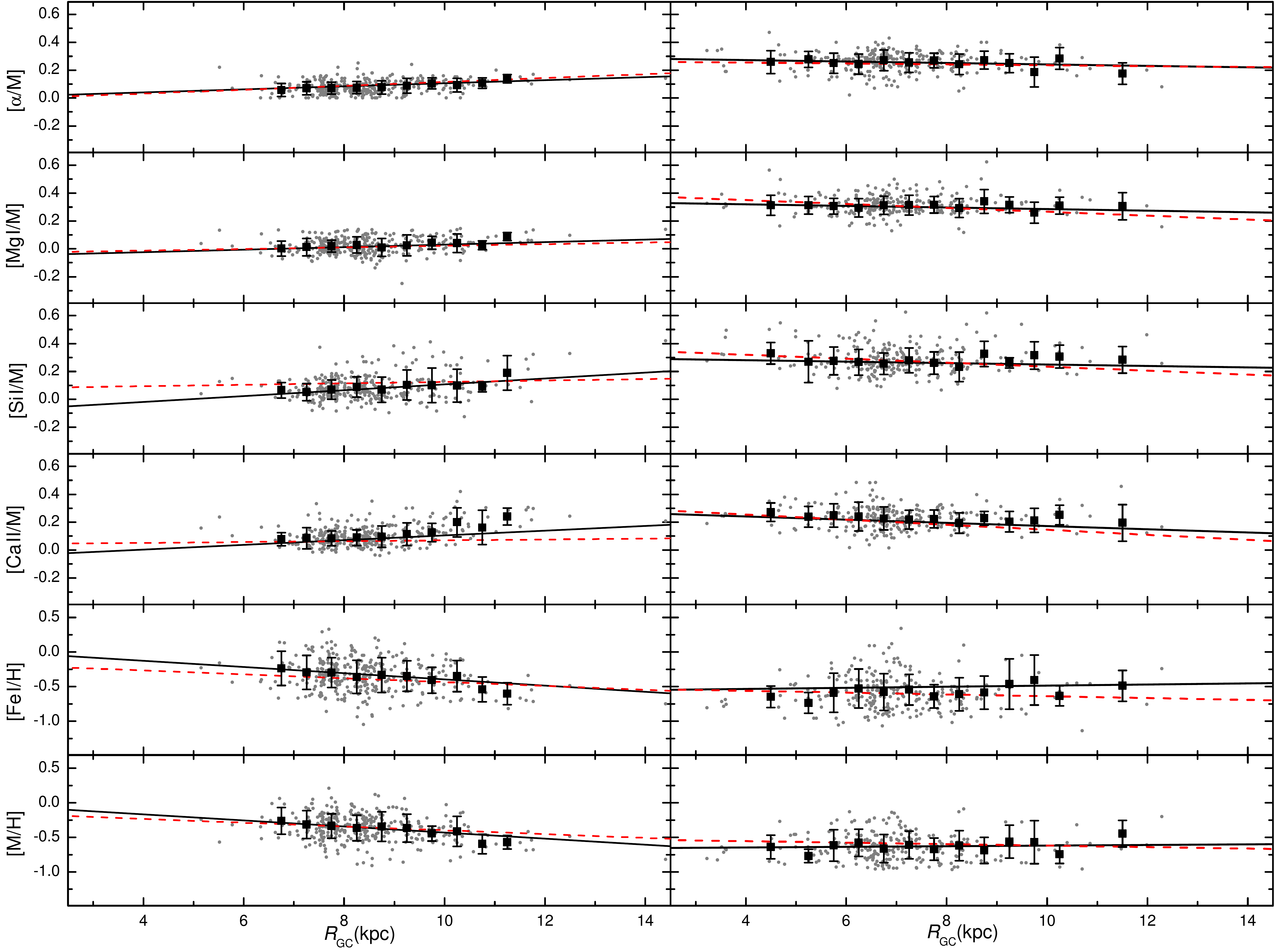}
 \caption{
 Abundances in the thin-disc (left column) and thick-disc (right column) stars of the \textit{main} sample as a function of galactocentric radii. 
 The lines (solid black for \textit{main} and dashed red for clean \textit{samples}) mark the linear fit to the data.
 Grey points mark the abundances of stars. Black squares show the mean value of stars in $R_{\rm GC}$ bins with standard deviations as error bars  (every bin contains at least 8 points).
  }
 \label{fig:R_gradients}
 \end{figure*}

\subsection{Radial metallicity gradients in Galactic thin- and thick-disc samples}
\label{sec:radial_gradient}

 To minimize the chances that the thin-disc stars are contaminated with thick-disc stars, since the 
 thin-disc population is outnumbered by the thick-disc population far from the plane (see Fig.~\ref{fig:Z_HISTOGRAM}), we restricted the maximum height allowed for thin-disc sample. We defined this maximum height as the $|Z|$ at which the cumulative number of thin-disc stars is twice higher than that of the thick-disc stars. This condition is met at $|Z|$=0.608~kpc, so that the thin-disc sample considered here contains only stars chemically selected to be in the thin disc and with $|Z|<$0.608~kpc. Conversely, to limit the impact of the $R_{\rm GC}$-$|Z|$ correlation outlined above (a significant number of thick-disc stars outside the solar circle are located close to the midplane), the thick-disc sample is restricted to heights above $|Z|$=1.07~kpc (defined as the height at which the cumulative number of thick-disc stars is twice that of the thin-disc stars).

In Fig.~\ref{fig:R_gradients} we present the abundance distribution in the Galactic disc as a function of $R_{\rm GC}$ (see Table~\ref{tab:Radial_GRADIENTS} for numerical values and Appendix~\ref{sec:appendix1} for samples with fewer restrictions).
We plot the abundances of the \textit{main} sample with error bars together with
the linear fits of the \textit{main} and \textit{clean} samples.
Our \textit{main} sample covers the range from 4~to~12~kpc. 
The general recommended metallicity ([M/H]) and recommended iron abundances ([\ion{Fe}{I}/H]) are presented separately. The iron abundance shows a significant gradient in the thin disc.
The slopes of both \textit{main} ($-0.044\pm0.009$~dex~kpc$^{-1}$) and \textit{clean} samples ($-0.028\pm0.018$~dex~kpc$^{-1}$) are similar (within the error limits) and support each other. 
The values of the metallicity gradient slopes of the \textit{main} ($-0.045\pm0.012$~dex~kpc$^{-1}$) and \textit{clean} samples ($-0.028\pm0.012$~dex~kpc$^{-1}$) also support the iron abundance gradients.
Using the same GES data (and the same distances, $R_{\rm GC}$ and Z) and a subsample constructed slightly differently, \citet{Blanco2014a} derived a metallicity gradient of $-0.058\pm$0.008~dex~kpc$^{-1}$, also similar to ours within the errors.

On the other hand, we find that the gradients in the thick disc are consistent with zero, within errors,
from both GES recommended metallicities \citep[in agreement with][]{Blanco2014a} and iron abundances, for both the \textit{main} and \textit{clean} samples
Hence, we find that the radial metallicity gradient flattens when moving from thin to thick discs. 

Th radial chemical gradients of the thick- and thin-discs have rarely been studied separately.
The recent works of \citet{Coskunoglu2012}, \citet{Bilir2012}, and \citet{Cheng2012a} provide 
metallicity gradients from large data samples.
\citet{Coskunoglu2012} and \citet{Bilir2012} kinematically separated the thin and thick discs in the RAVE survey and provided metallicity gradients separately for giant and dwarf stars for most probable thin-disc stars 
(giants: $-0.043\pm$0.005~dex~kpc$^{-1}$; dwarfs: $-0.033\pm$0.007~dex~kpc$^{-1}$; red clump: $-0.041\pm$0.003~dex~kpc$^{-1}$) and most probable thick-disc stars 
(giants: $-0.016\pm$0.011~dex~kpc$^{-1}$; dwarfs: $-0.010\pm$0.009~dex~kpc$^{-1}$; red clump: 0.017$\pm$0.008~dex~kpc$^{-1}$).
\citet{Boeche2013} found $\Delta$[Fe/H]/$\Delta R_{\rm GC}$=$-0.065\pm$0.003~dex~kpc$^{-1}$
for Z$_{\rm max}<$0.4~kpc and no slope for Z$_{\rm max}>$0.8~kpc. 
Thus we show that our results support the conclusions of \citet{Coskunoglu2012} 
even though we used a pure chemical division into thin and thick discs.
An analysis of metallicity gradients from the SEGUE survey was presented by \citet{Cheng2012a} and 
shows similar results. Their sample is divided into slices in $|Z|$. A slope of $-0.066\pm$0.037~dex~kpc$^{-1}$ is derived for the range 0.15$<|Z|<$0.25~kpc
where thin-disc stars dominate the sample, whereas the slope flattens farther from the Galactic plane 
until $\Delta$[Fe/H]/$\Delta R_{\rm GC}$=~0.003$\pm$0.006~dex~kpc$^{-1}$ in the 1.0$<|Z|<$1.5~kpc slice, where thick-disc objects dominates.
\citet{APrieto2006}, \citet{Juric2008}, and \citet{Katz2011} also found 
a significant negative gradient for thin-disc stars and no radial metallicity gradient at vertical heights $|Z|>$~1.0~kpc
in samples outside of the solar neighbourhood.
Metallicity gradients from the APOGEE survey were presented by \citet{Anders2013} and show similar results for Galactocentric metallicity gradients. Smilarly to SEGUE, their sample is also divided into $|Z|$ slices. 
Using what they defined as their \textit{Gold} sample, these authors found a slope of $\Delta$[Fe/H]/$\Delta R_{\rm GC}=-0.074\pm0.01$~dex~kpc$^{-1}$ in the range 0.0$<|Z|<$0.4~kpc (where thin-disc stars dominate) that flattens around $|Z|=~1.0$~kpc and changes sign to $\Delta$[Fe/H]/$\Delta R_{\rm GC}= 0.049\pm0.008$~dex~kpc$^{-1}$ in the range 1.5$<|Z|<$3.0~kpc  (where thick-disc stars dominate). 
Another study within GES by \citet{Bergemann2014}, based on the higher quality UVES spectra of far fewer stars, found a metallicty gradient of $\Delta$[Fe/H]/$\Delta R_{\rm GC}=-0.068\pm0.016$~dex~kpc$^{-1}$ for $|Z|<300$~pc.

Cepheids are very useful tools for deriving the radial gradients of the thin disc {\it at present times} (the ages of cepheid variables are younger than 200~Myr). The detected radial metallicity gradients range from $-0.029~dex~kpc^{-1}$ by \citet{Andrievsky2002} to $-0.052~dex~kpc^-1 \pm$0.003 \citep{Lemasle2008, Pedicelli2009}
and are compatible with our data for the thin-disc selected sample, although our sample is most probably significantly older than the Cepheids.
Owing to the sizable errors in stellar parameters in our sample, we were unable to quantify the individual stellar ages. However, taking into account that the selection window is centred on the old turn-off and does not cover hot and young stars, we can expect our stars to be older than 2~Gyr.
In addition, the majority of thin-disc stars should not be older than 7~Gyr. We therefore roughly expect that the our thin-disc sample contains stars of various ages between these limits and is not much older than 5~Gyr in the mean.

Based on GES UVES~spectra, which allow ages to be determined reliably, \citet{Bergemann2014} showed that the ages of their [Mg/Fe]-poor part of the sample are mostly younger than 5~Gyr.
The evolution of the disc in the past 5~Gyr has been slow, therefore it is expected, that the gradients of the thin disc are compatible with those obtained from the Cepheids.

Open-cluster ages range from younger than a Gyr to $\sim$9~Gyr 
and have therefore been used to examine the time evolution of the disc metallicity gradient.
Thus, data from the most recent studies of open clusters can be better used for a comparison with our sample. \citet{Andreuzzi2011} used high-resolution spectroscopy data and found an
overall metallicity gradient of $-$0.05~dex~kpc$^{-1}$, which agrees with our data.
However, some other works found a steeper gradient, especially when clusters from only the inner disc were considered. From the OCCAM survey data \citet{Frinchaboy2013} found an 
overall $\Delta$[Fe/H]/$\Delta R_{\rm GC}=-0.09\pm$0.03~dex~kpc$^{-1}$.
\citet{Magrini2009} reported $\Delta$[Fe/H]/$\Delta R_{\rm GC}$ to be $-0.053\pm$0.029~dex~kpc$^{-1}$ for ages younger than 0.8~Gyr, $-0.094\pm$0.008~dex~kpc$^{-1}$ for clusters up to 0.8--4~Gyr old, and $-0.091\pm$0.008~dex~kpc$^{-1}$ for clusters older than 4~Gyr. This latter gradient would be incompatible with our results.
\citet{Magrini2009} and \citet{Andreuzzi2011} found steeper gradients only for $R_{\rm GC}$ up to 12~kpc and a flat distribution beyond 12~kpc. 
GES also collects data for open clusters, with optimal sampling in age, [Fe/H], $R_{\rm GC}$ and mass; this will allow (among other things) studying the gradient and its evolution with homogeneous abundance determinations for open-cluster stars (e.g. \citet{Magrini2014}).

\subsection{Radial $\alpha$-elemental abundance gradients in Galactic thin and thick discs}
\label{sec:radial_gradient_alpha}

The first four panels of Fig.~\ref{fig:R_gradients} are dedicated to 
$\alpha$-elements radial gradients.
First we provide a galactocentric distribution of generally recommended [$\alpha$/M] ratios 
from \citet{Blanco2014b}, and then [\ion{Mg}{I}/M], [\ion{Si}{I}/M], and [\ion{Ca}{II}/M] (\ion{Ca}{I} and Ti are omitted because the errors on the abundance of these elements are higher in our \textit{main} sample). The values of the slopes are reported in Table~\ref{tab:Radial_GRADIENTS} for the \textit{main} and \textit{clean} samples. We find flat positive slopes of the [$\alpha$/M] and individual $\alpha$-elements in the thin-disc samples and flat ($\sim$2 sigma) negative slopes in the thick disc samples.
The slopes of the \textit{clean} sample  mostly confirm the slopes of the
\textit{main} sample. However, fewer thin-disc stars are located far from the solar radius
in the \textit{clean} sample, therefore \textit{clean} sample
slopes suffer larger errors (and are in general shallower) than the slopes of the \textit{main} sample.
There are very few published elemental radial gradients for field stars separated into thin- and 
thick-disc components, or as a function of $|Z|$. 
\citet{Boeche2013} derived radial slopes of magnesium, silicon, and aluminium
for samples from the RAVE survey as a function of $|Z_{\rm max}|$:
the thick-disc ($|Z_{\rm max}|$ range where the thick disc dominates) shows shallower negative slopes than the thin disc ($|Z_{\rm max}|$ range where the thin disc dominates); the change in slope for each $\alpha$-element between the thin- and thick-disc is about $\sim-0.014$ to $-0.018$. 
From APOGEE survey data, \citet{Anders2013} obtained a flat gradient in the thin disc-dominated-range 0.0$<|Z|<$0.4~kpc and a significant negative gradient of $\alpha$ elements in the thick-disc-dominated range 1.5$<|Z|<$3.0~kpc ($\Delta[\alpha/M]/\Delta R_{\rm GC}=-0.023\pm0.002$~dex~kpc$^{-1}$).
This differential behaviour between the thin and thick-disc agrees qualitatively and quantitatively with our data, even though the values for the gradients are quite different in our work and \citet{Boeche2013} or \citet{Anders2013}. The samples are of course quite different (different selection, different distance estimate methods, etc.), and  
the radial extents of RAVE and APOGEE are shorter than ours (($R_{\rm GC}$~=~4-12~kpc for GES vs. $R_{\rm GC}$~=~4.5-9.5~kpc for RAVE or $R_{\rm GC}$~=~6-11~kpc for APOGEE).
The UVES-based GES study by \citet{Bergemann2014} measured a magnesium gradient $\Delta$[\ion{Mg}{I}/Fe]/$\Delta R_{\rm GC}= 0.021\pm0.016$~dex~kpc$^{-1}$ for $|Z|<300$~pc, which also qualitatively agrees with our gradient.

The thin-disc gradients can also be compared with the works based on cepheid variables.
Slightly positive gradients of individual alpha elements seems to agree with the work by \citet{Luck2011}, who found positive gradients for individual $\alpha$-elements ($\Delta$[Mg/Fe]/$\Delta$$R_{\rm GC}$=0.014,
$\Delta$[Si/Fe]/$\Delta$$R_{\rm GC}$=0.014, $\Delta$[Ca/Fe]/$\Delta$$R_{\rm GC}$=0.021).
On the other hand, \citet{Lemasle2013} found slopes of
individual [$\alpha$/Fe] close to 0, assuming the derived
metallicity gradient derived by \citet{Lemasle2008} from the same data.

Open clusters seem also to agree with flat positive
gradients of [$\alpha$/Fe]  (\citealt{Yong2005, Jacobson2009, Magrini2009, Pancino2010, Yong2012}).
\citet{Yong2012} found positive slopes of $\Delta$[$\alpha$/Fe]/$\Delta$$R_{\rm GC}$, 
$\Delta$[Si/Fe]/$\Delta$$R_{\rm GC}$ and $\Delta$[Ca/Fe]/$\Delta$$R_{\rm GC}$
(0.01$\pm$0.01~dex~kpc$^{-1}$, 0.02$\pm$0.01~dex~kpc$^{-1}$ and 0.01$\pm$0.00~dex~kpc$^{-1}$ respectively), but no slope of $\Delta$[Mg/Fe]/$\Delta$$R_{\rm GC}$ (0.00$\pm$0.01~dex~kpc$^{-1}$).
\citet{Sestito2008} found no obvious trends, with the only exception of Ca, for which they found a significant positive slope 0.08$\pm$0.02.

\begin{table}
\caption{Radial abundance gradients in thick- and thin-disc stars ($\Delta X/\Delta R_{\rm GC}$),
errors on the slopes ($\epsilon$), and number of stars in the corresponding sub samples (n).
}
%\vspace{0.2cm}
\begin{tabular}{lcccccc}
\hline \hline
\smallskip
Line & $\frac{\Delta X}{\Delta R_{\rm GC}}$ & $\epsilon$ & n & $\frac{\Delta X}{\Delta R_{\rm GC}}$ & $\epsilon$ & n\\
 & $\rm \frac{dex}{kpc}$ & $\rm \frac{dex}{kpc}$  &   & $\rm \frac{dex}{kpc}$ & $\rm \frac{dex}{kpc}$ &  \\

\hline
  & \textit{main}  &   &   & \textit{clean}  &  &  \\

\hline
\multicolumn{6}{c}{Thin disc} \\ 
$[\alpha$/M] &	0.011	&	0.002	&	341	&	0.014	&	0.004	&	138	\\
$[\ion{Mg}{I}$/M] &	0.009	&	0.003	&	341	&	0.008	&	0.005	&	138	\\
$[\ion{Si}{I}$/M] &	0.021	&	0.004	&	341	&	0.015	&	0.009	&	138	\\
$[\ion{Ca}{II}$/M] &	0.017	&	0.003	&	341	&	0.009	&	0.005	&	138	\\
$[{\rm M}$/H]&	-0.045	&	0.012	&	341	&	-0.028	&	0.012	&	138	\\
$[\ion{Fe}{I}$/H] &	-0.044	&	0.009	&	341	&	-0.028	&	0.018	&	138	\\

\hline
\multicolumn{6}{c}{Thick disc} \\ 
$[\alpha$/M] &	-0.005	&	0.001	&	273	&	-0.005	&	0.006	&	57	\\
$[\ion{Mg}{I}$/M] &	-0.006	&	0.003	&	273	&	-0.014	&	0.005	&	57	\\
$[\ion{Al}{I}$/M] &	-0.004	&	0.002	&	273	&	-0.016	&	0.015	&	57	\\
$[\ion{Si}{I}$/M] &	-0.005	&	0.004	&	273	&	-0.014	&	0.006	&	57	\\
$[\ion{Ca}{II}$/M] &-0.011	&	0.004	&	273	&	-0.018	&	0.008	&	57	\\
$[{\rm M}$/H] &	0.005	&	0.005	&	273	&	-0.011	&	0.019	&	57	\\
$[\ion{Fe}{I}$/H] &	0.008	&	0.007	&	273	&	-0.021	&	0.029	&	57	\\

\hline
% & & Total Sample & 198,000 \\
 \label{tab:Radial_GRADIENTS}

\end{tabular}
\end{table}  

 \begin{figure*}[!htb]
 \centering
 \includegraphics[scale=0.32]{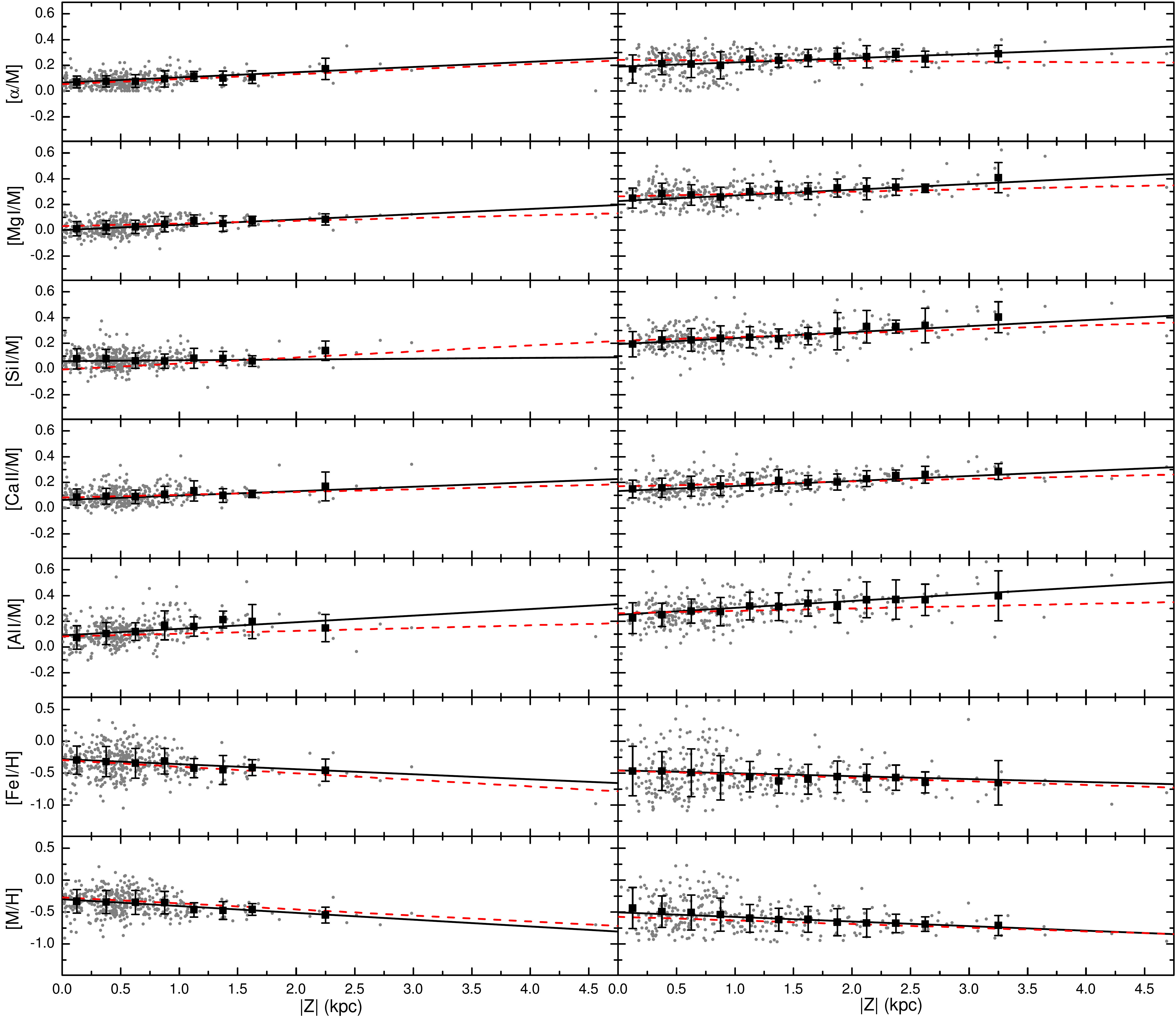}
 \caption{
 Vertical abundance distributions of thin-disc (left column) and thick-disc (right column) stars. 
 The full squares and error bars show the mean abundance and dispersion in bins of $|Z|$ (every bin contains at least 8 points).
 The lines (solid black for \textit{main} and dashed red for clean \textit{samples}) mark the linear fit to the data.
  }
 \label{fig:Z_gradients}
 \end{figure*}

\subsection{Vertical metallicity gradients in Galactic thin and thick discs}
\label{sec:vertical_gradient}
 
Because of the $R_{\rm GC}$-$|Z|$ correlation outlined in Sec.~\ref{sec:geometrical_extent_discs}, we restricted both the thin- and thick-disc samples to galactocentric radii close to the solar circle ($7 \leq R_{\rm GC}\leq 9$kpc). For the thin disc, the $R_{\rm GC}$-$|Z|$ correlation is weak, but for the thick disc, this additional selection prevents the numerous stars at high $Z$ in the inner disc from  completely dominating the high $Z$ bins, and conversely, the low $Z$ bins from being populated mostly by stars outside the solar circle. We checked that because there is no radial metallicity gradient in our thick-disc sample, ignoring this cautionary selection would have had only a marginal effect on the vertical gradient.

Figure~\ref{fig:Z_gradients} (bottom two rows) shows the vertical metallicity gradients of the thin and thick discs. The values of the slopes are reported in Table~\ref{tab:Vertical_GRADIENTS} (see Appendix~\ref{sec:appendix1} for samples with fewer restrictions), and
are clearly negative in both the \textit{main} and \textit{clean} subsamples
for $|Z|$  up to 5~kpc above/below the plane. The slopes in [\ion{Fe}{I}/H] and overall metallicity agree very well. 
The metallicity gradient in the thick disc is about 0.04$\pm$0.01~dex~kpc$^{-1}$ shallower than in the thin disc. 
The metallicity is very flat up to $|Z|$=1~kpc in both discs, and it is above this $|Z|$ that the gradients are significant.

\citet{Bilir2012} kinematically selected thin- from thick-disc stars in RAVE, and their sample spans up to around 2.8~kpc from the Galactic plane.
The results of \citet{Bilir2012} show a continuous metallicity gradient in the thin disc of $-0.109\pm$0.008~dex~kpc$^{-1}$, 
which is consistent with our values. Their thick-disc vertical gradient ($-0.034\pm$0.003~dex~kpc$^{-1}$), which is slightly shallower than our thick-disc vertical gradient.
\citet{Boeche2013} observed a slow metallicity decrease in the range $0<|Z_{\rm max}|<1.0$, where the-thin disc is dominant, compatible with the picture derived here, and a strong metallicity decline for $|Z_{\rm max}|>1.5$, where the thick disc is dominant in their sample.
\citet{Ruchti2011} also derived an iron abundance vertical gradient 
for metal-poor thick-disc stars ($-0.09\pm$0.05~dex~kpc$^{-1}$), which is consistent with the gradient determined here.

\subsection{Vertical $\alpha$-elemental and aluminium abundance gradients in Galactic thin and thick discs}
\label{sec:vertical_gradient_alpha}

This is one of the few efforts to estimate the vertical gradients 
of $\alpha$-element abundances (see Table~\ref{tab:Vertical_GRADIENTS}).
We detect positive gradients of $\alpha$-elements (relative to metallicity) in the thin- and thick-disc from the \textit{main} and \textit{clean} samples.
We also find a significant vertical [\ion{Al}{I}/M] abundance gradient, similarly steep in both discs (see~Fig.~\ref{fig:Z_gradients}).

Elemental abundance vertical gradients reported in the literature
are very few. Usually, authors did not separate the thin- and thick-disc and 
investigated common gradients, for example \citet{Boeche2013} showed 
positive [Mg/Fe] and [Si/Fe] vs.  $|Z_{\rm max}|$ gradients for a mixture of thin- and thick-disc stars.
However, it is evident from their Fig.~7 that [Mg/Fe] and [Si/Fe] grow much faster with $|Z_{\rm max}|$ in the range $|Z_{\rm max}|=1.4$ to 2.0 than at lower $|Z_{\rm max}|$, which is expected when moving from thin-disc-dominated to thick-disc-dominated regimes.
In agreement with our results, \citet{Ruchti2011} found
a flat vertical gradient in $\alpha$ enhancement (up to 0.03$\pm$0.02~dex~kpc$^{-1}$)
in their sample of metal-poor thick-disc stars.
They detected vertical gradients of [Mg/Fe] (0.03$\pm$0.02~dex~kpc$^{-1}$) and [Si/Fe] (0.02$\pm$0.01~dex~kpc$^{-1}$),
but no slope for [Ca/Fe].

There are no published works that describe the aluminium abundance as a function of vertical distance from the Galactic plane, separated into thin- and thick-disc components. However, 
\citet{Boeche2013} showed a steady increase of [Al/Fe] in the thin-disc-dominated range of $|Z_{\rm max}|$ and also in $|Z_{\rm max}|>$1, where the thick disc dominates.
The similarity of the vertical aluminium abundance gradients to those of $\alpha$-elements confirms that aluminium and  $\alpha$-elements have a similar origin. Indeed, aluminium abundances in the thin and thick discs show clearly distinct trends.

\begin{table}
\caption{Vertical abundance gradients in thick- and thin-disc stars ($\Delta X/\Delta |Z|$),
errors on the slopes ($\epsilon$), and number of stars in the corresponding sub samples (n).}
\label{tab:Vertical_GRADIENTS}
%\vspace{0.2cm}
\begin{tabular}{lcccccc}
\hline \hline
\smallskip
Line & $\frac{\Delta X}{\Delta |Z|}$ & $\epsilon$ & n & $\frac{\Delta X}{\Delta |Z|}$ & $\epsilon$ & n\\
 & $\rm \frac{dex}{kpc}$ & $\rm \frac{dex}{kpc}$  &   & $\rm \frac{dex}{kpc}$ & $\rm \frac{dex}{kpc}$ &  \\

\hline
  & \textit{main}  &   &   & \textit{clean}  &  &  \\

\hline
\multicolumn{6}{c}{Thin disc} \\ 
$[\alpha$/M]      &	0.041	&	0.004	&	397	&	0.036	&	0.006	&	185	\\
$[\ion{Mg}{I}$/M]  &	0.041	&	0.006	&	397	&	0.044	&	0.008	&	185	\\
$[\ion{Al}{I}$/M]  &	0.051	&	0.006	&	389	&	0.020	&	0.008	&	185	\\
$[\ion{Si}{I}$/M]  &	0.006	&	0.008	&	387	&	0.018	&	0.011	&	185	\\
$[\ion{Ca}{II}$/M] &	0.034	&	0.010	&	397	&	0.039	&	0.013	&	185	\\
$[{\rm M}$/H]      &	-0.079	&	0.013	&	397	&	-0.070	&	0.015	&	185	\\
$[\ion{Fe}{I}$/H]  &	-0.107	&	0.009	&	397	&	-0.057	&	0.016	&	185	\\

\hline
\multicolumn{6}{c}{Thick disc} \\ 
$[\alpha$/M]      &	0.033	&	0.002	&	334	&	0.011	&	0.005	&	104	\\
$[\ion{Mg}{I}$/M]  &	0.044	&	0.004	&	334	&	0.024	&	0.008	&	104	\\
$[\ion{Al}{I}$/M]  &	0.054	&	0.004	&	319	&	0.041	&	0.006	&	104	\\
$[\ion{Si}{I}$/M]  &	0.047	&	0.005	&	322	&	0.026	&	0.010	&	104	\\
$[\ion{Ca}{II}$/M] &	0.039	&	0.007	&	334	&	0.018	&	0.011	&	104	\\
$[{\rm M}$/H]      &	-0.046	&	0.010	&	334	&	-0.013	&	0.010	&	104	\\
$[\ion{Fe}{I}$/H]  &	-0.072	&	0.006	&	334	&	-0.037	&	0.016	&	104	\\

\hline
% & & Total Sample & 198,000 \\

\end{tabular}
\end{table}

\subsection{Implications for Galactic disc formation}

Thanks to the increasing extent of surveys that now reach far in galactocentric radius and height above the plane, any 
formation model of the disc must be able to account not only for local properties of the thin and thick disc, but also for their radial and vertical distributions. 
Such a model would have to explain not only the dynamical properties 
of the disc, but also its chemical distributions.

We here analysed a large number of stars, widely spread in the Galaxy, from which we 
derived vertical (0~to~3.5~kpc) and radial (4~to~12~kpc) chemical gradients for two chemically separated populations that we assigned to the thin and thick discs.
We recall that 
\citet{Blanco2014a} studied the kinematics for the same sample, separated into thin and thick disc in a similar manner, and found that (i) the difference in rotational velocities ($\rm V_{\phi}$), vertical and azimuthal velocity dispersions ($\sigma_z$, $\sigma_\phi$) and orbital eccentricities between the two populations support the existence of two (chemically separated) populations with different origins (see also their Fig.~12, where the gradient in $\rm V_{\phi}$ upon changing the chemical composition of stars is visible, independently of the separation itself); (ii) this distinction is smallest above [M/H]$=-0.25$ where the two populations may be related or harder to distinguish from each other;
(iii) a vertical dependence of $\rm V_{\phi}$ vs. $|Z|$ ($\rm V_{\phi}$ decreasing far from the plane) is detected in the thick disc; (iii) a very clear positive trend of $\rm V_{\phi}$ with metallicity is seen in the thick disc, while the thin disc in contrast shows a weak negative trend; (iv) orbital eccentricities correlate with metallicity in the thin, but not in the thick disc.

Taken together with the bi-modality of the $\alpha$-elements observed in our GES iDR1 sample, this clear kinematical distinction between the thin- and thick-discs sample, strongly argues in favour of two distinct populations. 
This is somewhat in contrast with the recent work of \citet{Bovy2012a}, who claimed that the SEGUE survey, after it was corrected for the survey selection effects, is compatible with a single Galactic disc with continuously changing characteristics over time: mono-abundance (chemically selected) sub-populations of the Galactic disc show a continuum of structural \citep{Bovy2012b} and kinematical \citep{Bovy2012c} properties and not two distinct structures. 
It is beyond the scope of this paper, which only uses a small fraction of what GES will provide to examine  the survey selection effects in detail, but these effects will have to be taken into account in the future. We would like to note, however, that our chemical separation is based on higher accuracy and more numerous abundances (we recall that SEGUE data are based on low-resolution spectra). This should help to define cleaner chemically distinct (or mono-abundance) samples, which is reflected for example by the fact that our thin and thick discs co-exist in wide metallicity range, whereas the $\alpha$-rich population of SEGUE hardly overlap the $\alpha$-rich population in metallicity.
Below we examine the impact of our chemically separated thin- and thick-disc stars as if they were indeed two populations with different origins. 
The study of thin-disc gradients can provide strong evidence for the mechanism of Galaxy formation (e.g. \citealt{Chiappini1997, Alibes2001a, Alibes2001b, Cescutti2007}). Our elemental abundance radial gradients agree with the models of \citet{Cescutti2007}, which assumed an inside-out build-up of the disc on a time-scale of 7~Gyr in the solar neighbourhood (see \citealt{Matteucci1989, Cescutti2007}).
The inside-out scenario is also supported by \citet{Pilkington2012}, who examined radial and vertical metallicity gradients of several galaxies using a suite of disc-galaxy hydrodynamical simulations. They found most of the models predict radial gradients today that are consistent with those observed in late-type discs, but they evolve to this self-similarity in different fashions, although each adheres to classical inside-out growth. 
On the other hand, is it interesting to note that \citet[][based on the local sample of \citealt{Adibekyan2012}]{Haywood2013}, advocated an outside-in formation of the disc, based on the age-metallicity relations of the thin and thick discs. The metal-poor thin disc is found by \citet{Haywood2013} to be as old and as $\alpha$-rich as the youngest part of the thick disc, but of higher metallicity and rotational velocities, which argues for this population to originate in the outer parts of the Galaxy from where they would have migrated inwards. We note here that, the metal-poor part of our thin disc (with metallicities below $-0.6$) seems to be confined to regions outside the solar circle, which agrees with this scenario, although we cannot probe whether these stars are indeed old. 
This scenario also allows for a positive [$\alpha$/Fe] radial gradient.

For the thick-disc, the situation is even more complicated, since the different thick-disc formation scenarios do not necessarily provide clear predictions for the radial gradient in the thick-disc today. Some would be ruled out if a gradient \emph{does exist}, but the current results of no radial metallicity gradient in the thick-disc do not help to rule out any of these formation scenarios. 
This is discussed at length for example in \citet{Cheng2012a}, who, in agreement with our results, obtained a flat metallicity gradient in the thick disc (selected as stars far from the midplane). Here we add to the flat radial metallicity gradient (Sec.~\ref{sec:radial_gradient}) a mild negative [$\alpha$/M] gradient (as traced by Mg, Si, and Ca, see Sec.~\ref{sec:radial_gradient_alpha}), and define these gradients on chemically selected thick-disc stars, allowing us to also trace the vertical behaviour of this population, which is found to be negative in metallicity (Sec.~\ref{sec:vertical_gradient}) and positive in $\alpha$-elements (Sec.~\ref{sec:vertical_gradient_alpha}). 

Among the four main thick-disc building scenarios mentioned in Sec.\ref{sec:introduction} (i) vertical heating of a pre-existing primary disc by a minor merger, (ii) formation by an early accretion of gas (gas-rich merger or gas flows) forming the thick disc \textit{in situ}, (iii) direct accretion of small satellites, or (iv) formation of a puffed-up structure by mere radial migration (churning and blurring) of stars in the disc), some expect vertical gradients in metallicity and $\alpha$-enrichment
while others do not. The predictions of scenarios (i) and (iv) depend heavily on the chemical structure of the pre-existing (primary) disc, while to the first order at least, scenarios (ii) and (iii) predict no gradients in the thick disc, and would be ruled out based on the $\alpha$ gradients observed here, as well as some of the kinematical arguments presented in \citet{Blanco2014a} ($\rm V_{\phi}$ correlation with metallicity in the thick disc for example). Below we comment in detail on some recent efforts to model some of these scenarios chemo-dynamically.

The dynamical heating of a primary disc by violent merger \citep[e.g.][]{Kazantzidis2008, Villalobos2008, Qu2011} of a smaller satellite can heat the stars into thick-disc-like scale heights.
If vertical chemical gradients were present in the primary disc, they would be preserved: 
the least-bound disc objects, which tend to be more metal poor and older, 
achieve higher scale heights and can create the observed negative vertical metallicity gradient, and possibly a positive gradients in $\alpha$-enrichment.
This scenario was explicitly modelled by \citet{Bekki2011}.
The vertical metallicity gradient from their model is expected to be stronger 
in regions closer to the Galactic centre and to slowly becom shallower 
in the outer parts of the disc.
\citet{Bekki2011} furthermore showed that the initial vertical metallicity gradient
of the first-generation Galactic thin disc (FGTD) can change owing to dynamical influences of minor
mergers and the stellar bar.
When they introduced no metallicity gradient in their model FGTD,
they later observed a negative vertical metallicity gradient in the thick disc, 
which was steeper in the inner regions ($R_{\rm GC}$$<$7 kpc).
This scenario is compatible with the observations discussed here.

\citet{Minchev2014} developed a very interesting chemodynamical model where the formation and evolution of the disc allows for stellar migration, triggered by mergers in the early epochs, and then by secular processes (bar and spiral arms). In their model, the vigorous migration induced by early mergers is mandatory to explain the velocity dispersion of old stars in the solar neighbourhood. The chemistry part of the model is a thin-disc chemical evolution model, with a disc forming inside out, but the resulting model shows features that resemble what would be classified as a thick-disc. The gradients that we measured here broadly agree with their predictions, which match the flattening of the radial metallicity gradient with distance from the midplane particularly well, and change of slope of [$\alpha$/M] vs. galactocentric distances from a positive gradient for stars close to the mid-plane to a shallow negative gradient for stars far from the plane. These two features are the result of the 
lack of young stars far from the plane in the outer disc. On the other hand, an interesting feature of the model is that the radial gradients of mono-age populations are constant at all heights. It is beyond the present accuracy of GES iDR1 GIRAFFE data to measure accurate ages (it may become feasible with the Gaia catalogue), but if one takes the alpha enrichment as a proxy for age, one would expect that the metallicity gradient of $\alpha$-rich and $\alpha$-poor stars to be the same. The current $R_{\rm GC}$ and $Z$ coverage of GES iDR1, and their correlation (see Sec.~\ref{sec:geometrical_extent_discs}), does not allow performing this test fully yet, but this should certainly be within reach of future data releases of the survey.

The radial migration processes studied by \citet{Schonrich2009a, Schonrich2009b} and \citet{Roskar2008} 
examine the continuous radial and vertical re-arragement (radial mixing via blurring and churning) by triggered spiral-arms and molecular clouds. 
The metal-poorer and older stars are more affected by heating and achieve higher $|Z|$,
thus producing negative metallicity and small positive $\alpha$-enrichment gradients.

The dominant channel that has given rise to the population that we call the thick disc is therefore not very clear yet.
Our results, in agreement with results of several other studies,
show that our chemically-separated thick disc hosts vertical gradients in metallicity and [$\alpha$/M] that could agree qualitatively with scenario (i) of a violent dynamical heating of a primary disc, while scenario (iv) could also potentially reproduce the observed gradients. We also note that the orbital eccentricity distribution for the GES iDR1 thick-disc sample \citep[see][]{Blanco2014a} supports the heating scenario.

\section{Summary and concluding remarks}
\label{sec:conclusions}

Elemental abundances directly reflect the processes of the stellar formation and thus the history of a given stellar population.
The main contributors to the interstellar medium are supernovae type-Ia, type-II, and AGB stars.
The final picture of stellar abundances is predetermined
by a complex interplay: which supernovae type was predominant 
in the region of the stellar nursery, what was the typical mass of SN-II in the region, 
when the mechanism of SN-Ia switched on, and how much the region was contaminated by SN-Ia ejecta.
Up to now, most high-resolution spectroscopic studies of the Galactic disc(s) were confined to objects in the Solar vicinity. Our study extended the volume in which detailed elemental abundances are measured, using the GIRAFFE observations from the Gaia-ESO first internal data release (see figure~9 in \citealt{Blanco2014a}) for $\sim$2\,000 FGK dwarfs and giants. 
We showed that the general behaviour of element-to-iron ratios is very similar to those described by other authors in the solar neighbourhood (\citealt{Neves2009, Adibekyan2012, Ishigaki2013} etc.). 
The $\alpha$-elements show a bimodal distribution \citep[see also][]{Blanco2014a}, among which [\ion{Mg}{I}/M] has the most prominent separation, which we used to chemically define two populations that we attributed to the thin- and thick-discs. These divided thin- and thick-discs showed separations in most other $\alpha$-elements (Si, Ti, Ca, and the globally recommended [$\alpha$/M] parameter determined together with stellar parameters in GES iDR1).
Calcium showed the least clear separation, as expected because it has a stronger contribution from SN-Ia than other $\alpha$-elements.
Moreover, that dispersions for most of elements are compatible with the measurement errors and the the intrinsic scatter is low or even non-detectable.

Although GES iDR1 contains relatively few metal-poor stars (102), we found the distinction between the two halo populations advocated by \citet{Nissen2010}, and agree with these authors that this difference is greater for magnesium-to-iron ratios than for calcium or silicon, possibly for the same reason as above. 
According to these authors, the unexpected $\alpha$-poor stars could be
the former members of accreted satellite galaxies
where SN-Ia and SN-II yields were different, as is seen in the studies of the 
smaller satellite galaxies \citep[see for example][and references therein]{Venn2004,Letarte2006,Tolstoy2009,Swaelmen2013}.

Using the distances from \citet{Blanco2014a}, we also showed that the GES iDR1 (GIRAFFE) data cover a wide range in galacto-centric radius and distance to the mid-plane, allowing us to probe chemical gradients in an extended Galactic region. As expected, we found that the population identified as the thick disc is more extended vertically than the thin disc \citep[see for example][and reference therein]{Boeche2013, Blanco2014a}, which is expected for a population with a larger scale height \citep[e.g.][]{Gilmore1983} that is also more centrally concentrated towards the inner Galaxy. This indicates at a shorter scale length for this population, similarly to the findings by \citet{Bensby2011}, \citet{Cheng2012b}, or \citet{Bovy2012b}. We then took into account the identified correlation between $R_{\rm GC}$ and $|Z|$ in the sample (which might blur the radial and vertical signatures) and derived the radial and vertical gradients in metallicity, four $\alpha$-element abundances (Mg, Si, Ca, Ti), and aluminium.

As found by \citet{Blanco2014a}, a radial metallicity gradient is present in the thin disc, with a slope similar to that of the younger tracers such as cepheids \citep{Lemasle2008, Pedicelli2009}, but also to the gradient measured by other large surveys such as SEGUE or RAVE for field stars close to the Galactic plane \citep{Cheng2012a, Boeche2013}; this gradient is confirmed in our own iron measurements. 
Positive radial individual $\alpha$-over-iron (Mg/Fe, Si/Fe, Ca/Fe) gradients were also observed in the thin-disc, matching the gradients observed in cepheids by some authors \citep{Luck2011} and open clusters (\citealt{Magrini2009, Pancino2010, Yong2012}), but at variance with the gradients observed by \citet{Boeche2013} in RAVE. 
The thin disc also hosts a negative vertical metallicity gradient in the solar cylinder \citep[similar to][from a kinematically selected thin-disc sample in RAVE]{Bilir2012}, accompanied by positive [$\alpha$/Fe] and [Al/Fe] gradients, compatible with a thin-disc where older stars (where [$\alpha$/Fe] is taken as a proxy for age) are kinematically hotter than younger thin-disc stars. 
 
The thick-disc, on the other hand, presented no radial metallicity gradient, in agreement with the SEGUE (\citealt{Cheng2012a}) or RAVE (\citealt{Boeche2013}) samples that were selected far from the mid-plane, and a shallower (negative) vertical metallicity gradient than the thin disc. The thick disc displays shallow $\alpha$-element radial gradients detected at the $\sim 2 \sigma$ level, in the opposite sense as in the thin disc (decreasing [$\alpha$/Fe] towards large $R_{\rm GC}$), and positive vertical [$\alpha$/Fe] and [Al/Fe] gradients.

Together with \citet{Blanco2014a}, this is the first effort to describe the detailed chemical structure of the Galactic disc from the Gaia-ESO Survey first internal data-release of GIRAFFE spectra. Clearly future GES data releases will bring an order of magnitude better  statistics and a more isotropic set of lines of sight and will allow a much more complete description of the structure and origin of the Galactic components. Furthermore, combining the larger GES GIRAFFE sample with the more detailed UVES counterpart of the survey will yield particularly interesting insights.

\begin{acknowledgements}
\v{S}.~Mikolaitis, V.~Hill, A.~Recio-Blanco and P. de Laverny acknowledge the the support of the French Agence Nationale de la Recherche, under contract ANR-2010-BLAN- 0508-01OTP, and the Programme National de Cosmologie et Galaxies.
Based on data products from observations made with ESO Telescopes at the La Silla Paranal Observatory under programme \textbf{ID 188.B-3002}. This work was partly supported by the European Union FP7 programme through ERC grant number 320360 and by the Leverhulme Trust through grant RPG-2012-541. We acknowledge the support from INAF and Ministero dell' Istruzione, dell' Universit\`a' e della Ricerca (MIUR) in the form of the grant "Premiale VLT 2012". The results presented here benefit from discussions held during the Gaia-ESO workshops and conferences supported by the ESF (European Science Foundation) through the GREAT Research Network Programme. This research has made use of SIMBAD (operated at CDS, Strasbourg, France), and NASA's Astrophysics Data System.  T.~Bensby was funded by grant No. 621-2009-3911 from The Swedish Research Council. S.~G.~Sousa acknowledges support from the Funda\c{c}\~ao para a Ci\^encia e Tecnologia, FCT (Portugal) in the form of the fellowship SFRH/BPD/47611/2008.

\end{acknowledgements}

\bibliographystyle{aa} % style aa.bst

\begin{thebibliography}{115}
\expandafter\ifx\csname natexlab\endcsname\relax\def\natexlab#1{#1}\fi

\bibitem[{{Abadi} {et~al.}(2003){Abadi}, {Navarro}, {Steinmetz}, \&
  {Eke}}]{Abadi2003}
{Abadi}, M.~G., {Navarro}, J.~F., {Steinmetz}, M., \& {Eke}, V.~R. 2003, \apj,
  597, 21

\bibitem[{{Adibekyan} {et~al.}(2011){Adibekyan}, {Santos}, {Sousa}, \&
  {Israelian}}]{Adibekyan2011}
{Adibekyan}, V.~Z., {Santos}, N.~C., {Sousa}, S.~G., \& {Israelian}, G. 2011,
  \aap, 535, L11

\bibitem[{{Adibekyan} {et~al.}(2012){Adibekyan}, {Sousa}, {Santos}, {Delgado
  Mena}, {Gonz{\'a}lez Hern{\'a}ndez}, {Israelian}, {Mayor}, \&
  {Khachatryan}}]{Adibekyan2012}
{Adibekyan}, V.~Z., {Sousa}, S.~G., {Santos}, N.~C., {et~al.} 2012, \aap, 545,
  A32

\bibitem[{{Alib{\'e}s} {et~al.}(2001{\natexlab{a}}){Alib{\'e}s}, {Labay}, \&
  {Canal}}]{Alibes2001b}
{Alib{\'e}s}, A., {Labay}, J., \& {Canal}, R. 2001{\natexlab{a}}, arXiv:0107016

\bibitem[{{Alib{\'e}s} {et~al.}(2001{\natexlab{b}}){Alib{\'e}s}, {Labay}, \&
  {Canal}}]{Alibes2001a}
{Alib{\'e}s}, A., {Labay}, J., \& {Canal}, R. 2001{\natexlab{b}}, \aap, 370,
  1103

\bibitem[{{Allende Prieto} {et~al.}(2004){Allende Prieto}, {Barklem},
  {Lambert}, \& {Cunha}}]{APrieto2004}
{Allende Prieto}, C., {Barklem}, P.~S., {Lambert}, D.~L., \& {Cunha}, K. 2004,
  \aap, 420, 183

\bibitem[{{Allende Prieto} {et~al.}(2006){Allende Prieto}, {Beers}, {Wilhelm},
  {Newberg}, {Rockosi}, {Yanny}, \& {Lee}}]{APrieto2006}
{Allende Prieto}, C., {Beers}, T.~C., {Wilhelm}, R., {et~al.} 2006, \apj, 636,
  804

\bibitem[{{Alvarez} \& {Plez}(1998)}]{Alvarez1998}
{Alvarez}, R. \& {Plez}, B. 1998, \aap, 330, 1109

\bibitem[{{Anders} {et~al.}(2013){Anders}, {Chiappini}, {Santiago},
  {Rocha-Pinto}, {Girardi}, {da Costa}, {Maia}, {Steinmetz}, {Minchev},
  {Schultheis}, {Boeche}, {Miglio}, {Montalb{\'a}n}, {Schneider}, {Beers},
  {Cunha}, {Allende Prieto}, {Balbinot}, {Bizyaev}, {Brauer}, {Brinkmann},
  {Frinchaboy}, {Garc{\'{\i}}a P{\'e}rez}, {Hayden}, {Hearty}, {Holtzman},
  {Johnson}, {Kinemuchi}, {Majewski}, {Malanushenko}, {Malanushenko},
  {Nidever}, {O'Connell}, {Pan}, {Robin}, {Schiavon}, {Shetrone}, {Skrutskie},
  {Smith}, {Stassun}, \& {Zasowski}}]{Anders2013}
{Anders}, F., {Chiappini}, C., {Santiago}, B.~X., {et~al.} 2013,
  arXiv:1311.4549

\bibitem[{{Andreuzzi} {et~al.}(2011){Andreuzzi}, {Bragaglia}, {Tosi}, \&
  {Marconi}}]{Andreuzzi2011}
{Andreuzzi}, G., {Bragaglia}, A., {Tosi}, M., \& {Marconi}, G. 2011, \mnras,
  412, 1265

\bibitem[{{Andrievsky} {et~al.}(2002){Andrievsky}, {Kovtyukh}, {Luck},
  {L{\'e}pine}, {Bersier}, {Maciel}, {Barbuy}, {Klochkova}, {Panchuk}, \&
  {Karpischek}}]{Andrievsky2002}
{Andrievsky}, S.~M., {Kovtyukh}, V.~V., {Luck}, R.~E., {et~al.} 2002, \aap,
  381, 32

\bibitem[{{Andrievsky} {et~al.}(2008){Andrievsky}, {Spite}, {Korotin}, {Spite},
  {Bonifacio}, {Cayrel}, {Hill}, \& {Fran{\c c}ois}}]{Andrievsky2008}
{Andrievsky}, S.~M., {Spite}, M., {Korotin}, S.~A., {et~al.} 2008, \aap, 481,
  481

\bibitem[{{Baumueller} \& {Gehren}(1997)}]{Baumueller1997}
{Baumueller}, D. \& {Gehren}, T. 1997, \aap, 325, 1088

\bibitem[{{Bekki} \& {Tsujimoto}(2011)}]{Bekki2011}
{Bekki}, K. \& {Tsujimoto}, T. 2011, \apj, 738, 4

\bibitem[{{Bensby} {et~al.}(2011){Bensby}, {Alves-Brito}, {Oey}, {Yong}, \&
  {Mel{\'e}ndez}}]{Bensby2011}
{Bensby}, T., {Alves-Brito}, A., {Oey}, M.~S., {Yong}, D., \& {Mel{\'e}ndez},
  J. 2011, \apjl, 735, L46

\bibitem[{{Bensby} {et~al.}(2003){Bensby}, {Feltzing}, \&
  {Lundstr{\"o}m}}]{Bensby2003}
{Bensby}, T., {Feltzing}, S., \& {Lundstr{\"o}m}, I. 2003, \aap, 410, 527

\bibitem[{{Bensby} {et~al.}(2004){Bensby}, {Feltzing}, \&
  {Lundstr{\"o}m}}]{Bensby2004}
{Bensby}, T., {Feltzing}, S., \& {Lundstr{\"o}m}, I. 2004, \aap, 415, 155

\bibitem[{{Bensby} {et~al.}(2005){Bensby}, {Feltzing}, {Lundstr{\"o}m}, \&
  {Ilyin}}]{Bensby2005}
{Bensby}, T., {Feltzing}, S., {Lundstr{\"o}m}, I., \& {Ilyin}, I. 2005, \aap,
  433, 185

\bibitem[{{Bensby} {et~al.}(2014){Bensby}, {Feltzing}, \& {Oey}}]{Bensby2014}
{Bensby}, T., {Feltzing}, S., \& {Oey}, M.~S. 2014, \aap, 562, A71

\bibitem[{{Bergemann}(2011)}]{Bergemann2011}
{Bergemann}, M. 2011, \mnras, 413, 2184

\bibitem[{{Bergemann} {et~al.}(2013){Bergemann}, {Kudritzki}, {W{\"u}rl},
  {Plez}, {Davies}, \& {Gazak}}]{Bergemann2013}
{Bergemann}, M., {Kudritzki}, R.-P., {W{\"u}rl}, M., {et~al.} 2013, \apj, 764,
  115

\bibitem[{{Bergemann} {et~al.}(2014){Bergemann}, {Ruchti}, {Serenelli},
  {Feltzing}, {Alvez-Brito}, {Asplund}, {Bensby}, {Gruiters}, {Heiter}, {Korn},
  {Lind}, {Marino}, {Jofre}, {Nordlander}, {Ryde}, {Gilmore}, {Randich},
  {Ferguson}, {Jeffries}, {Micela}, {Negueruela}, {Prusti}, {Rix}, {Vallenari},
  {Alfaro}, {Allende Prieto}, {Bragaglia}, {Koposov}, {Pancino},
  {Recio-Blanco}, {Smiljanic}, {Walton}, {Costado}, {Franciosini}, {Hill},
  {Lardo}, {de Laverny}, {Magrini}, {Maiorca}, {Masseron}, {Morbidelli},
  {Sacco}, {Kordopatis}, \& {Tautvaisiene}}]{Bergemann2014}
{Bergemann}, M., {Ruchti}, G., {Serenelli}, A., {et~al.} 2014, arXiv:1401.4437

\bibitem[{{Bilir} {et~al.}(2012){Bilir}, {Karaali}, {Ak}, {{\"O}nal}, {Da{\v
  g}tekin}, {Yontan}, {Gilmore}, \& {Seabroke}}]{Bilir2012}
{Bilir}, S., {Karaali}, S., {Ak}, S., {et~al.} 2012, \mnras, 421, 3362

\bibitem[{{Boeche} {et~al.}(2013){Boeche}, {Siebert}, {Piffl}, {Just},
  {Steinmetz}, {Sharma}, {Kordopatis}, {Gilmore}, {Chiappini}, {Williams},
  {Grebel}, {Bland-Hawthorn}, {Gibson}, {Munari}, {Siviero}, {Bienaym{\'e}},
  {Navarro}, {Parker}, {Reid}, {Seabroke}, {Watson}, {Wyse}, \&
  {Zwitter}}]{Boeche2013}
{Boeche}, C., {Siebert}, A., {Piffl}, T., {et~al.} 2013, \aap, 559, A59

\bibitem[{{Bournaud} {et~al.}(2009){Bournaud}, {Elmegreen}, \&
  {Martig}}]{Bournaud2009}
{Bournaud}, F., {Elmegreen}, B.~G., \& {Martig}, M. 2009, \apjl, 707, L1

\bibitem[{{Bovy} {et~al.}(2012{\natexlab{a}}){Bovy}, {Rix}, \&
  {Hogg}}]{Bovy2012b}
{Bovy}, J., {Rix}, H.-W., \& {Hogg}, D.~W. 2012{\natexlab{a}}, \apj, 751, 131

\bibitem[{{Bovy} {et~al.}(2012{\natexlab{b}}){Bovy}, {Rix}, \&
  {Hogg}}]{Bovy2012a}
{Bovy}, J., {Rix}, H.-W., \& {Hogg}, D.~W. 2012{\natexlab{b}}, \apj, 751, 131

\bibitem[{{Bovy} {et~al.}(2012{\natexlab{c}}){Bovy}, {Rix}, {Hogg}, {Beers},
  {Lee}, \& {Zhang}}]{Bovy2012c}
{Bovy}, J., {Rix}, H.-W., {Hogg}, D.~W., {et~al.} 2012{\natexlab{c}}, \apj,
  755, 115

\bibitem[{{Brook} {et~al.}(2004){Brook}, {Kawata}, {Gibson}, \&
  {Freeman}}]{Brook2004}
{Brook}, C.~B., {Kawata}, D., {Gibson}, B.~K., \& {Freeman}, K.~C. 2004, \apj,
  612, 894

\bibitem[{{Cabrera-Lavers} {et~al.}(2005){Cabrera-Lavers}, {Garz{\'o}n}, \&
  {Hammersley}}]{CLavers2005}
{Cabrera-Lavers}, A., {Garz{\'o}n}, F., \& {Hammersley}, P.~L. 2005, \aap, 433,
  173

\bibitem[{{Carretta} {et~al.}(2013){Carretta}, {Gratton}, {Bragaglia},
  {D'Orazi}, \& {Lucatello}}]{Carretta2013}
{Carretta}, E., {Gratton}, R.~G., {Bragaglia}, A., {D'Orazi}, V., \&
  {Lucatello}, S. 2013, \aap, 550, A34

\bibitem[{{Casetti-Dinescu} {et~al.}(2011){Casetti-Dinescu}, {Girard},
  {Korchagin}, \& {van Altena}}]{Casetti-Dinescu2011}
{Casetti-Dinescu}, D.~I., {Girard}, T.~M., {Korchagin}, V.~I., \& {van Altena},
  W.~F. 2011, \apj, 728, 7

\bibitem[{{Cescutti} {et~al.}(2007){Cescutti}, {Matteucci}, {Fran{\c c}ois}, \&
  {Chiappini}}]{Cescutti2007}
{Cescutti}, G., {Matteucci}, F., {Fran{\c c}ois}, P., \& {Chiappini}, C. 2007,
  \aap, 462, 943

\bibitem[{{Chen} {et~al.}(2000){Chen}, {Nissen}, {Zhao}, {Zhang}, \&
  {Benoni}}]{Chen2000}
{Chen}, Y.~Q., {Nissen}, P.~E., {Zhao}, G., {Zhang}, H.~W., \& {Benoni}, T.
  2000, \aaps, 141, 491

\bibitem[{{Cheng} {et~al.}(2012{\natexlab{a}}){Cheng}, {Rockosi}, {Morrison},
  {Lee}, {Beers}, {Bizyaev}, {Harding}, {Malanushenko}, {Malanushenko},
  {Oravetz}, {Pan}, {Schlesinger}, {Schneider}, {Simmons}, \&
  {Weaver}}]{Cheng2012b}
{Cheng}, J.~Y., {Rockosi}, C.~M., {Morrison}, H.~L., {et~al.}
  2012{\natexlab{a}}, \apj, 752, 51

\bibitem[{{Cheng} {et~al.}(2012{\natexlab{b}}){Cheng}, {Rockosi}, {Morrison},
  {Sch{\"o}nrich}, {Lee}, {Beers}, {Bizyaev}, {Pan}, \&
  {Schneider}}]{Cheng2012a}
{Cheng}, J.~Y., {Rockosi}, C.~M., {Morrison}, H.~L., {et~al.}
  2012{\natexlab{b}}, \apj, 746, 149

\bibitem[{{Chiappini} {et~al.}(1997){Chiappini}, {Matteucci}, \&
  {Gratton}}]{Chiappini1997}
{Chiappini}, C., {Matteucci}, F., \& {Gratton}, R. 1997, \apj, 477, 765

\bibitem[{{Co{\c s}kuno{\v g}lu} {et~al.}(2012){Co{\c s}kuno{\v g}lu}, {Ak},
  {Bilir}, {Karaali}, {{\"O}nal}, {Yaz}, {Gilmore}, \&
  {Seabroke}}]{Coskunoglu2012}
{Co{\c s}kuno{\v g}lu}, B., {Ak}, S., {Bilir}, S., {et~al.} 2012, \mnras, 419,
  2844

\bibitem[{{Edvardsson} {et~al.}(1993){Edvardsson}, {Andersen}, {Gustafsson},
  {Lambert}, {Nissen}, \& {Tomkin}}]{Edvardsson1993}
{Edvardsson}, B., {Andersen}, J., {Gustafsson}, B., {et~al.} 1993, \aap, 275,
  101

\bibitem[{{Frinchaboy} {et~al.}(2013){Frinchaboy}, {Thompson}, {Jackson},
  {O'Connell}, {Meyer}, {Zasowski}, {Majewski}, {Chojnowksi}, {Johnson},
  {Allende Prieto}, {Beers}, {Bizyaev}, {Brewington}, {Cunha}, {Ebelke}, {Elia
  Garc{\'{\i}}a P{\'e}rez}, {Hearty}, {Holtzman}, {Kinemuchi}, {Malanushenko},
  {Malanushenko}, {Marchante}, {M{\'e}sz{\'a}ros}, {Muna}, {Nidever},
  {Oravetz}, {Pan}, {Schiavon}, {Schneider}, {Shetrone}, {Simmons}, {Snedden},
  {Smith}, \& {Wilson}}]{Frinchaboy2013}
{Frinchaboy}, P.~M., {Thompson}, B., {Jackson}, K.~M., {et~al.} 2013, \apjl,
  777, L1

\bibitem[{{Fuhrmann}(1998)}]{Fuhrmann1998}
{Fuhrmann}, K. 1998, \aap, 338, 161

\bibitem[{{Fuhrmann}(2004)}]{Fuhrmann2004}
{Fuhrmann}, K. 2004, Astronomische Nachrichten, 325, 3

\bibitem[{{Fuhrmann}(2011)}]{Fuhrmann2011}
{Fuhrmann}, K. 2011, \mnras, 414, 2893

\bibitem[{{Gehren} {et~al.}(2004){Gehren}, {Liang}, {Shi}, {Zhang}, \&
  {Zhao}}]{Gehren2004}
{Gehren}, T., {Liang}, Y.~C., {Shi}, J.~R., {Zhang}, H.~W., \& {Zhao}, G. 2004,
  \aap, 413, 1045

\bibitem[{{Gehren} {et~al.}(2006){Gehren}, {Shi}, {Zhang}, {Zhao}, \&
  {Korn}}]{Gehren2006}
{Gehren}, T., {Shi}, J.~R., {Zhang}, H.~W., {Zhao}, G., \& {Korn}, A.~J. 2006,
  \aap, 451, 1065

\bibitem[{{Gilmore} {et~al.}(2012){Gilmore}, {Randich}, {Asplund}, {Binney},
  {Bonifacio}, {Drew}, {Feltzing}, {Ferguson}, {Jeffries}, {Micela},
  {Negueruela}, {Prusti}, {Rix}, {Vallenari}, {Alfaro}, {Allende-Prieto},
  {Babusiaux}, {Bensby}, {Blomme}, {Bragaglia}, {Flaccomio}, {Fran{\c c}ois},
  {Irwin}, {Koposov}, {Korn}, {Lanzafame}, {Pancino}, {Paunzen},
  {Recio-Blanco}, {Sacco}, {Smiljanic}, {Van Eck}, \& {Walton}}]{Gilmore2012}
{Gilmore}, G., {Randich}, S., {Asplund}, M., {et~al.} 2012, The Messenger, 147,
  25

\bibitem[{{Gilmore} \& {Reid}(1983)}]{Gilmore1983}
{Gilmore}, G. \& {Reid}, N. 1983, \mnras, 202, 1025

\bibitem[{{Gratton} {et~al.}(2003{\natexlab{a}}){Gratton}, {Carretta},
  {Claudi}, {Lucatello}, \& {Barbieri}}]{Gratton2003a}
{Gratton}, R.~G., {Carretta}, E., {Claudi}, R., {Lucatello}, S., \& {Barbieri},
  M. 2003{\natexlab{a}}, \aap, 404, 187

\bibitem[{{Gratton} {et~al.}(2003{\natexlab{b}}){Gratton}, {Carretta},
  {Desidera}, {Lucatello}, {Mazzei}, \& {Barbieri}}]{Gratton2003b}
{Gratton}, R.~G., {Carretta}, E., {Desidera}, S., {et~al.} 2003{\natexlab{b}},
  \aap, 406, 131

\bibitem[{{Gratton} {et~al.}(2000){Gratton}, {Carretta}, {Matteucci}, \&
  {Sneden}}]{Gratton2000}
{Gratton}, R.~G., {Carretta}, E., {Matteucci}, F., \& {Sneden}, C. 2000, \aap,
  358, 671

\bibitem[{{Grevesse} {et~al.}(2007){Grevesse}, {Asplund}, \&
  {Sauval}}]{Grevesse2007}
{Grevesse}, N., {Asplund}, M., \& {Sauval}, A.~J. 2007, \ssr, 130, 105

\bibitem[{{Gustafsson} {et~al.}(2008){Gustafsson}, {Edvardsson}, {Eriksson},
  {J{\o}rgensen}, {Nordlund}, \& {Plez}}]{Gustafsson2008}
{Gustafsson}, B., {Edvardsson}, B., {Eriksson}, K., {et~al.} 2008, \aap, 486,
  951

\bibitem[{{Hayden} {et~al.}(2013){Hayden}, {Holtzman}, {Bovy}, {Majewski},
  {Allende Prieto}, {Beers}, {Cunha}, {Frinchaboy}, {Garc{\'{\i}}a P{\'e}rez},
  {Girardi}, {Hearty}, {Johnson}, {Lee}, {Nidever}, {Schiavon}, {Schlesinger},
  {Schneider}, {Schultheis}, {Shetrone}, {Smith}, {Zasowski}, {Bizyaev},
  {Feuillet}, {Hasselquist}, {Kinemuchi}, {Malanushenko}, {Malanushenko},
  {O'Connell}, {Pan}, \& {Stassun}}]{Hayden2013}
{Hayden}, M.~R., {Holtzman}, J.~A., {Bovy}, J., {et~al.} 2013, arXiv:1311.4569

\bibitem[{{Haywood} {et~al.}(2013){Haywood}, {Di Matteo}, {Lehnert}, {Katz}, \&
  {G{\'o}mez}}]{Haywood2013}
{Haywood}, M., {Di Matteo}, P., {Lehnert}, M.~D., {Katz}, D., \& {G{\'o}mez},
  A. 2013, \aap, 560, A109

\bibitem[{{Heiter} {et~al.}(2014){Heiter}, {Soubiran}, {Netopil}, \&
  {Paunzen}}]{Heiter2014}
{Heiter}, U., {Soubiran}, C., {Netopil}, M., \& {Paunzen}, E. 2014, \aap, 561,
  A93

\bibitem[{{Ishigaki} {et~al.}(2013){Ishigaki}, {Aoki}, \&
  {Chiba}}]{Ishigaki2013}
{Ishigaki}, M.~N., {Aoki}, W., \& {Chiba}, M. 2013, \apj, 771, 67

\bibitem[{{Ivezi{\'c}} {et~al.}(2008){Ivezi{\'c}}, {Sesar}, {Juri{\'c}},
  {Bond}, {Dalcanton}, {Rockosi}, {Yanny}, {Newberg}, {Beers}, {Allende
  Prieto}, {Wilhelm}, {Lee}, {Sivarani}, {Norris}, {Bailer-Jones}, {Re
  Fiorentin}, {Schlegel}, {Uomoto}, {Lupton}, {Knapp}, {Gunn}, {Covey},
  {Smith}, {Miknaitis}, {Doi}, {Tanaka}, {Fukugita}, {Kent}, {Finkbeiner},
  {Munn}, {Pier}, {Quinn}, {Hawley}, {Anderson}, {Kiuchi}, {Chen}, {Bushong},
  {Sohi}, {Haggard}, {Kimball}, {Barentine}, {Brewington}, {Harvanek},
  {Kleinman}, {Krzesinski}, {Long}, {Nitta}, {Snedden}, {Lee}, {Harris},
  {Brinkmann}, {Schneider}, \& {York}}]{Ivezic2008}
{Ivezi{\'c}}, {\v Z}., {Sesar}, B., {Juri{\'c}}, M., {et~al.} 2008, \apj, 684,
  287

\bibitem[{{Jacobson} {et~al.}(2009){Jacobson}, {Friel}, \&
  {Pilachowski}}]{Jacobson2009}
{Jacobson}, H.~R., {Friel}, E.~D., \& {Pilachowski}, C.~A. 2009, \aj, 137, 4753

\bibitem[{{Jofr{\'e}} {et~al.}(2014){Jofr{\'e}}, {Heiter}, {Soubiran},
  {Blanco-Cuaresma}, {Worley}, {Pancino}, {Cantat-Gaudin}, {Magrini},
  {Bergemann}, {Gonz{\'a}lez Hern{\'a}ndez}, {Hill}, {Lardo}, {de Laverny},
  {Lind}, {Masseron}, {Montes}, {Mucciarelli}, {Nordlander}, {Recio Blanco},
  {Sobeck}, {Sordo}, {Sousa}, {Tabernero}, {Vallenari}, \& {Van
  Eck}}]{Jofre2014}
{Jofr{\'e}}, P., {Heiter}, U., {Soubiran}, C., {et~al.} 2014, \aap, 564, A133

\bibitem[{{Juri{\'c}} {et~al.}(2008){Juri{\'c}}, {Ivezi{\'c}}, {Brooks},
  {Lupton}, {Schlegel}, {Finkbeiner}, {Padmanabhan}, {Bond}, {Sesar},
  {Rockosi}, {Knapp}, {Gunn}, {Sumi}, {Schneider}, {Barentine}, {Brewington},
  {Brinkmann}, {Fukugita}, {Harvanek}, {Kleinman}, {Krzesinski}, {Long},
  {Neilsen}, {Nitta}, {Snedden}, \& {York}}]{Juric2008}
{Juri{\'c}}, M., {Ivezi{\'c}}, {\v Z}., {Brooks}, A., {et~al.} 2008, \apj, 673,
  864

\bibitem[{{Katz} {et~al.}(2011){Katz}, {Soubiran}, {Cayrel}, {Barbuy}, {Friel},
  {Bienaym{\'e}}, \& {Perrin}}]{Katz2011}
{Katz}, D., {Soubiran}, C., {Cayrel}, R., {et~al.} 2011, \aap, 525, A90

\bibitem[{{Kazantzidis} {et~al.}(2008){Kazantzidis}, {Bullock}, {Zentner},
  {Kravtsov}, \& {Moustakas}}]{Kazantzidis2008}
{Kazantzidis}, S., {Bullock}, J.~S., {Zentner}, A.~R., {Kravtsov}, A.~V., \&
  {Moustakas}, L.~A. 2008, \apj, 688, 254

\bibitem[{{Kordopatis} {et~al.}(2013){Kordopatis}, {Hill}, {Irwin}, {Gilmore},
  {Wyse}, {Tolstoy}, {de Laverny}, {Recio-Blanco}, {Battaglia}, \&
  {Starkenburg}}]{Kordopatis2013}
{Kordopatis}, G., {Hill}, V., {Irwin}, M., {et~al.} 2013, \aap, 555, A12

\bibitem[{{Kordopatis} {et~al.}(2011){Kordopatis}, {Recio-Blanco}, {de
  Laverny}, {Gilmore}, {Hill}, {Wyse}, {Helmi}, {Bijaoui}, {Zoccali}, \&
  {Bienaym{\'e}}}]{Kordopatis2011}
{Kordopatis}, G., {Recio-Blanco}, A., {de Laverny}, P., {et~al.} 2011, \aap,
  535, A107

\bibitem[{{Lee} {et~al.}(2011){Lee}, {Beers}, {An}, {Ivezi{\'c}}, {Just},
  {Rockosi}, {Morrison}, {Johnson}, {Sch{\"o}nrich}, {Bird}, {Yanny},
  {Harding}, \& {Rocha-Pinto}}]{Lee2011}
{Lee}, Y.~S., {Beers}, T.~C., {An}, D., {et~al.} 2011, \apj, 738, 187

\bibitem[{{Lemasle} {et~al.}(2013){Lemasle}, {Fran{\c c}ois}, {Genovali},
  {Kovtyukh}, {Bono}, {Inno}, {Laney}, {Kaper}, {Bergemann}, {Fabrizio},
  {Matsunaga}, {Pedicelli}, {Primas}, \& {Romaniello}}]{Lemasle2013}
{Lemasle}, B., {Fran{\c c}ois}, P., {Genovali}, K., {et~al.} 2013, \aap, 558,
  A31

\bibitem[{{Lemasle} {et~al.}(2008){Lemasle}, {Fran{\c c}ois}, {Piersimoni},
  {Pedicelli}, {Bono}, {Laney}, {Primas}, \& {Romaniello}}]{Lemasle2008}
{Lemasle}, B., {Fran{\c c}ois}, P., {Piersimoni}, A., {et~al.} 2008, \aap, 490,
  613

\bibitem[{{Letarte} {et~al.}(2006){Letarte}, {Hill}, {Jablonka}, {Tolstoy},
  {Fran{\c c}ois}, \& {Meylan}}]{Letarte2006}
{Letarte}, B., {Hill}, V., {Jablonka}, P., {et~al.} 2006, \aap, 453, 547

\bibitem[{{Luck} \& {Lambert}(2011)}]{Luck2011}
{Luck}, R.~E. \& {Lambert}, D.~L. 2011, \aj, 142, 136

\bibitem[{{Maciel} \& {Costa}(2010)}]{Maciel2010}
{Maciel}, W.~J. \& {Costa}, R.~D.~D. 2010, in IAU Symposium, Vol. 265, IAU
  Symposium, ed. K.~{Cunha}, M.~{Spite}, \& B.~{Barbuy}, 317--324

\bibitem[{{Magrini} {et~al.}(2014){Magrini}, {Randich}, {Romano}, {Friel},
  {Bragaglia}, {Smiljanic}, {Jacobson}, {Vallenari}, {Tosi}, {Spina}, {Donati},
  {Maiorca}, {Cantat-Gaudin}, {Sordo}, {Bergemann}, {Damiani}, {Tautvai{\v
  s}ien*error*{\.e}}, {Blanco-Cuaresma}, {Jim{\'e}nez-Esteban}, {Geisler},
  {Mowlavi}, {Munoz}, {San Roman}, {Soubiran}, {Villanova}, {Zaggia},
  {Gilmore}, {Asplund}, {Feltzing}, {Jeffries}, {Bensby}, {Koposov}, {Korn},
  {Flaccomio}, {Pancino}, {Recio-Blanco}, {Sacco}, {Costado}, {Franciosini},
  {Jofre}, {de Laverny}, {Hill}, {Heiter}, {Hourihane}, {Jackson}, {Lardo},
  {Morbidelli}, {Lewis}, {Lind}, {Masseron}, {Prisinzano}, \&
  {Worley}}]{Magrini2014}
{Magrini}, L., {Randich}, S., {Romano}, D., {et~al.} 2014, \aap, 563, A44

\bibitem[{{Magrini} {et~al.}(2009){Magrini}, {Sestito}, {Randich}, \&
  {Galli}}]{Magrini2009}
{Magrini}, L., {Sestito}, P., {Randich}, S., \& {Galli}, D. 2009, \aap, 494, 95

\bibitem[{{Mashonkina} {et~al.}(2012){Mashonkina}, {Ryabtsev}, \&
  {Frebel}}]{Mashonkina2012}
{Mashonkina}, L., {Ryabtsev}, A., \& {Frebel}, A. 2012, \aap, 540, A98

\bibitem[{{Mashonkina} {et~al.}(2007){Mashonkina}, {Vinogradova}, {Ptitsyn},
  {Khokhlova}, \& {Chernetsova}}]{Mashonkina2007}
{Mashonkina}, L.~I., {Vinogradova}, A.~B., {Ptitsyn}, D.~A., {Khokhlova},
  V.~S., \& {Chernetsova}, T.~A. 2007, Astronomy Reports, 51, 903

\bibitem[{{Matteucci} \& {Francois}(1989)}]{Matteucci1989}
{Matteucci}, F. \& {Francois}, P. 1989, \mnras, 239, 885

\bibitem[{{McWilliam}(1997)}]{McWilliam1997}
{McWilliam}, A. 1997, \araa, 35, 503

\bibitem[{{Minchev} {et~al.}(2013){Minchev}, {Chiappini}, \&
  {Martig}}]{Minchev2013}
{Minchev}, I., {Chiappini}, C., \& {Martig}, M. 2013, \aap, 558, A9

\bibitem[{{Minchev} {et~al.}(2014){Minchev}, {Chiappini}, \&
  {Martig}}]{Minchev2014}
{Minchev}, I., {Chiappini}, C., \& {Martig}, M. 2014, arXiv:1401.5796

\bibitem[{{Minchev} \& {Famaey}(2010)}]{Minchev2010}
{Minchev}, I. \& {Famaey}, B. 2010, \apj, 722, 112

\bibitem[{{Neves} {et~al.}(2009){Neves}, {Santos}, {Sousa}, {Correia}, \&
  {Israelian}}]{Neves2009}
{Neves}, V., {Santos}, N.~C., {Sousa}, S.~G., {Correia}, A.~C.~M., \&
  {Israelian}, G. 2009, \aap, 497, 563

\bibitem[{{Nissen} \& {Schuster}(1997)}]{Nissen1997}
{Nissen}, P.~E. \& {Schuster}, W.~J. 1997, \aap, 326, 751

\bibitem[{{Nissen} \& {Schuster}(2010)}]{Nissen2010}
{Nissen}, P.~E. \& {Schuster}, W.~J. 2010, \aap, 511, L10

\bibitem[{{Nissen} \& {Schuster}(2011)}]{Nissen2011}
{Nissen}, P.~E. \& {Schuster}, W.~J. 2011, \aap, 530, A15

\bibitem[{{Nordstr{\"o}m} {et~al.}(2004){Nordstr{\"o}m}, {Andersen},
  {Holmberg}, {J{\o}rgensen}, {Mayor}, \& {Pont}}]{Nordstrom2004}
{Nordstr{\"o}m}, B., {Andersen}, J., {Holmberg}, J., {et~al.} 2004, \pasa, 21,
  129

\bibitem[{{Pancino} {et~al.}(2010){Pancino}, {Carrera}, {Rossetti}, \&
  {Gallart}}]{Pancino2010}
{Pancino}, E., {Carrera}, R., {Rossetti}, E., \& {Gallart}, C. 2010, \aap, 511,
  A56

\bibitem[{{Pedicelli} {et~al.}(2009){Pedicelli}, {Bono}, {Lemasle}, {Fran{\c
  c}ois}, {Groenewegen}, {Lub}, {Pel}, {Laney}, {Piersimoni}, {Romaniello},
  {Buonanno}, {Caputo}, {Cassisi}, {Castelli}, {Leurini}, {Pietrinferni},
  {Primas}, \& {Pritchard}}]{Pedicelli2009}
{Pedicelli}, S., {Bono}, G., {Lemasle}, B., {et~al.} 2009, \aap, 504, 81

\bibitem[{{Pilkington} {et~al.}(2012){Pilkington}, {Few}, {Gibson}, {Calura},
  {Michel-Dansac}, {Thacker}, {Moll{\'a}}, {Matteucci}, {Rahimi}, {Kawata},
  {Kobayashi}, {Brook}, {Stinson}, {Couchman}, {Bailin}, \&
  {Wadsley}}]{Pilkington2012}
{Pilkington}, K., {Few}, C.~G., {Gibson}, B.~K., {et~al.} 2012, \aap, 540, A56

\bibitem[{{Prochaska} {et~al.}(2000){Prochaska}, {Naumov}, {Carney},
  {McWilliam}, \& {Wolfe}}]{Prochaska2000}
{Prochaska}, J.~X., {Naumov}, S.~O., {Carney}, B.~W., {McWilliam}, A., \&
  {Wolfe}, A.~M. 2000, \aj, 120, 2513

\bibitem[{{Qu} {et~al.}(2011){Qu}, {Di Matteo}, {Lehnert}, \& {van
  Driel}}]{Qu2011}
{Qu}, Y., {Di Matteo}, P., {Lehnert}, M.~D., \& {van Driel}, W. 2011, \aap,
  530, A10

\bibitem[{{Quinn} {et~al.}(1993){Quinn}, {Hernquist}, \&
  {Fullagar}}]{Quinn1993}
{Quinn}, P.~J., {Hernquist}, L., \& {Fullagar}, D.~P. 1993, \apj, 403, 74

\bibitem[{{Ram{\'{\i}}rez} {et~al.}(2013){Ram{\'{\i}}rez}, {Allende Prieto}, \&
  {Lambert}}]{Ramirez2013}
{Ram{\'{\i}}rez}, I., {Allende Prieto}, C., \& {Lambert}, D.~L. 2013, \apj,
  764, 78

\bibitem[{{Ram{\'{\i}}rez} {et~al.}(2012){Ram{\'{\i}}rez}, {Mel{\'e}ndez}, \&
  {Chanam{\'e}}}]{Ramirez2012}
{Ram{\'{\i}}rez}, I., {Mel{\'e}ndez}, J., \& {Chanam{\'e}}, J. 2012, \apj, 757,
  164

\bibitem[{{Randich} {et~al.}(2013){Randich}, {Gilmore}, \& {Gaia-ESO
  Consortium}}]{Randich2013}
{Randich}, S., {Gilmore}, G., \& {Gaia-ESO Consortium}. 2013, The Messenger,
  154, 47

\bibitem[{{Recio-Blanco} {et~al.}(2014{\natexlab{a}}){Recio-Blanco}, {Allende
  Prieto}, \& et~al.}]{Blanco2014b}
{Recio-Blanco}, A., {Allende Prieto}, C., \& et~al. 2014{\natexlab{a}}, \aap

\bibitem[{{Recio-Blanco} {et~al.}(2006){Recio-Blanco}, {Bijaoui}, \& {de
  Laverny}}]{Blanco2006}
{Recio-Blanco}, A., {Bijaoui}, A., \& {de Laverny}, P. 2006, \mnras, 370, 141

\bibitem[{{Recio-Blanco} {et~al.}(2014{\natexlab{b}}){Recio-Blanco}, {de
  Laverny}, {Kordopatis}, {Helmi}, {Hill}, {Gilmore}, {Wyse}, {Adibekyan},
  {Randich}, {Asplund}, {Feltzing}, {Jeffries}, {Micela}, {Vallenari},
  {Alfaro}, {Allende Prieto}, {Bensby}, {Bragaglia}, {Flaccomio}, {Koposov},
  {Korn}, {Lanzafame}, {Pancino}, {Smiljanic}, {Jackson}, {Lewis}, {Magrini},
  {Morbidelli}, {Prisinzano}, {Sacco}, {Worley}, {Hourihane}, {Bergemann},
  {Costado}, {Heiter}, {Joffre}, {Lardo}, {Lind}, \& {Maiorca}}]{Blanco2014a}
{Recio-Blanco}, A., {de Laverny}, P., {Kordopatis}, G., {et~al.}
  2014{\natexlab{b}}, \aap, 567, A5

\bibitem[{{Ro{\v s}kar} {et~al.}(2008){Ro{\v s}kar}, {Debattista}, {Quinn},
  {Stinson}, \& {Wadsley}}]{Roskar2008}
{Ro{\v s}kar}, R., {Debattista}, V.~P., {Quinn}, T.~R., {Stinson}, G.~S., \&
  {Wadsley}, J. 2008, \apjl, 684, L79

\bibitem[{{Ruchti} {et~al.}(2011){Ruchti}, {Fulbright}, {Wyse}, {Gilmore},
  {Bienaym{\'e}}, {Bland-Hawthorn}, {Gibson}, {Grebel}, {Helmi}, {Munari},
  {Navarro}, {Parker}, {Reid}, {Seabroke}, {Siebert}, {Siviero}, {Steinmetz},
  {Watson}, {Williams}, \& {Zwitter}}]{Ruchti2011}
{Ruchti}, G.~R., {Fulbright}, J.~P., {Wyse}, R.~F.~G., {et~al.} 2011, \apj,
  737, 9

\bibitem[{{Schlesinger} {et~al.}(2012){Schlesinger}, {Johnson}, {Rockosi},
  {Lee}, {Morrison}, {Sch{\"o}nrich}, {Allende Prieto}, {Beers}, {Yanny},
  {Harding}, {Schneider}, {Chiappini}, {da Costa}, {Maia}, {Minchev},
  {Rocha-Pinto}, \& {Santiago}}]{Schlesinger2012}
{Schlesinger}, K.~J., {Johnson}, J.~A., {Rockosi}, C.~M., {et~al.} 2012, \apj,
  761, 160

\bibitem[{{Sch{\"o}nrich} \& {Binney}(2009{\natexlab{a}})}]{Schonrich2009a}
{Sch{\"o}nrich}, R. \& {Binney}, J. 2009{\natexlab{a}}, \mnras, 396, 203

\bibitem[{{Sch{\"o}nrich} \& {Binney}(2009{\natexlab{b}})}]{Schonrich2009b}
{Sch{\"o}nrich}, R. \& {Binney}, J. 2009{\natexlab{b}}, \mnras, 399, 1145

\bibitem[{{Sestito} {et~al.}(2008){Sestito}, {Bragaglia}, {Randich},
  {Pallavicini}, {Andrievsky}, \& {Korotin}}]{Sestito2008}
{Sestito}, P., {Bragaglia}, A., {Randich}, S., {et~al.} 2008, \aap, 488, 943

\bibitem[{{Smiljanic} {et~al.}(2014){Smiljanic}, {Korn}, \&
  et~al.}]{Smiljanic2014}
{Smiljanic}, R., {Korn}, A., \& et~al. 2014, submitted

\bibitem[{{Smiljanic} {et~al.}(2009){Smiljanic}, {Pasquini}, {Bonifacio},
  {Galli}, {Gratton}, {Randich}, \& {Wolff}}]{Smiljanic2009}
{Smiljanic}, R., {Pasquini}, L., {Bonifacio}, P., {et~al.} 2009, \aap, 499, 103

\bibitem[{{Steinmetz}(2012)}]{Steinmetz2012}
{Steinmetz}, M. 2012, Astronomische Nachrichten, 333, 523

\bibitem[{{Tan} \& {Zhao}(2011)}]{Tan2011}
{Tan}, K. \& {Zhao}, G. 2011, \apjl, 738, L33

\bibitem[{{Tolstoy} {et~al.}(2009){Tolstoy}, {Hill}, \& {Tosi}}]{Tolstoy2009}
{Tolstoy}, E., {Hill}, V., \& {Tosi}, M. 2009, \araa, 47, 371

\bibitem[{{Valenti} \& {Piskunov}(1996)}]{Valenti1996}
{Valenti}, J.~A. \& {Piskunov}, N. 1996, \aaps, 118, 595

\bibitem[{{Van der Swaelmen} {et~al.}(2013){Van der Swaelmen}, {Hill},
  {Primas}, \& {Cole}}]{Swaelmen2013}
{Van der Swaelmen}, M., {Hill}, V., {Primas}, F., \& {Cole}, A.~A. 2013, \aap,
  560, A44

\bibitem[{{Venn} {et~al.}(2004){Venn}, {Irwin}, {Shetrone}, {Tout}, {Hill}, \&
  {Tolstoy}}]{Venn2004}
{Venn}, K.~A., {Irwin}, M., {Shetrone}, M.~D., {et~al.} 2004, \aj, 128, 1177

\bibitem[{{Villalobos} \& {Helmi}(2008)}]{Villalobos2008}
{Villalobos}, {\'A}. \& {Helmi}, A. 2008, \mnras, 391, 1806

\bibitem[{{Yanny} {et~al.}(2009){Yanny}, {Rockosi}, {Newberg}, {Knapp},
  {Adelman-McCarthy}, {Alcorn}, {Allam}, {Allende Prieto}, {An}, {Anderson},
  {Anderson}, {Bailer-Jones}, {Bastian}, {Beers}, {Bell}, {Belokurov},
  {Bizyaev}, {Blythe}, {Bochanski}, {Boroski}, {Brinchmann}, {Brinkmann},
  {Brewington}, {Carey}, {Cudworth}, {Evans}, {Evans}, {Gates}, {G{\"a}nsicke},
  {Gillespie}, {Gilmore}, {Nebot Gomez-Moran}, {Grebel}, {Greenwell}, {Gunn},
  {Jordan}, {Jordan}, {Harding}, {Harris}, {Hendry}, {Holder}, {Ivans},
  {Ivezi{\v c}}, {Jester}, {Johnson}, {Kent}, {Kleinman}, {Kniazev},
  {Krzesinski}, {Kron}, {Kuropatkin}, {Lebedeva}, {Lee}, {French Leger},
  {L{\'e}pine}, {Levine}, {Lin}, {Long}, {Loomis}, {Lupton}, {Malanushenko},
  {Malanushenko}, {Margon}, {Martinez-Delgado}, {McGehee}, {Monet}, {Morrison},
  {Munn}, {Neilsen}, {Nitta}, {Norris}, {Oravetz}, {Owen}, {Padmanabhan},
  {Pan}, {Peterson}, {Pier}, {Platson}, {Re Fiorentin}, {Richards}, {Rix},
  {Schlegel}, {Schneider}, {Schreiber}, {Schwope}, {Sibley}, {Simmons},
  {Snedden}, {Allyn Smith}, {Stark}, {Stauffer}, {Steinmetz}, {Stoughton},
  {SubbaRao}, {Szalay}, {Szkody}, {Thakar}, {Sivarani}, {Tucker}, {Uomoto},
  {Vanden Berk}, {Vidrih}, {Wadadekar}, {Watters}, {Wilhelm}, {Wyse}, {Yarger},
  \& {Zucker}}]{Yanny2009}
{Yanny}, B., {Rockosi}, C., {Newberg}, H.~J., {et~al.} 2009, \aj, 137, 4377

\bibitem[{{Yoachim} \& {Dalcanton}(2006)}]{Yoachim2006}
{Yoachim}, P. \& {Dalcanton}, J.~J. 2006, \aj, 131, 226

\bibitem[{{Yong} {et~al.}(2012){Yong}, {Carney}, \& {Friel}}]{Yong2012}
{Yong}, D., {Carney}, B.~W., \& {Friel}, E.~D. 2012, \aj, 144, 95

\bibitem[{{Yong} {et~al.}(2005){Yong}, {Carney}, \& {Teixera de
  Almeida}}]{Yong2005}
{Yong}, D., {Carney}, B.~W., \& {Teixera de Almeida}, M.~L. 2005, \aj, 130, 597

\end{thebibliography}

\clearpage

\appendix

\section{Description of radial and vertical gradients in different subsamples}
\label{sec:appendix1}

To derive our gradients we used the subsamples described in~Sects.~\ref{sec:radial_gradient} and \ref{sec:vertical_gradient}.
To clean the sample to determine radial gradients with as little cross-contamination between the two discs, we made restriction in $|Z|$ ($0.0~{\rm kpc}<|Z|<0.608~{\rm kpc}$ for the thin-disc and $|Z|>1.07~{\rm kpc}$ for the thick-disc), and we did not consider stars whose abundances were too close to our separation function (closer than the typical error in \ion{Mg}{I}), introducing a gap $\pm0.05$~dex around the separation. 
Similarly, for vertical gradients we restricted the sample in $R_{\rm GC}$, using only stars from the Solar cylinder ($7~{\rm kpc}<R_{\rm GC}<9~{\rm kpc}$). For vertical gradients we also used the gap $\pm0.05$~dex around the separation function for the two discs.

We checked that the gradients we derived were not sensitive to these sample selection refinements, and the gradients of recommended [$\alpha$/M], [Mg/M] and [M/H] abundances for the various selections are presented in Tables~\ref{tab:Radial_GRADIENTS_other_samples} and~\ref{tab:Vertical_GRADIENTS_other_samples}. There is little variation of the gradients value upon the exact sample selection, which validates our choice of using the cleanest samples for the final gradients reported in the paper.

\begin{table}
\caption{Radial abundance gradients in thick- and thin-disc stars ($\Delta X/\Delta R_{\rm GC}$)
errors on the slopes ($\epsilon$) and number of stars in the corresponding sub samples (n).
}
%\vspace{0.2cm}
\begin{tabular}{lcccccc}
\hline \hline
\smallskip
Line & $\frac{\Delta X}{\Delta R_{\rm GC}}$ & $\epsilon$ & n & $\frac{\Delta X}{\Delta R_{\rm GC}}$ & $\epsilon$ & n\\
 & $\rm \frac{dex}{kpc}$ & $\rm \frac{dex}{kpc}$  &   & $\rm \frac{dex}{kpc}$ & $\rm \frac{dex}{kpc}$ &  \\
\hline
  & \textit{Thin disc}  &   &   & \textit{Thick disc}  &  &  \\
\hline
\multicolumn{6}{c}{\textit{Main} sample} \\ 
$[\alpha$/M] &	0.006	&	0.001	&	879	&	-0.012	&	0.002	&	935	\\
$[\ion{Mg}{I}$/M] &	0.007	&	0.001	&	879	&	-0.011	&	0.002	&	935	\\
$[{\rm M}$/H] &	-0.041	&	0.005	&	879	&	 0.008	&	0.006	&	935	\\
\hline
\multicolumn{6}{c}{\textit{Main} sample and $|Z|$ restriction} \\ 
$[\alpha$/M] &	0.010	&	0.002	&	484	&	-0.007	&	0.003	&	382	\\
$[\ion{Mg}{I}$/M] &	0.008	&	0.003	&	484	&	-0.005	&	0.003	&	382	\\
$[{\rm M}$/H] &   -0.048	&	0.008	&	484	&	 0.003	&	0.006	&	382	\\
\hline
\multicolumn{6}{c}{\textit{Main} sample and gap restriction} \\ 
$[\alpha$/M] &	0.005	&	0.001	&	540	&	-0.006	&	0.003	&	610	\\
$[\ion{Mg}{I}$/M] &	0.009	&	0.002	&	540	&	-0.007	&	0.002	&	610	\\
$[{\rm M}$/H] &	-0.039	&	0.006	&	540	&	0.000	&	0.007	&	610	\\
\hline
\multicolumn{6}{c}{\textit{Main} sample, $|Z|$ and gap restrictions, dwarfs only} \\ 
$[\alpha$/M] 	&	0.011	&	0.003	&	322	&	-0.005	&	0.004	&	208	\\
$[\ion{Mg}{I}$/M] &	0.005	&	0.004	&	322	&	-0.004	&	0.004	&	208	\\
$[{\rm M}$/H] 	&	-0.042	&	0.010	&	322	&	0.005	&	0.011	&	208	\\
\hline
 \label{tab:Radial_GRADIENTS_other_samples}
 
\end{tabular}
\end{table}

\begin{table}
\caption{Vertical abundance gradients in thick- and thin-disc stars ($\Delta X/\Delta |Z|$),
errors on the slopes ($\epsilon$), and number of stars in the corresponding sub samples (n).}
%\vspace{0.2cm}
\begin{tabular}{lcccccc}
\hline \hline
\smallskip
Line & $\frac{\Delta X}{\Delta |Z|}$ & $\epsilon$ & n & $\frac{\Delta X}{\Delta |Z|}$ & $\epsilon$ & n\\
 & $\rm \frac{dex}{kpc}$ & $\rm \frac{dex}{kpc}$  &   & $\rm \frac{dex}{kpc}$ & $\rm \frac{dex}{kpc}$ &  \\
\hline
& \textit{Thin disc}  &   &   & \textit{Thick disc}  &  &  \\
\hline
\multicolumn{6}{c}{\textit{Main} sample} \\ 
$[\alpha$/M] &	0.033	&	0.003	&	879	&	 0.043	&	0.004	&	935	\\
$[\ion{Mg}{I}$/M] &	0.035	&	0.004	&	879	&	 0.047	&	0.003	&	935	\\
$[{\rm M}$/H] &	-0.104	&	0.012	&	879	&	-0.066	&	0.010	&	935	\\
\hline
\multicolumn{6}{c}{\textit{Main} sample and $R_{\rm GC}$ restriction} \\ 
$[\alpha$/M]     &	0.034	&	0.005	&	430	&	 0.044	&	0.006	&	476	\\
$[\ion{Mg}{I}$/M]&	0.040	&	0.008	&	430	&	 0.051	&	0.006	&	476	\\
$[{\rm M}$/H]     &   -0.084	&	0.018	&	430	&	-0.068	&	0.014	&	476	\\
\hline
\multicolumn{6}{c}{\textit{Main} sample and gap restriction} \\ 
$[\alpha$/M] &	0.030	&	0.004	&	539	&	 0.038	&	0.004	&	606	\\
$[\ion{Mg}{I}$/M] &	0.035	&	0.005	&	539	&	 0.043	&	0.003	&	606	\\
$[{\rm M}$/H] &	-0.087	&	0.015	&	539	&	-0.044	&	0.010	&	606	\\
\hline
\multicolumn{6}{c}{\textit{Main} sample, $R_{\rm GC}$ and gap restrictions, dwarfs only} \\ 
$[\alpha$/M] 	&	0.041	&	0.006	&	389	&	 0.037	&	0.007	&	315	\\
$[\ion{Mg}{I}$/M] &	0.041	&	0.006	&	389	&	 0.041	&	0.007	&	315	\\
$[{\rm M}$/H] 	&	-0.082	&	0.020	&	389	&	-0.074	&	0.016	&	315	\\
\hline
% & & Total Sample & 198,000 \\
 \label{tab:Vertical_GRADIENTS_other_samples}

\end{tabular}
\end{table}

\end{document}